%% file: template.tex
\documentclass{article}
\PassOptionsToPackage{colorlinks,citecolor=blue,linkcolor=blue,urlcolor=blue}{hyperref}

\usepackage{arxiv}

\usepackage[utf8]{inputenc} % allow utf-8 input
\usepackage[T1]{fontenc}    % use 8-bit T1 fonts
\usepackage{url}            % simple URL typesetting
\usepackage{booktabs}       % professional-quality tables
\usepackage{amsfonts}       % blackboard math symbols
\usepackage{nicefrac}       % compact symbols for 1/2, etc.
\usepackage{microtype}      % microtypography
\usepackage{lipsum}		% Can be removed after putting your text content
\usepackage{graphicx}
\usepackage{natbib}

\usepackage{doi}

\usepackage{threeparttable}
\usepackage[section]{placeins} 
\usepackage{rotating}
\usepackage{caption}
\usepackage{tabularx}
\usepackage{algorithm2e}
\usepackage{bbm}
\usepackage{subcaption} % Required for subfigures
\newcommand{\PreserveBackslash}[1]{\let\temp=\\#1\let\\=\temp}
\newcolumntype{C}[1]{>{\PreserveBackslash\centering}p{#1}}
\newcolumntype{Y}{>{\centering\arraybackslash}X}
\usepackage{amsmath}
\usepackage{bibunits}   % load in preamble

\usepackage{etoolbox}
\AtBeginEnvironment{figure}{\centering}
\title{Engineering Social Networks: How Initial Group Assignment Shapes Student Social Interactions}

\author{
Johanna Einsiedler \\
    University of Basel; University of Copenhagen \\
	\texttt{johanna.einsiedler@sodas.ku.dk} \\
    \And
Nikolaj Arpe Harmon \\
University of Copenhagen \\
\And
Dadvid Dreyer Lassen \\
University of Copenhage \\
\And 
Andreas Bjerre-Nielsen \\
University of Copenhagen
}

\renewcommand{\shorttitle}{Engineering Social Networks}

\hypersetup{
pdftitle={A template for the arxiv style},
pdfsubject={q-bio.NC, q-bio.QM},
pdfauthor={David S.~Hippocampus, Elias D.~Striatum},
pdfkeywords={First keyword, Second keyword, More},
}

\begin{document}
\maketitle

\begin{abstract}
 A large literature uses exogenous variation to estimate how assignment to classrooms or other groups shapes social networks. Yet most of these analyses remain dyadic, treating each link in isolation, even though ties often form through triadic closure, as a friend of a friend also becomes a friend. Using fine-grained data on phone calls, text messages, physical co-location, and social-media ties, we estimate the network formation effects of randomly assigning first-year university students to classrooms and to smaller social groups. To analyze explicitly whether group assignment interact with triadic closure, we use our random assignment to estimate a subgraph generated model of network formation. Accounting for triadic closure turns out to be crucial. For social groups in particular, group assignment affects network formation almost entirely by inducing additional triadic closure. Estimates ignoring triadic closure can thus yield misleading predictions about the network effects and benefits of group assignment policies.
\end{abstract}

Code and synthetic data are available at: \url{https://github.com/johanna-einsiedler/engineering-social-networks}
\begin{bibunit}[ecta-fullname]
\section{Introduction}
Our lives are profoundly shaped by the social networks we are embedded in, which channel access to opportunities, information, and social capital and thereby affect behavior at both the individual and aggregate level \citep{jackson_social_2008,easley_networks_2010}.
Existing work finds causal social influence of peers and social links on fundamental aspects of human life, including academic achievement \citep{booij_ability_2017, epple_chapter_2011, mehta_time-use_2019},  social behavior \citep{duncan_peer_2005,bursztyn_social_2017}, and occupational choice \citep{marmaros_peer_2002}. As a result, policymakers often seek to intervene in these network formation processes to further various policy goals. %Successfully achieving such goals, however, requires an understanding of how social networks form and how they are affected by interventions. 
Existing causal studies evaluate how organized groups affect whether two people connect, but by treating each potential tie in isolation they say nothing on how groups affect higher-order structure - in particular triadic closure, where shared connections draw people together - even though that structure is what makes networks cohesive.

Social links do not form at random. Researchers have identified three mechanisms behind who connects with whom \citep{rivera2010dynamics}. 
First, social foci---such as communal activities, non-private physical spaces, and social events---provide the opportunity for interaction and relationship building \citep{feld_focused_1981}. 
Second, homophily describes the tendency of individuals to associate with those who are similar in various characteristics \citep{mcpherson_birds_2001}, which in turn shapes social integration and positive spillovers \citep{golub2012homophily,bramoulle2012homophily,bjerre2020assortative}. 
Third, triadic closure posits that two individuals are more likely to form a link if they share at least one mutual connection (thereby forming a triangle) \citep{Watts1998,kossinets_empirical_2006}, 
which plays an essential role in fostering cooperation and trust as well as coordinating behavior \citep{jackson2012social, breza2019social}.
In practice, many policy interventions targeting networks operate by reshaping social foci within people's private, professional or academic lives. Examples of such interventions are the random assignment process of college students to dorm rooms (e.g. \citeauthor{sacerdote_peer_2001}, \citeyear{sacerdote_peer_2001}) or the allocation of high school students to classrooms (e.g. \citeauthor{graham_teacher--classroom_2020}, \citeyear{graham_teacher--classroom_2020}). To design social-foci interventions effectively, it is necessary to  understand not only how changes in social foci causally affect network formation directly, but how it propagates through triadic closure - since the same intervention can build scattered ties or knit dense clusters depending on which channel dominates.

In this paper, we estimate the causal effect of two policy interventions that aim to change social networks in higher education by modifying social foci. When arriving for their first semester at the Technical University of Denmark (DTU), we randomly assigned students within each study program to different academic classrooms for their teaching assistant (TA) sessions, with approximately 30 students in each classroom. Independently, we also randomly assigned the same students to social groups of around seven within each program, formed to welcome them to university. Crucially, we do not only estimate the overall effect of each intervention on the propensity to connect. We separate that effect into two channels: the direct effect of a shared focus on whether a given pair links, and its effect through triadic closure, whereby a shared focus creates the mutual connections that pull further pairs together. To our knowledge, no prior study identifies these two channels separately.

To estimate these effects of interest, we overcome several key challenges that pervade the literature on network formation. First, to separate causal effects from endogenous selection, we leverage the random assignment of the 2013 cohort, which we implemented with DTU's administration and student organizations. Because late arrivals and non-random regrouping meant assignment did not perfectly determine final membership, we measure students' realized groups and instrument them with their randomly assigned groups. Second, to reliably measure social connections, we use rich smartphone-based data from the Copenhagen Network Study \citep{sapiezynski_interaction_2019}. This allows us to measure students' actual social interactions via phone calls, messages, Facebook friendship and time spent together physically. 
Third, separating a direct link from one induced by triadic closure is not just a measurement problem but an identification problem: the two channels are bundled in any observed network. We address this by embedding a subgraph generated model (SUGM) of network formation by \citet{chandrasekhar_network_res} in our randomized design through a control function. %By adapting the SUGM to our setting, we extend our random assignment design to estimate triadic closure effects alongside direct link formation.
This lets us recover the causal effect of shared group membership on direct linking and on triadic closure separately - turning the SUGM, which describes higher-order formation, into a tool for causal identification of it.

We start our analysis by estimating the overall causal effect of changing social foci on students' social interactions in a standard dyadic framework. Our random group assignment procedure generates exogenous variation in whether a pair of students is assigned to the same social group or classroom. We use this assignment to instrument realized shared membership in a two-stage least squares (2SLS) framework. We find strong effects of social groups on interactions. Students assigned to the same social group in their first semester meet 2.9 more times per week in their second semester than students who were not assigned to the same group. Similarly, students in the same social group talk on the phone for 17 more seconds per week, send 1.3 more text messages, and are 4.4 times as likely to be Facebook friends. Moreover, looking at time horizons beyond the second semester reveals that most of the effects are declining but persistent. The effect of classrooms is weaker and limited to fewer dimensions. Students in the same classroom in their first semester have an average weekly call duration that is about two seconds longer, with the effect being significant at the 5\% level. By contrast, the effects on Facebook friendship, weekly SMS count (0.11 higher), and weekly physical meetings are positive but small and not statistically significant at the 10\% level.

Then, we apply principal component analysis to combine data on physical meetings, phone calls, and text messages into a single binary measure of which students have formed links with each other. Based on this, we estimate that two students placed in the same social group are 32 percentage points more likely to form a link (on a baseline of 9\%). For classrooms, the effect is 3 percentage points (on a baseline of 8\%). These estimates capture the total effect on linking. They do not, on their own, reveal whether shared foci build ties directly or by inducing closure - the decomposition we turn to next.

Next, we examine triadic closure effects. First, we adapt the SUGM framework of network formation to leverage random assignment for identification in a similar way as in the dyadic 2SLS framework. In a SUGM, a link can arise two ways. It may form directly between a pair of individuals as a function of their characteristics and circumstances. Or it may be the byproduct of triadic closure when two individuals each link to a third individual. As noted by \citet{chandrasekhar_network_res}, the latter implies that the model is able to capture triadic closure effects, while remaining parsimonious and tractable. We adapt their SUGM model to allow for the standard identification issues around non-random sorting into groups and classrooms and then show how our random group assignment can be used to estimate causal effects using an instrumental variable and a control function approach. In the adapted model, shared membership can affect both whether a pair links directly and whether any triad they belong to closes into a triangle. Results show that social group effects indeed interact strongly with triadic closure. Being in the same social group substantially increases the likelihood that three students form a triangle. The estimated effect on pairwise linking formation, by contrast, is small, statistically insignificant and if anything negative. Together these suggest that shared group membership shapes the networks primarily by drawing clusters of students together rather than strengthening pairwise connectivity - though the direct channel is estimated imprecisely enough that we cannot rule out a modest positive effect. For classrooms, results are not precise enough to reliably disentangle pairwise link and triadic closure effects, however, estimates point primarily to effects on pairwise link formation with triadic closure effects playing a limited role.

Finally, we use the estimated dyadic and SUGMs to simulate counterfactual networks ($N=500$ draws per model) and to quantify how accounting for triadic closure changes substantive conclusions. The simulations show that the two models can produce similar link density while implying very different higher-order structure: relative to the dyadic specification, the SUGM generates substantially more triangles and higher clustering. We then translate these structural differences into welfare differences using a parsimonious utility framework in which students benefit from direct links and from spillovers through mutual friends. When spillover strength is small, the models yield similar welfare predictions; as spillovers become more important, the SUGM predicts higher average utility. 
Taken together, the simulation exercise shows that triadic closure is not only a statistical feature of formation but also a substantively consequential mechanism for evaluating interventions that shift social foci. In particular, policies that change classroom or social-group composition can have effects that operate primarily through the creation (or suppression) of clustered friendship structures---and these effects will be understated if researchers rely on models and estimates that ignore triadic closure.
%Overall, this papers results show... XXX SUMMARY AND IMPLICATIONS

Overall our paper relates and contributes to a large literature on network formation and peer effects. Previous work has documented the importance of social foci---such as spatial proximity and organizational grouping---in driving the formation of social ties. For instance, \cite{marmaros2006friendships} show that dormitory proximity plays a pivotal role in establishing friendships among college students, while \cite{hallinan1985ability} provide evidence that ability grouping influences student interactions. These studies underscore that initial group composition is crucial for the subsequent structure of social networks. Recent advances in measurement have further enriched our understanding of network formation by overcoming systematic error in survey data, caused by e.g. social desirability bias or incomplete knowledge. Mobile phone data, for example, have been used to infer friendship network structure reliably \citep{onnela2007structure, eagle2009inferring, stopczynski2014measuring, stadtfeld2015partnership}. These studies illustrate that modern data sources can capture the dynamics and link strengths within networks, thereby providing a more nuanced view of how early interactions emerge and evolve.

Our approach contributes to the literature on measuring the impact of social foci on how social networks form. We leverage random assignment to groups which allows us to isolate the causal effect of initial social contact on formation of group structures. This design is similar in spirit to the methodology employed by \cite{sacerdote_peer_2001}, who exploit random roommate assignments to uncover peer effects, and to the structural framework advanced by \cite{griffith2024random} for assessing counterfactual treatment effects under non-random peer influences. The main novelty relative to this work is that we study not only the overall propensity to form links, but also the impact on triadic closure.
%Moreover, by drawing on broader network theories \cite{jackson_social_2008, easley_networks_2010}, we conceptualize social foci as key organizational determinants that structure early network formation.

Methodologically, our paper contributes by showing how  subgraph-based network formation models can be combined with random assignment (or other exogenous variation) to credibly establish causal effects. This enables us to synthesize diverse strands of literature---from the micro-foundations of network ties \citep{mcpherson_birds_2001, feld_focused_1981} to the causal identification of peer effects \citep{sacerdote_peer_2001, kremer_peer_2003}---and demonstrates that early, exogenously induced groups have lasting effects on the structure of social networks.

Beyond immediate interactions, a growing literature has emphasized that early peer contacts yield persistent spillovers on academic achievement and socio-economic trajectories. For example, \cite{goette2012impact} document how minimal group settings can influence collective behavior, while \cite{altmejd2021brother} and \cite{barrios2022neighbors}  demonstrate that siblings and neighbors socially influence college major choices across countries. Our findings resonate with these studies by showing that the peer networks forged through random assignment not only affect initial interactions but also have lasting consequences for the structure of social networks.
Our work further points to the importance of non-linear peer effects stemming from the subgroup formation within organizationally assigned groups. 

\section{Data and setting}

%%%

Our analysis focuses on the social networks of new undergraduate students at the Technical University of Denmark (DTU).  DTU offers various undergraduate and graduate educational programs in civil engineering. First-year students at DTU enter one of several bachelor programs or diploma programs, ranging from Biomedical Engineering, to Cyber Technology and Mechanical Engineering. 

Upon joining DTU, new students are assigned to two key social foci by the university administration and student organizations: first, within each study program, DTU administration splits students  into two or more \emph{teaching assistant classroom} groups containing around 30 students each. All teaching assistant (TA) sessions take place within these assigned groups. Second, in collaboration with the administration, an organization of older students groups new students into introductory \emph{social groups}, each consisting of about 7 students from the same study program. Throughout the first semester, these groups meet with a senior ``group mentor'' and conduct various social introduction activities such as e.g. an inaugural weekend trip. Participation in this social group system is voluntary, however, a majority of students sign up to join a social group.\footnote{Out of the 1,372 students that we have data on, we observe 1,069 joining a social group during the first semester.}

Both classroom and introductory social group assignments at DTU are prime examples of the types of social foci typically leveraged for policy interventions. By manipulating assignment of different students to these groups, policy makers may be able to shape the social networks that students form. The aim of our analysis is to understand the causal effects of such group assignments on the properties of the students' social networks.

\subsection{Random assignment in the 2013 cohort}

A key concern when estimating the causal effects of assignment to classrooms or social groups is the possibility of endogenous selection. Even when assignment is formally done by administrators or student organizations, students with certain unobservable characteristics can often be more (or less) likely to end up in the same group. The direct effect of such unobservables can bias the estimated effects of group assignment.

To overcome this challenge, our analysis exploits random assignment to groups. In mid-August 2013, a few weeks prior to the start of the semester, we assisted the administration and student groups at DTU in randomly assigning incoming students to TA classrooms and introductory social groups. In both cases the randomization of students was performed conditional on gender to comply with existing rules about the potential gender mix within groups.\footnote{Specifically, the administration and student organization impose certain restriction on the minimum number of female students that should be assigned to a given group or classroom.} For the social groups, randomization was additionally conditioned on whether the student had dietary restrictions, as this was requested by the student organization.\footnote{To ease planning of the various social activities, students with dietary restrictions are always grouped together when forming groups.} We account for this conditioning in our analysis (see Section \ref{sec:overresults}). Balance tests across socio-demographic characteristics show no significant differences between groups (see Supplemental Material, Section A), supporting successful randomization.

Importantly, the final classroom and social group assignment used in practice may differ from our assigned groups for two reasons: First, our randomization was applied to the list of students who had signed up with the DTU administration and/or requested  membership in a social group by mid-August. Late sign-ups were thus not part of our randomization. Second, the student organization specifically reserved the right to make minor adjustments to the social groups if the need arose. We account for both of these sources of imperfect compliance in our analysis.

\subsection{Administrative data}

The basic data set for our analysis was obtained from the university administration and covers students enrolled in the 2013 cohort. In addition to information about which study program the student is enrolled in, this data includes basic demographic information, such as sex and age. We supplement this data with information on the randomized classroom and social group assignment discussed above. Further, we obtained updated data from the university administration in 2024, containing information on the actual classroom and social group that each student was placed in during the academic year.\footnote{The administrative data on classrooms was reported in a different format than the original classroom assignment and contained more detailed information on all courses a student has been enrolled in during their first semester. We thus took a conservative approach and marked every pair of students as being members of the same classroom that shared at least one course. We provide a robustness check using a different definition in the Supplemental Material (Section E).}

\subsection{Measuring networks and social interactions: The Copenhagen Network Study}

To construct a comprehensive and reliable measure of networks, we leverage a unique data set on social interactions generated by the Copenhagen Network Study (CNS) \citep{stopczynski2014measuring,sapiezynski_interaction_2019}. Over 700 students studying at DTU in 2013 participated in the study. Participants of this study were handed out Android-based Google Nexus 4 smartphones and asked to install data collection software from Google Play Store. The experiment was GDPR compliant and registered with the Danish authorities (for a more comprehensive discussion and detailed information on the data collection process see \citet{sapiezynski_interaction_2019}).
 We use three kinds of social interaction data collected as part of CNS: 

\textit{Bluetooth data.} The smartphones were set to request and receive responses from all nearby Bluetooth discoverable devices every five minutes. These responses included unique identifiers of devices within a maximum range of 10 meters that were connected to the study participants. For each pair of users, we counted a maximum of one meeting per 15-minute interval. In cases where Bluetooth sensors detected users in close proximity multiple times within the same 15-minute period, only the first instance was recorded as a meeting. Further, the location of each meeting was recorded and subsequently categorized as being on or off the DTU campus. Additionally, based on the administrative class scheduling data obtained from DTU, we computed indicators for individual class attendance, based on whether individuals  are physically present in the classrooms \citep{bjerre2020negative}.

\textit{Calls and short messages.} Metadata on calls and short messages (SMS) were obtained from the smartphones on a daily basis. Participants were required to reveal their phone number and could thus be matched to the call logs. For the case of SMS, each record consists of the unique identification numbers of the sender and receiver as well as a timestamp, specifying the time of the event. Similarly for calls, identification numbers of the person calling and the person receiving the call were logged, the timestamp when the call started and the call duration in seconds.

\textit{Facebook data.} In addition, the majority of students voluntarily opted in to allow data collection from Facebook. A snapshot of the students friendship network was collected at the end of the observation period.
It should be noted that the data were collected in 2013, when Facebook was considerably more popular as a social medium and SMS was used far more frequently as a primary communication channel than today.

\subsection{Samples}

Our analysis relies on two different main samples, one aimed at estimating the effects of classroom assignment and one aimed at estimating the effects of social groups. We describe their construction below, while Table \ref{tab:overview} provides an overview of sample totals.

\subsubsection{Social Groups Sample}
As per the official data obtained from the DTU administration in mid-August, 1,372 students enrolled in the 2013 DTU undergraduate cohort. Since fewer students had signed up for a social group by mid-August 2013, however, restricting to students that were part of our social group randomization cuts the sample to 616. To analyze the effect of actual group membership, we restrict attention to 606 students who we also see in the final social group assignment data from fall semester 2013. To be able to construct data on social interactions and networks, we also restrict to students who participated in the Copenhagen Network Study ($n=223$). Finally, since our analysis studies group assignment and network formation \emph{within} each study program, we restrict attention to study programs where we observe at  least 10 students. This leaves us with a final sample of 171 students for use in analyzing social groups.

\subsubsection{Classroom Sample}
The construction of our analysis sample for studying classrooms proceeds analogously to above. However, here we have more complete data, with random assignment information being available for the full sample and actual classroom membership information available for 901 students. After imposing all the other restrictions described above, we are left with a sample of 276 students which we use to analyze the effects of classrooms.

\input{tables/01_overview}

\subsection{Measuring social linking}

\subsubsection{Dyadic data structure and social outcomes}
\label{sec:dyadic_data_structure}
For our main analysis, we adopt a dyadic data structure; individual observations are pairs of students within the same study program and we aim to examine whether being in the same social group or classroom affects social interactions between the two students constituting the pair.

Accordingly, from each of our two student-level analysis samples we form all possible pairs, $ij$, where students $i$ and $j$  are from the same study program $s$.\footnote{The two social foci we analyze,  classrooms and social groups, are only ever shared by students from the same study program. Comparisons across students from different study programs would thus never be meaningful in our analysis.} In each of the resulting dyadic data sets, we further construct the variables $SameGroup_{ijs}$ as a dummy variables for whether $i$ and $j$ are in the same social group or classroom, respectively.  Analogously, we construct the variable $SameAssignedGroup_{ijs}$ as a dummy for whether the pair initially had been assigned to be in the same social group/classroom in our randomization. As noted earlier, $SameGroup_{ijs}$ and $SameAssignedGroup_{ijs}$ may differ because of imperfect compliance.

Our main outcomes of interest are the social interactions and network formation, as measured based on calls, SMS, physical co-presence and Facebook friendship. 
For easier interpretation, we compute weekly averages for each of the four semesters in our data\footnote{\textbf{Semester 1 - Fall Semester 2013:}  1\textsuperscript{st} September 2013 - 31\textsuperscript{st} of January 2014; \textbf{Semester 2 - Spring Semester 2014:}  1\textsuperscript{st} February 2014- 31\textsuperscript{st} of July 2014; \textbf{Semester 3 - Fall Semester 2014:}  1\textsuperscript{st} August 2014- 31\textsuperscript{st} of January 2015; \textbf{Semester 4 - Spring Semester 2015:} 1\textsuperscript{st} February 2015- 31\textsuperscript{st} of July 2015 }. For each continuous outcome variable, this is done by summing all observed pairwise interactions and dividing by the number of weeks in the semester.

We construct the following variables for all pairs with the relevant data available: (1) average weekly call duration in minutes, (2) average weekly SMS count, (3) average weekly meetings outside the DTU campus and (4) a binary indicator for whether a given pair became friends on Facebook before the end of the observation period or not.

In our main analysis, we focus on interactions in students' second semester at DTU, i.e. the Spring semester of 2014. We do this to avoid social interactions that happen mechanically due to scheduled activities in the social group and/or around the actual TA sessions taking place in the first semester. In the Supplemental Material (Section ), we examine effects on other semesters to see whether the effects on interactions persist at longer time horizons.

\subsubsection{Composite measure of social links}
\label{sec:summary_measure}
In addition to estimating the effect of group membership on different types of social interactions, we are also interested in providing an overall measure of the effect on networks and link formation. To achieve this, we construct a composite measure that captures social interaction across multiple channels and establish a criterion for determining whether a given pair has formed a link.

To do this, we first use principal component analysis (PCA) to aggregate the three continuous measures of pairwise social interactions into a single aggregate continuous measure. We then use this aggregate to define which students are linked.

For our main analysis, we code a pair of students as linked if their score on the aggregate interaction measure is above the 90\textsuperscript{th} percentile within their study program. 
Table \ref{tab:network_stats} provides a summary of the resulting network structures across study programs. Since we only consider pairs within the same study program, each study program effectively constitutes its own network. Across the 20 study programs where we have  all necessary information for at least 10 individuals, we observe average degrees between 2.3 to 6.9, with a median of 3.9. This indicates that our cutoff aligns well with the notion of a social circle of having a ``handful'' of friends within the study program. The maximum of the observed diameters is 6, meaning that two individuals within the same study program are at maximum separated by 6 other individuals. The minimum average path length is 1.3, compared to a maximum of 2.7, showing that some networks are considerably more dispersed than others.  It should be noted that, because the dataset only includes students who participated in the study, interactions with individuals outside the sample are not observed. Consequently, the resulting network represents only a partial view of the students' social environment and network statistics might be subject to bias \citep{KOSSINETS2006247}.
We provide further detail on the construction of the link measure, detailed statistics for every study program and robustness checks for this in the Supplemental Material (Section B).

\input{tables/02_network_stats}

\section{Overall effects of group assignment on interactions and link formation}
\label{sec:overresults}

% Regression
%
%

We start by estimating the overall effect of shared membership in a classroom or social group using a simple dyadic specification. We do this in the context of the following simple linear regression, applied to our dyadic samples of student pairs from the same study program:
\begin{equation} \label{eq:naive_dyadic}
Y_{ijs} = \beta_0 + \beta_1 SameGroup_{ijs} + v_{ijs}
\end{equation}
The outcome $Y_{ijs}$ here is some measure of pairwise social interactions and/or link formation between students $i$ and $j$. $SameGroup_{ijs}$ is either our dummy variable for being in the same TA classroom or the same social group. The coefficient of interest is $\beta_1$ which measures the causal effect of shared group membership on the outcome variable. Note that, as written, the model assumes that the causal effect is homogeneous. If targeting a Local Average Treatment Effect (LATE) interpretation, however, our main instrumental variable approach can allow for heterogeneous treatment effects in the standard way \citep{ImbensAngrist1994}.

Potential endogenous selection into groups means that simple estimation of \eqref{eq:naive_dyadic} using ordinary least squares (OLS) is likely to give a biased estimate of $\beta_1$. Instead we rely on the variation generated by our random assignment in an instrumental variables setup. Recall that $SameAssignedGroup_{ijs}$ is a dummy for whether $i$ and $j$  were assigned to the same group in our randomization. Under the realistic assumption that our assignment influenced the actual groups formed, $SameAssignedGroup_{ijs}$ should be a relevant instrument  for $SameGroup_{ijs}$ in \eqref{eq:naive_dyadic}. 

To be a valid instrument, $SameAssignedGroup_{ijs}$ also needs to satisfy the necessary exclusion restriction, i.e. being assigned to the same group must not affect interactions other than through the actual group membership. 
Despite stemming from a randomization, there are two concerns: First, randomization was done conditional on some observed characteristics. Second, our randomization did not directly randomize the dummy $SameAssignedGroup_{ijs}$ but randomized all students in a given study program into a number of different groups. Thus, while on an individual level group assignment is random conditional on covariates, as pointed out forcefully by \cite{borusyak_nonrandom_2023}, such a randomization need not guarantee exogeneity of $SameAssignedGroup_{ijs}$ in \eqref{eq:naive_dyadic}.\footnote{A very concrete problem in our setting is the size differences across study program. In a study program with many students, a given pair of students is systematically less likely to end up in the same group under our randomization; this makes $SameAssignedGroup_{ijs}$ systematically related to the size of $i$ and $j$'s study line. At the same time, study lines  may well differ in the extent of social interactions and network formation.}

A convenient way to address both concerns is to use the ``instrument recentering'' approach proposed by \cite{borusyak_nonrandom_2023}. Intuitively, this means adjusting the observed value of the instrument to account for the baseline probability of two individuals being assigned to the same group by chance. For a more formal explanation see Supplemental Material (Section C).
To implement this practically, we re-run our random assignment procedure 1,000 times and  compute the share of realizations where each pair $ijs$ is assigned to the same group (see Section C of the Supplemental Material for the exact randomization algorithm and distribution of group assignment probabilities). We then subtract this average from $SameAssignedGroup_{ijs}$ to arrive at a \emph{recentered} instrumental variable which we denote by $\widetilde{SameAssignedGroup_{ijs}}$. We use this as our instrument in a two-stage least squares (2SLS) estimation of \eqref{eq:naive_dyadic}, with the following first stage:
\begin{equation} \label{eq:1st}
SameGroup_{ijs} = \gamma_0 +  \gamma_1 \widetilde{SameAssignedGroup_{ijs}} + u_{ijs} 
\end{equation}

Estimating linear dyadic specifications like the ones above using 2SLS is common in the literature, see e.g. \cite{ Algan2023, harmon_peer_2019}. Briefly foreshadowing later results, we note that a numerically equivalent way to estimate $\beta_1$ using random assignment here would be to use a control function approach and include estimated residuals from \eqref{eq:1st} as controls in \eqref{eq:naive_dyadic}. We adopt such an approach later when estimating our SUGM model.

For inference, we need to address the concern that errors terms are likely to be correlated across pairs of students who share a member. To account for this, we use dyad robust standard errors  based on \cite{samii_cluster-robust_2015} for our main specification. Supplemental Material (Section E) presents alternative results using bootstrapped standard errors based on resampling study programs. Results are almost identical to those obtained using dyad robust standard errors. \footnote{Dyadic robust errors may fail to account for all dependencies in the data under second order network effects and/or triadic closure effects. Additionally, the bootstrap results ensure direct comparability with the triadic closure results presented later which use the same bootstrap approach for inference.} As an additional robustness check we also estimate our main specification with study program fixed-effects and a different definition of TA classroom membership. For the fixed-effects results change only slight, the alternative classroom definition results in a smaller sample and results in similar effect size with slightly higher standard errors (see Supplemental Material, Section E).

\subsection{Effect on measures of link formation}

\begin{figure}[h!]
\centering

 \includegraphics[width=.5\textwidth]{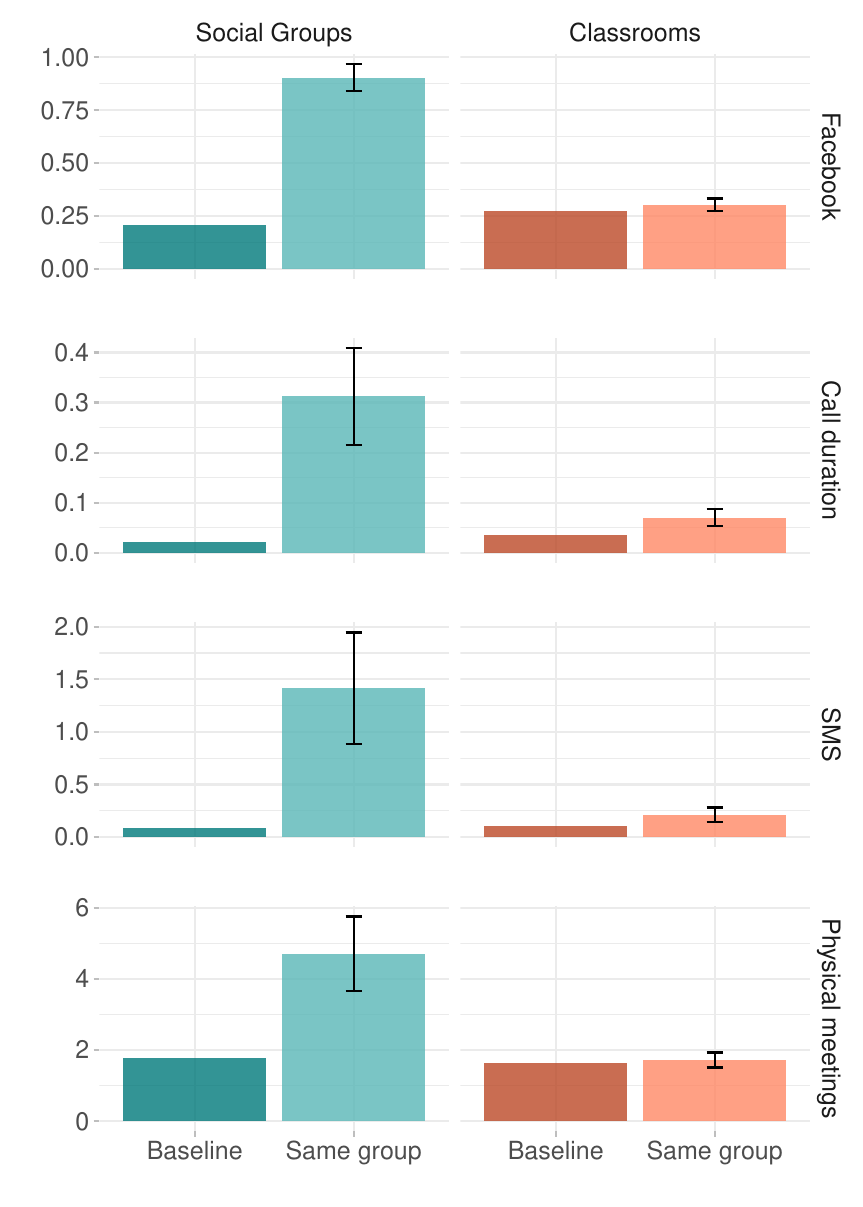}
\caption{Average weekly interactions for pairs not sharing a social group vs. pairs sharing a social group (left panel) as well as for pairs not sharing a classroom vs. pairs sharing a classroom (right panel), based on 2SLS with recentered instrument. }
\label{fig:iv_results}
\end{figure}

Figure \ref{fig:iv_results} shows the results for 2SLS estimates of the effects of group membership on interactions, based on the recentered instrument. The leftmost column of (teal) bars shows results for social groups, while the rightmost column of (red) bars shows results for classrooms. Within each row and column, the first bar shows the predicted \emph{baseline} levels of social interactions among pairs not in the same group ($\beta_0$), while the second shows the predicted levels of social interaction among pairs who are part of the \emph{same group} ($\beta_0+\beta_1$). Error bars reflect the standards error on the difference between baseline and same group pairs (the standard error on $\beta_1$). For completeness, Table \ref{tab:iv_results} reports numerical estimates. In the Supplemental Material (Section D) we additionally report the 2SLS specifications reduced form and first stage estimates.\footnote{Unsurprisingly, for both social groups and classrooms, the first stage relationship between shared group membership and (recentered) assigned shared members is strong. For social groups, the first stage coefficient on the instrument is around 0.76. For classrooms it is around 0.96. Accordingly, as shown in Table \ref{tab:iv_results}, the usual first-stage F-statistics for instrument strength are large.}

For social group membership we find significant effects across all forms of interactions. Over 90\% of students in our sample that were in the same social group eventually became Facebook friends, compared to a baseline of 21\% of students in the same study program but not the same social group. For SMS and call duration we find particularly large effects compared to the baseline. On average, members of the same social groups spend 19 seconds on the phone with each other every week, compared to the baseline of 1 second. We also find that members of the same social group text each other more; on average they exchange 1.4 SMS per week compared to only 0.08 at baseline. Finally, students in the same social group are also much more likely to meet physically outside of campus, 4.7 times per week as opposed to only 1.8 times for baseline pairs not in the same group.\footnote{When it comes to physical interactions, we restrict our main analysis to interactions taking place outside of the university campus, as we are mainly interested in the effect of shared group membership on interactions not enforced through study related activities (e.g. group work), we chose to exclude any meetings happening on the university campus for our main specification. However, one could easily argue that a significant part of friendship-induced meetings could also happen on campus (e.g. going for lunch in between classes, studying together). We thus report results for (1) the average number of weekly meetings outside class hours and (2) the overall average number of weekly meetings (i.e. including during classes) in the Supplemental Material (Section F).  Effect sizes are larger for these alternative specifications and exhibit identical significance patterns. The effects reported here can thus be seen as a conservative lower bound of the effect of group membership on physical meetings.}

\input{tables/03_iv_results}

For classrooms, the effects are less pronounced. We neither find an effect of classroom membership on the likelihood of becoming Facebook friends nor physical meetings off campus. We do, however, observe comparatively small but significant effects on call duration and SMS count. Pairs within a classroom spend an average of 6 seconds on the phone with each other per week (compared to 3 seconds for those not sharing a classroom) and exchange $\sim1$ SMS every month (compared to a baseline of 0.6).

Finally, in the Supplemental Material (Section F), we show that the estimated effects appear persistent. When looking at social interactions in later semesters we find that effect sizes do decline but for phone interactions the estimates remain significant.

\subsection{Effect on overall link formation}

In addition to measuring the overall effect of group membership on social interactions, we are also interested in providing a summary measure of the effects on link formation. The last column of Table \ref{tab:iv_results} shows the estimated effects of social group and classroom membership on our summary measure (using a cutoff at the 90\textsuperscript{th} percentile) of links. For social groups, being in the same group increases the likelihood of forming a link by 32 percentage points, corresponding to a 3.8 fold increase compared to the baseline of 8\%. For classrooms, the effect is less pronounced. While the effect is nonetheless significant, it is only 3 percentage points, thus roughly increasing the baseline likelihood by a third.\footnote{For robustness and transparency, in the Supplemental Material (Section B) we  report alternative results using different versions of our aggregate measure of links. Overall, effect sizes are similar.}

\section{Interactions between group assignment and triadic closure}

We finish our empirical analysis by examining possible interactions between our social foci group interventions and triadic closure. Doing so requires an extension to our empirical specification. Being based on dyadic analysis of pairs, the specifications in the previous section are well-suited to study the overall effects on link formation. By focusing only on pairs, however, the specifications are unable to account for triadic closure effects which inherently operate across three students simultaneously.

Empirical models of triadic closure effects can raise challenges of complexity and computational cost. To mitigate this, we rely on recent work by \citet{chandrasekhar_network_res} showing that so-called Subgraph Generated Models (SUGM) are able to fit triadic closure patterns well with a relatively small number of parameters and without incurring excessive computational cost. Specifically, we show how the same random-assignment variation used in the dyadic 2SLS framework can be to estimate the SUGM in an instrumental-variable control-function approach.

\subsection{A Subgraph Generated Model}

A simple way to understand SUGMs is that they model network formation and measurement \emph{as if} happening through a two-step process: In the first (unobserved) step, a number of different types of graphs form independently between individuals. In the simplest possible form with triadic closure, these subgraphs are just pairwise links and triangles consisting of links across three individuals. For each pair of individuals there is thus some probability that a pairwise link is formed. Separately, for each triad of individuals there is also some probability that a triangle of links form between them. These probabilities typically depend on the characteristics of the pair or triad, including whether they have any shared social foci such as being in the same social group or TA classroom.

In the second step, the final network is observed as the \emph{union} of the pairwise links that have formed \emph{plus} all links that are part of some triangle that has formed. In the observed data, a link between a given pair of individuals may thus reflect that a pairwise link formed in the first step and/or that some triangle involving this pair formed in the first step. 

\subsubsection{Effect of groups on link and triangle formation}

We now formally describe the SUGM that we estimate in our analysis. As in the previous sections, we analyze social groups and classrooms separately, e.g. we set up and estimate a separate SUGM for studying social groups and classrooms.\footnote{When analyzing the effect of social groups, the effect of classrooms will thus be subsumed in the flexible unobservable terms included in the model, and vice versa when analyzing the effects of classrooms.} In what follows below, the term ``group'' will thus refer to either social groups or classrooms.

First, in terms of data and notation, we extend our analysis data to not only contain all within-study program pairs $ij$ but also all within-study program triads of students $ijk$. Correspondingly, we let $Triangle_{ijks}$ be a dummy variable for whether students $i$, $j$ and $k$ from study line $s$ are all linked to each other based on our summary measure of links between students. Letting $\Omega_s$ denote all the students in study line $s$, we also introduce notation for the sets of all unique student pairs and student triads in the study line, $\Omega_s^L=\{(i,j)\in \Omega_s^2:i>j\}$ and $\Omega_s^T=\{(i,j,k)\in \Omega_s^3:i>j \,\wedge j>k \}$.

Second, to explicate the two-step structure of the SUGM, we let  $L_{ijs}^{direct}$ be a dummy variable for whether student $i$ and $j$ from study program $s$ formed a pairwise link in the first step of the SUGM. Similarly, let $T_{ijks}^{direct}$ be a dummy variable for whether the triad $ijk$ formed a triangle in the first step of the SUGM. Note that these variables are fundamentally unobserved. Additionally, as shorthand notation let $\boldsymbol{G}$ be a vector summarizing the group assignment of all students in the data.

In the first step of the SUGM, the likelihood that a given pair $ij$ forms a link or that a triad $ijk$ forms a triangle will depend on group memberships as well as a number of other factors and characteristics, which we summarize by the (unobserved) scalars, $\eta_{ijs}$ and $\eta_{ijks}$, respectively. We let $\boldsymbol{\eta}$ denote the vector of all such unobserved scalars across all pairs and triads in the data. We normalize these unobservables to be mean zero.\footnote{As we explicate below, the unobservables will enter into regression equations with intercepts ($\beta_0$ and $\theta_0$). Assuming that the error terms are mean zero thus simply normalizes the intercepts to a specific and easily interpretable value}

As a natural extension of the linear regression framework of the previous sections, we now assume that the likelihood of forming a pairwise link depends linearly on whether the pair is in the same group and on the scalar unobservable:
\begin{align}
E\left[L_{ijs}^{direct}|\boldsymbol{G},\boldsymbol{\eta}\right]=\beta_0 + \beta_1 SameGroup_{ijs} + \eta_{ijs} \label{eq:ldirect}
\end{align}
As before, $\beta_1$ is the first key coefficient of interest, representing the causal effect of $i$ and $j$ being in the same group on the likelihood of forming a link in the first step of the SUGM. Similarly, for the likelihood that a triad forms a triangle, we assume that it depends linearly on the number of pairs within the triad who are in the same group plus the scalar summarizing unobservables:
\begin{align} \label{eq:tdirect}
E\left[T_{ijks}^{direct}|\boldsymbol{G},\boldsymbol{\eta}\right]=\theta_0 + \theta_1 \sum_{\left(\iota,\kappa\right)\in\left\{ (i,j),(i,k),(j,k)\right\} } SameGroup_{\iota \kappa s} + \eta_{ijks}
\end{align}
The key coefficient of interest here is $\theta_1$ which is defined as the causal effect of having one more pair within the triad be in the same group. Switching from a situation where all three members in the triad are in different groups to a situation where two of the students in the triad are in the same group thus increases the likelihood of forming a triangle by $\theta_1$. Switching to a situation where they are all in the same group increases the likelihood of forming a triangle by $3 \theta_1$.\footnote{Note that mechanically, the number of pairs within a triad that are in the same group is either 0, 1 or 3. It is not possible for two pairs of students to be in the same group without having all three students in the triad in the same group.}

In order to impose that the model has the SUGM structure, we further impose that conditional on group memberships and scalar unobservables, the likelihood of forming a link is independent of the likelihood of forming a triangle that is 
\begin{align*} \label{eq:sugmstruc} 
\left\{ L_{ijs}^{direct} \right\}_{(i,j)\in \Omega_s^L} \cup  \left\{ T_{ijks}^{direct} \right\}_{(i,j,k)\in \Omega_s^T} \quad \textrm{ mutually independent given  } \left(\boldsymbol{G},\boldsymbol{\eta}\right)
\end{align*}

%NH NOTE ABOUT POST AI EDITING: Add mutual independence of the etas here (motivate askey to the SUGM)

Note that at this point, the assumptions above are just a normalization. Since we have imposed no restriction on the dependence of the scalar unobservables $\boldsymbol{\eta}$ across the different pairs or triads of students, the assumptions above can fit any unconditional dependence patterns in the unconditional likelihood of forming links and triangles. Assumptions further below will however impose additional restrictions.

In terms of the observed data, the final network is the union of the links and triangles formed in the first step. This means that in our observed network there will be a link between $i$ and $j$ if such a pairwise link formed in the first step \emph{or} if some triangle formed that contains both $i$ and $j$:
\begin{align}
Link_{ijs}=
\begin{cases}
    1, & \text{if} \quad L_{ijs}^{direct}=1 \quad \textrm{or} \quad T_{ijks}^{direct}=1 \textrm{ for some } k \\
   0,              & \text{otherwise}
\end{cases} 
\end{align}

Finally, to see that the SUGM model outlined above is a natural extension of the dyadic 2SLS framework of the preceding section, note that if we set $\theta_0,\theta_1,\eta_{ijks}=0$ (e.g. remove the possibility of forming triangles in the first step), then the SUGM collapses exactly to the linear regression \eqref{eq:naive_dyadic} from Section \ref{sec:overresults} with $\beta_1$ and $\beta_0$ having the same values and interpretation. 

If $\theta_0$, $\theta_1$ and $\eta_{ijks}$ are not all zero however, the model explicitly includes triadic closure effects and if $\theta_1>0$ in particular, these effects interact with shared group membership. The model then implies that the final network will contain a relatively large share of triangles of three interconnected students, compared to overall number of links between students. Moreover this pattern will be exacerbated among triads of students who are placed in the same group. Estimating $\theta_1$ will thus allow us to determine whether our social foci interventions on groups interact with triadic closure.

\subsubsection{Identification and estimation using random assignment}

Analogously to Section \ref{sec:overresults}, the fundamental challenge for identifying the causal effects of shared group membership, is that we typically would expect shared group membership to be systematically related to many potential other factors that affect link and triangle formation, e.g. $SameGroup_{ijs}$ will not be independent of $\eta_{ijs}$ and $\eta_{ijks}$.

As before, we solve this using the variation generated from our random group assignment. Since a SUGM with triadic closure inherently becomes non-linear, we integrate the instrumental variables approach using control functions. Our instrument will again be the recentered dummy for whether given pair of students have been assigned to the same group, $\widetilde{SameAssignedGroup_{ijs}}$.  The instrument first stage is assumed to be as before:
\begin{equation} \label{eq:1sttriad}
SameGroup_{ijs} = \gamma_0 +  \gamma_1 \widetilde{SameAssignedGroup_{ijs}} + u_{ijs} 
\end{equation}
As a matter of notation, we let $\boldsymbol{Z}$ and $\boldsymbol{u}$  denote the vectors stacking the value of the instrument and the first stage error, $u_{ijs}$, across all pairs of students.

 To accommodate the SUGM structure, we then proceed to impose slightly stronger versions of the standard instrumental variables assumptions. First, we restrict the first stage relationship between the instrument and actual group membership. Just as above, we assume that the coefficient on the instrument in the first stage is non-zero and that conditional on the value of the instrument, the error term in the first stage is mean zero:
\begin{align} \label{eq:triad1stass}
\gamma_1 \ne 0 \\
E\left[u_{ijs}|Z_{ijs}\right]=0
\end{align}

Next we invoke assumptions analogous to the exclusion restriction used previously but now strengthened to match the SUGM and allow for a control function approach. 

Let $\xi_{ijs}=\eta_{ijs}-E\left[\eta_{ijs}|\boldsymbol{u},\boldsymbol{Z}\right]$ and $\xi_{ijks}=\eta_{ijks}-E\left[\eta_{ijks}|\boldsymbol{u},\boldsymbol{Z}\right]$ denote the deviations of the unobservables from their mean conditional on the instruments and first stage errors. To permit a standard control function approach we next assume that this conditional mean does not depend on the instrument but only on the first stage errors corresponding to the involved pairs of students:
\begin{align}
E\left[\eta_{ijs}|\boldsymbol{Z},\boldsymbol{u}\right]=h_{L}(u_{ijs}) \\
E\left[\eta_{ijks}|\boldsymbol{Z},\boldsymbol{u}\right]=h_{T}(u_{ijs},u_{iks,}u_{jks})
\end{align}
The interpretation of these assumptions are that the first stage \eqref{eq:1sttriad} decomposes the variation in same group membership into a part stemming from the random assignment instrument, the term involving $\widetilde{SameAssignedGroup_{ijs}}$ and a part stemming from endogenous selection of certain pairs of students into the same group, $u_{ijs}$. The assumptions above then correspond to saying that the latter of these contains information about the mean of the endogenous link and triangle formation unobservables, $\eta_{ijs},\eta_{ijks}$ but the former does not.

Next, to preserve the SUGM structure within the control function framework, we impose that the deviations of the unobservables from their conditional mean are mutually independent conditional on the first stage errors and instruments:

\begin{align} \label{eq:sugmstrucCF} 
%  &  \left\{ \xi_{ijs}\right\} _{(i,j)\in \Omega_s^L} \cup \left\{ \xi_{ijks}\right\} _{(i,j,k)\in \Omega_s^T} \perp (\boldsymbol{u},\boldsymbol{Z}) \\
&\left\{ \xi_{ijs}\right\} _{(i,j)\in \Omega_s^L} \cup \left\{ \xi_{ijks}\right\} _{(i,j,k)\in \Omega_s^T} \quad \textrm{ mutually independent given  } \left(\boldsymbol{Z},\boldsymbol{u}\right)
\end{align}

The key feature of SUGM models that make them so empirically tractable is that they split link and triangle formation into processes that are independent. The assumption above ensures that this continues to hold also once we condition on the instruments and first stage errors.

\subsection{Estimation using control functions and non-linear least squares}

Together the assumptions above allow us to estimate the key causal parameters of interest, $\beta_1$ and $\theta_1$, using a two-step control function approach explained in detail in the appendix.

To conduct inference, we apply a (block) bootstrap approach and resample study programs 100 times. This corresponds to an assumption that the formation of links and triangles is independent across study lines but requires no additional restrictions on dependence within a study program. We show the distribution of the resulting effect sizes across draws in the Supplemental Material (Section H).

\clearpage

\subsection{Results}

Table \ref{tab:triad_results} shows estimates from the SUGM outlined above both for the effects of social groups and classrooms. For social groups, we see a significant effect of shared group membership on the likelihood of forming a triangle (in the first step of the SUGM). For a given triad of students, having two of them be in the same social group increases the likelihood of forming a triangle by 3.7 percentage points. If instead all three are in the same group, the likelihood increases by a total of 11.1 percentage points. Conversely, for a triad of students where none are in the same group, the estimate is in fact slightly negative, suggesting virtually no chance that a triangle forms. These results confirm that social group membership interacts very markedly with triadic closure. Indeed, looking at the effects of social group membership on link formation in the first step, the estimate is negative and insignificant. In other words, results suggest that the overall effects of social groups on link formation is driven primarily by the formation of triangles.

Turning to the results for classrooms, estimates are more noisy. For pairwise link formation, the estimated effect of classroom membership are sizable, suggesting that being in the same classroom positively impacts the likelihood that a given pair forms a link (in the first step of the SUGM). Conversely, the estimates suggest that classroom membership has no impact on the likelihood of forming a triangle. Taken together these results suggest that a classrooms affect network formation very differently than social groups, with classroom effects operating exclusively at the pairwise-level. At the same time, standard errors are large enough that none of the effects are significant. At conventional level of significance we are thus not able to draw any firm conclusions.

For our main specification we use a quadratic control function and give equal weights to both dyads and triangles in the optimization procedure. As explained in the appendix, another natural choice would be to weight triangles proportional to the ratio of dyads to triangles. We implement this specification a robustness check and find that effects are all in the same direction, of similar magnitude and at the same significance level as our main specification. Additionally, we also estimate the model with linear or cubic
control functions. This yields qualitatively similar results, with a positive and significant effect of shared group membership on triangle formation and no robust effect on direct link formation. All robustness checks are reported in the Supplemental Material (Section H).

\input{tables/04_triad_results}

\section{Simulations and welfare implications}

\subsection{Utility framework}
To interpret the importance of modeling triadic closure, we evaluate simulated networks through a welfare/utility lens in which individuals value (i) direct links and
(ii) spillovers that arise from being indirectly connected through mutual acquaintances. Let
$A$ denote the (undirected) adjacency matrix of the network, with $A_{ij}=1$ if $i$ and $j$
are linked and $A_{ij}=0$ otherwise, and let $d_i=\sum_{j\neq i}A_{ij}$ be node $i$'s degree.
We normalize the utility from a direct link to be $U_L$ per endpoint, so that direct-link
utility is
\begin{equation}
U^{(i)}_{\text{direct}}(A)=U_L\sum_{j\neq i} A_{ij}=U_L\cdot d_i .
\end{equation}

Spillovers depend on whether $i$ and $j$ have at least one common neighbor, i.e. whether
$(A^2)_{ij}>0$. We distinguish spillovers from \emph{open triads} (a friend-of-a-friend
relationship without a direct link, $A_{ij}=0$) and \emph{closed triads} (a direct link that is
embedded in at least one triangle, $A_{ij}=1$). Individual spillover utility is
\begin{equation}
U^{(i)}_{\text{indirect}}(A)
=
U_L \sum_{j\neq i}
\Big(
\alpha\,\mathbbm{1}\{(A^2)_{ij}>0\}\,(1-A_{ij})
+
\beta\,\mathbbm{1}\{(A^2)_{ij}>0\}\,A_{ij}
\Big),
\end{equation}
where $\alpha\in(0,1)$ captures the value of open-triad spillovers and $\beta>0$ captures the
additional value of closed-triad embeddedness. For convenience we also use the re-parameterization
$\gamma=\alpha+\beta$ (overall magnitude of spillovers) and a mixing parameter $\lambda$
that governs the relative weight placed on open- versus closed-triad spillovers (equivalently,
$\alpha$ and $\beta$ can be expressed as functions of $(\gamma,\lambda)$). Total utility is then
\begin{equation}
U_i(A)=U^{(i)}_{\text{direct}}(A)+U^{(i)}_{\text{indirect}}(A),
\qquad
\bar U(A)=\frac{1}{n}\sum_{i=1}^n U_i(A).
\end{equation}
This framework makes clear that if $\gamma=0$ (no spillovers), welfare depends only on the
number of links; if $\gamma>0$, welfare also depends on higher-order structure, in particular
the prevalence of triangles and friend-of-friend relationships.

\subsection{Simulation design}
To evaluate the implications of the different network estimation frameworks under the proposed utility model, we simulate  (i) a purely dyadic model (link formation driven only by pair-level terms), and
(ii) the SUGM specification that allows links to form either directly at the pair level or
indirectly as part of triangle formation (triadic closure). As our unconstrained estimation of the SUGM yields partially negative coefficients, we re-estimate the same model with the additional constraints that the intercepts and the sum of intercept and same group effects need to be positive. The resulting estimates are of the same sign. For social groups they only differ slightly in magnitude and show the same significance levels. For classrooms we find that under the constraints, the effect sizes are smaller and significant. They are reported in the Supplemental Material (Section H).

\begin{figure}[h]
    \centering
    \begin{subfigure}{0.42\textwidth}
        \centering
        \includegraphics[width=\linewidth]{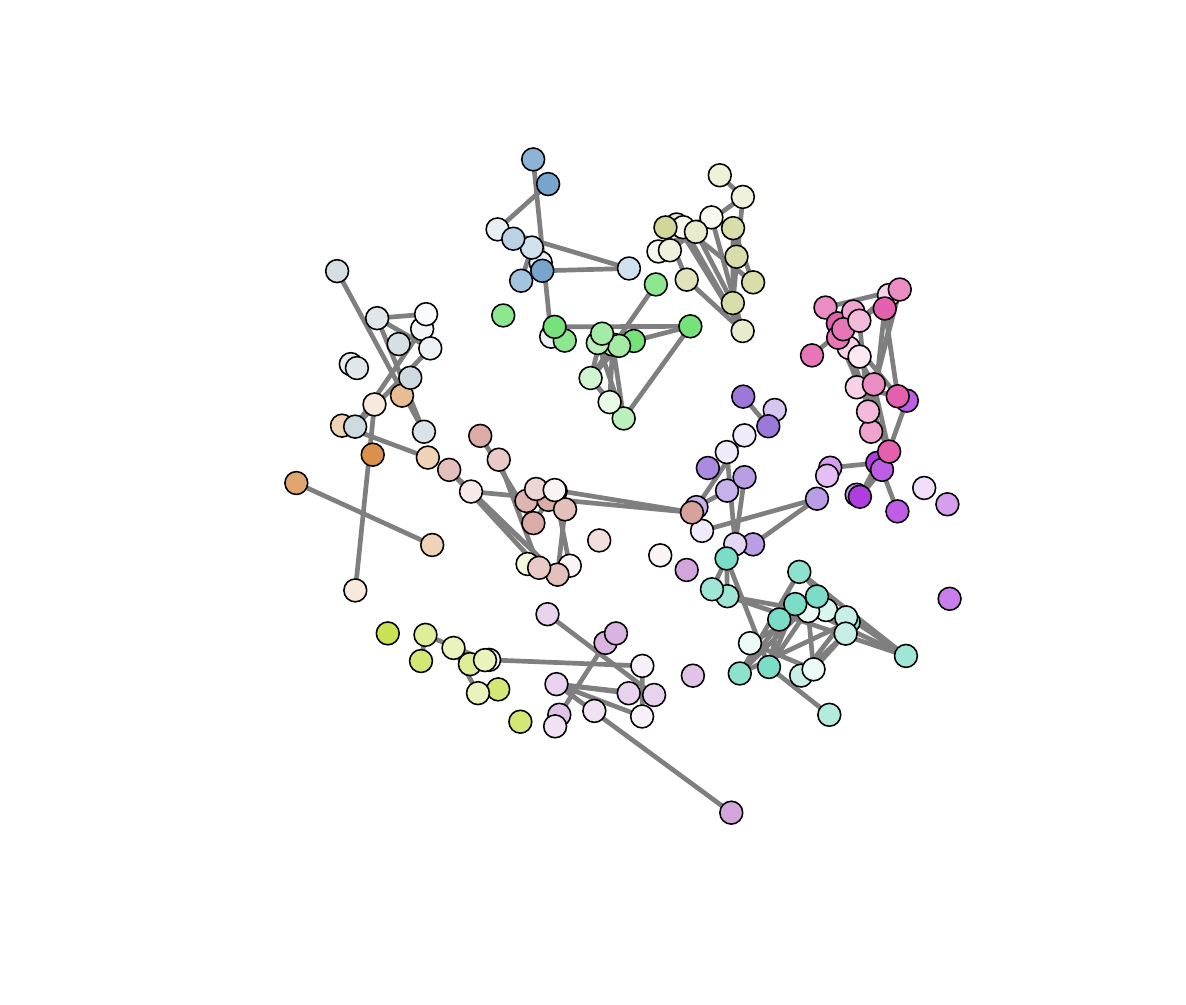}
        \caption{\centering Example of a simulated network from the dyadic model}
        \label{fig:network_no_triads}
    \end{subfigure}
    \hfill
    \begin{subfigure}{0.45\textwidth}
        \centering
        \includegraphics[width=\linewidth]{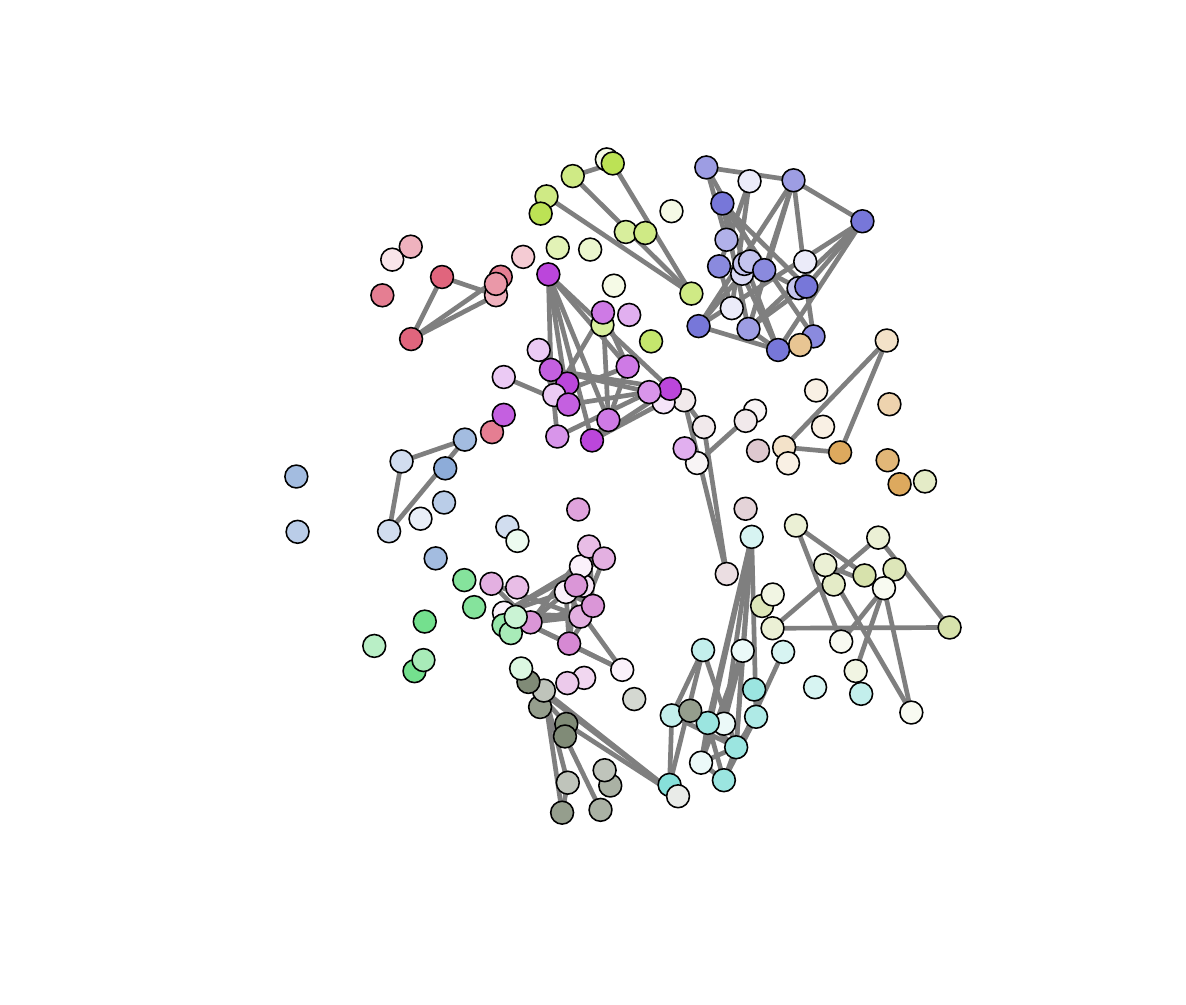}
        \caption{\centering Example of a simulated network from the SUGM model}
        \label{fig:network_triads}
    \end{subfigure}
    \caption{Examples of simulated networks. Color indicates studyline membership, shade indicates social group membership.}
    \label{fig:networks}
\end{figure}

For each model, we generate
$N=500$ simulated networks using the estimated parameters and the observed grouping
structure (study line membership and social group membership). Figure~\ref{fig:networks}
illustrates typical draws from each model. Below we comment on how they differ in terms of network structure and its consequences for welfare.

\subsection{Results}

\begin{table}[t]
\centering
\renewcommand{\arraystretch}{1.2}
\caption{Number of observed links and triangles averaged across simulations based on the dyadic model and the SUGM model. N=500.}
\label{tab:links_triangles}
\begin{tabular}{l|cc}
\hline
 & Avg. links & Avg. triangles \\
\hline
Dyadic Model & 130 &  9.52 \\
SUGM &  133 & 51.6\\
\hline
\end{tabular}

\end{table}

A first takeaway is that the two models generate networks with \emph{similar link density}
but \emph{very different clustering}. Table~\ref{tab:links_triangles} shows that the average
number of links is nearly identical across the two models (about 130 vs.\ 133), whereas the
SUGM produces far more triangles.
Thus, a dyadic model can match first-order moments (links) while substantially
under-predicting the higher-order structure that is central to triadic closure.

\begin{figure}[h]
    \centering
        \includegraphics[width=\linewidth]{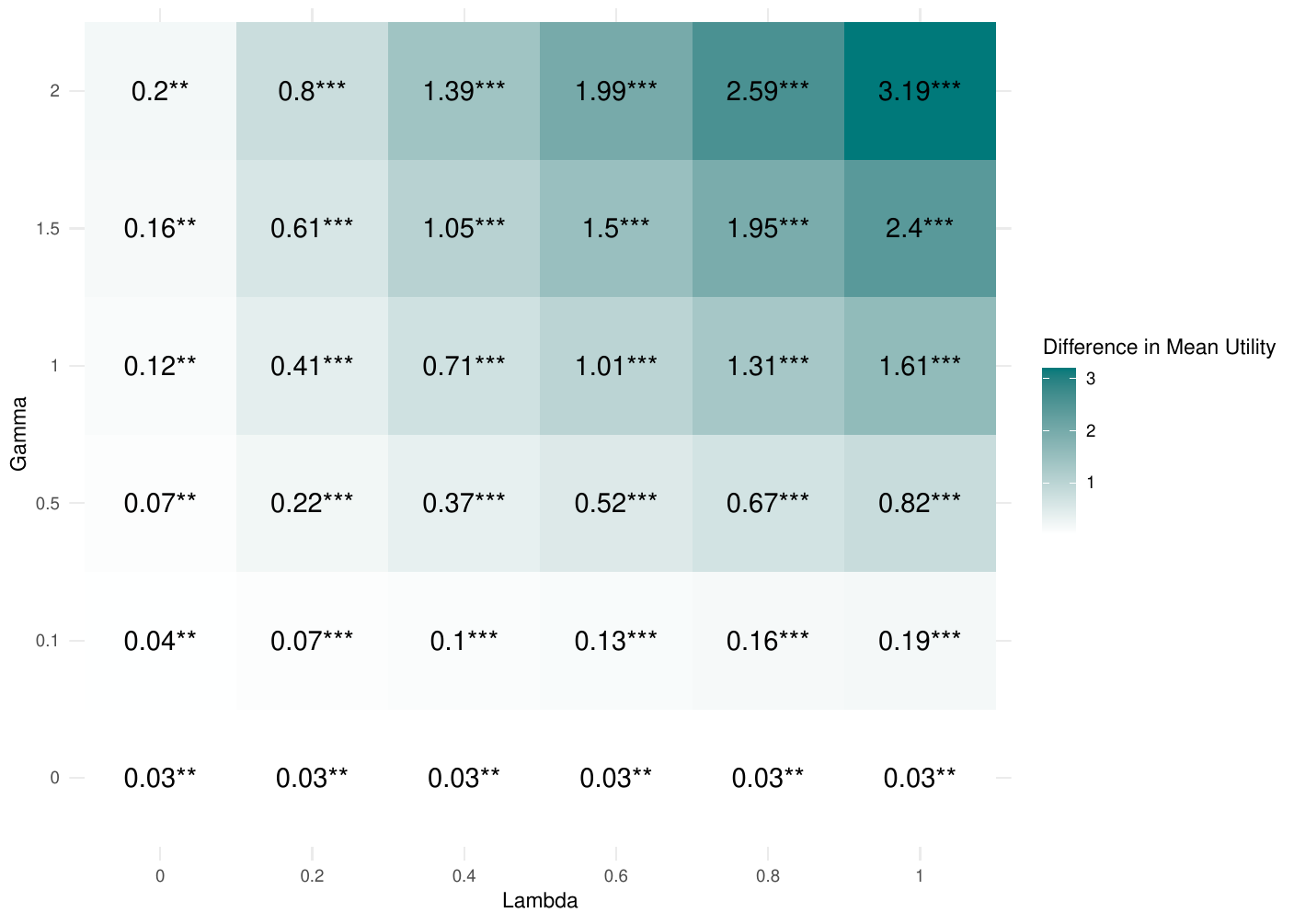}
\caption{Mean utility difference across $N=500$ simulated networks (SUGM minus dyadic), shown over the spillover strength $\gamma$ and the mixing parameter $\lambda$ that governs the relative weight on open- versus closed-triad spillovers.}
    \label{fig:heatmap_simulations}
         \end{figure}

We next translate these structural differences into welfare differences by computing $\bar U(A)$
for each simulated network and comparing the dyadic and SUGM predictions. Figure~\ref{fig:heatmap_simulations}
plots the mean difference in utility (SUGM minus dyadic) across simulations as a function of
spillover strength $\gamma$ and the spillover-mix parameter $\lambda$.
When spillovers are small (low $\gamma$), the welfare gap is close to zero: because both models generate similar numbers of links, they imply similar utility when
utility is dominated by direct connections. As spillovers become more important (higher
$\gamma$), the welfare gap increases sharply, reflecting that the SUGM generates substantially
more triangles and thus more opportunities for indirect benefits and embedded relationships.
Moreover, the gap varies systematically with $\lambda$, consistent with the fact that changing the
relative importance of open- versus closed-triad spillovers changes how much the additional
triadic structure produced by the SUGM is valued.

Overall, the simulations show that allowing for triadic closure can substantially change counterfactual welfare evaluations. In settings where policy objectives depend on clustering, cohesion, or spillovers through mutual
friends, dyadic models thus  risk understating the welfare consequences of interventions that operate
through group structure and cliques.

\section{Conclusion}

This paper evaluates how group membership---in small groups with a social purpose and larger groups with an academic purpose---affects interactions between first-year students at university.
To achieve this, we first exploit initial random group assignment and estimate a dyadic 2SLS model. Further we introduce a novel empirical strategy, estimating a subgraph generated model using random assignment. We use this model to simultaneously estimate causal effects on link and triad formation.

Our findings contribute to a better understanding of multiple aspects of the driving forces of social network formation. First, we study how group membership in a university setting affects actual interactions between students. We find large, significant and persistent effects for social groups with around 7 members each and small effects of mixed significance for classrooms with a size of around 30 students each. Based on these results, it becomes clear that assignment to social groups constitutes a potent intervention for engineering friendship circles, at least under the conditions studied by us.

Second, we find evidence that social groups primarily influence interactions through the formation of small cliques (i.e., triads) rather than just pairwise connections. In contrast, our estimates for classrooms are noisier but suggest no significant effect on triangle formation, indicating that network formation processes operate very differently across group types.

Third, our simulation results show that incorporating triadic closure is not merely a modeling refinement but can meaningfully change welfare and policy conclusions. Networks simulated from the dyadic and SUGM specifications can exhibit similar link density yet markedly different clustering, with the SUGM generating substantially more triangles. When we evaluate these counterfactual networks using a simple utility framework that allows for spillovers through mutual friends, the implied welfare differences grow with the strength of spillovers. This highlights that interventions such as social-group assignment may improve outcomes primarily by fostering clustered friendship structures and the associated indirect benefits, effects that would be understated by models that ignore triangle formation.

%Jointly, our results suggest that assigning students to small groups with a primarily social purpose is an effective measure to influence the structure of friendship networks. Further, researchers should be cautious in using group membership as a proxy for friendship as our results have shown that the effect of group membership on actual interactions can be vary greatly, depending on group structure and size. Lastly, policy makers should consider potential interactions of social foci with triadic closure when planning interventions. 

Overall, our results suggest that assigning students to small groups with a primarily social purpose can be an effective way to shape the structure of friendship networks. At the same time, several limitations should be considered when interpreting these findings. First, participation in the study is voluntary, raising the possibility of self-selection into the sample. Students who choose to participate may differ systematically from non-participants, for example in their sociability or engagement with university life, which could affect the observed interaction patterns. Second, the external validity of our results may be limited by the specific institutional context of the study. The data come from a relatively homogeneous cohort of first-year students at a Danish technical university in 2013, and the relatively low levels of homophily we observe may partly reflect this setting. Third, our empirical approach relies on independence assumptions inherent to the SUGM framework, which abstract from potentially richer forms of dependence in network formation beyond links and triangles.

These considerations caution against extrapolating the quantitative magnitudes of our estimates to other contexts. Nevertheless, the randomized assignment of students to groups provides a rare opportunity to study causal drivers of network formation in a natural institutional environment. By combining this design with a framework that explicitly incorporates triadic closure, our analysis highlights how institutional structures can shape not only the prevalence of social ties but also the clustered architecture of friendship networks. More broadly, our results underscore the importance of accounting for higher-order network structure when evaluating interventions that aim to influence social interactions.

%\input{tables/network_stats}
%%%%%%%%%%%%%%%%%%%%%%%%%%%%%%%%%%%%%%%%%%%%%%
%% Single Appendix:            %%
%%%%%%%%%%%%%%%%%%%%%%%%%%%%%%%%%%%%%%%%%%%%%%

\clearpage

\section*{Appendix - Details of control function estimator of the SUGM model} \label{app}

This section provides the derivations underlying the control function estimator for the SUGM model.

In addition to the setup and assumptions from  the main text, we use the following additional notation: We define $L_{ijs}^{incidental}$ as a dummy variables for whether a triangle has formed in the first step which implies that there is a link between $i$ and $j$ from study program $s$ in the final network:
%NH NOTE ABOUT POST AI EDITING: Check that these two definitions have been edited ocrrectly and are now right.
\begin{align}
L_{ijs}^{incidental}=
\begin{cases}
    1, & \text{if} \quad T_{ijks}^{direct}=1 \textrm{ for some } k \\
   0,              & \text{otherwise}
\end{cases} 
\end{align}
Below we will refer to this also as the pairwise link forming \emph{incidentally}.

Similarly,we define $T_{ijks}^{incidental}$ as a dummy for whether a triangle has formed incidentally between $i$, $j$ and $k$ from study line $s$. This occurs if each of the links in the triangle has either formed directly or has formed as part of some other triangle:
\begin{align}
T_{ijks}^{incidental}=
\begin{cases}
    1, & \text{if} \quad  \forall\,\left(\iota \kappa\right)\in\left\{ (i,j),(i,k),(j,k)\right\}:\, L_{\iota \kappa s}^{direct}=1\vee\max_{\mu\in\varOmega_{s}\setminus\{i,j,k\}}T_{\iota \kappa \mu s}^{direct}=1 \\
   0,              & \text{otherwise}
\end{cases} 
\end{align}

Finally, to shorten notation we let $Z_{ijs}$ denote the (recentered) instrument and $G_{ijs}$ denote the pairwise same group dummy: 
$$Z_{ijs}=\widetilde{SameAssignedGroup_{ijs}}$$
$$G_{ijs}=SameGroup_{ijs}$$
and also introduce shorthand notation for the conditional probabilities that links and triangles form directly:
\begin{align}
p_{ijs}^L=P\left(L_{ijs}^{direct}=1|\boldsymbol{G},\boldsymbol{\eta}\right) \\
p_{ijks}^T=P\left(  T_{ijks}^{direct}=1|\boldsymbol{G},\boldsymbol{\eta}\right)
\end{align}

\subsection{Conditional probability of forming a link or triangle in the final network}

We start by deriving expressions for the conditional probability of forming (seeing) pairwise links and triangles in the final observed network.

First we note that the (conditional) probability that a pairwise link
forms incidentally is equal to the probability that at least one triangle
containing the pair forms directly. Working step by step we can then
write:

\begin{align*}
P\left(L_{ijs}^{incidental}=1|\boldsymbol{G},\boldsymbol{\eta}\right) & =P\left(\max_{k\in\varOmega_{s}\setminus\left\{ i,j\right\} }T_{ijks}^{direct}=1|\boldsymbol{G},\boldsymbol{\eta}\right)\\
 & =1-P\left(\max_{k\in\varOmega_{s}\setminus\left\{ i,j\right\} }T_{ijks}^{direct}=0|\boldsymbol{G},\boldsymbol{\eta}\right)\\
 & =1-\underset{k\in\varOmega_{s}\setminus\left\{ i,j\right\} }{\Pi}P\left(T_{ijks}^{direct}=0|\boldsymbol{G},\boldsymbol{\eta}\right)\\
 & =1-\underset{k\in\varOmega_{s}\setminus\left\{ i,j\right\} }{\Pi}\left(1-p_{ijks}^T\right)
\end{align*}

The second line simply writes things in terms of the probability that
none of the involved triangles form. The third line uses the independence underlying the SUGM model, \eqref{eq:sugmstruc}.

With this we can characterize the (conditional) likelihood of observing
a given link in the data. It happens for sure if the link forms directly,
but even if the link does not form directly, it could form incidentally:

\begin{align}
P\left(L_{ijs}=1|\boldsymbol{G},\boldsymbol{\eta}\right)= & p_{ijs}^L+\left(1-p_{ijs}^L\right)P\left(L_{ijs}^{incidental}=1|\boldsymbol{G},\boldsymbol{\eta}\right)\nonumber \\
 = & p^L_{ijs}+\left(1-p^L_{ijs}\right)\left(1-\underset{k\in\varOmega_{s}\setminus\left\{ i,j\right\} }{\Pi}\left(1-p^T_{ijks}\right)\right) \label{eq:Ldirobs}
\end{align}

Next we go through similar steps for the conditional probability of observing
a triangle in the data. A triangle forming incidentally means that
each of its three links must form directly or as part of some other
triangle forming directly. Whether these links form and/or these other
triangles form however are independent events.\footnote{Note a key fact here is that each other triangle can overlap with
at most one link in the triangle under consideration (because two
triangles either share 0, 1 or 3 links).} Going through steps similar to the ones used just above gives us:

\begin{align*}
P\Big( & T_{ijks}^{incidental}=1|\boldsymbol{G},\boldsymbol{\eta}\Big)=\\
&P\left(\forall\,\left(\iota \kappa\right)\in\left\{ (i,j),(i,k),(j,k)\right\}:\, L_{\iota \kappa s}^{direct}=1\vee\max_{\mu\in\varOmega_{s}\setminus\{i,j,k\}}T_{\iota \kappa \mu s}^{direct}=1 |\boldsymbol{G},\boldsymbol{\eta}\right)\\
 & =\underset{\left(\iota \kappa\right)\in\left\{ (i,j),(i,k),(j,k)\right\} }{\Pi}P\left(L_{\iota \kappa s}^{direct}=1\vee\max_{\mu\in\varOmega_{s}\setminus\{i,j,k\}}T_{\iota \kappa \mu s}^{direct}=1|\boldsymbol{G},\boldsymbol{\eta}\right)\\
 & =\underset{\left(\iota \kappa\right)\in\left\{ (i,j),(i,k),(j,k)\right\} }{\Pi}\left(1-P\left(L_{\iota \kappa s}^{direct}=0\wedge\max_{\mu\in\varOmega_{s}\setminus\{i,j,k\}}T_{\iota \kappa \mu s}^{direct}=0|\boldsymbol{G},\boldsymbol{\eta}\right)\right)\\
 & =\underset{\left(\iota \kappa\right)\in\left\{ (i,j),(i,k),(j,k)\right\} }{\Pi}\left(1-P\left(L_{\iota \kappa s}^{direct}=0|\boldsymbol{G},\boldsymbol{\eta}\right)\underset{\mu\in\varOmega_{s}\setminus\{i,j,k\}}{\Pi}P\left(T_{\iota \kappa \mu s}^{direct}=0|\boldsymbol{G},\boldsymbol{\eta}\right)\right)\\
 & =\underset{\left(\iota \kappa\right)\in\left\{ (i,j),(i,k),(j,k)\right\} }{\Pi}\left(1-\left(1-p^L_{\iota \kappa s}\right)\underset{\mu\in\varOmega_{s}\setminus\{i,j,k\}}{\Pi}\left(1-p^T_{\iota \kappa \mu s}\right)\right)
\end{align*}

The second and fourth line uses the independence in \eqref{eq:sugmstruc}.
The other lines simply write out complementary probabilities.

With this expression can then write the probability of observing a
given triangle by noting that it will be observed if formed directly
but otherwise could still be observed because it formed incidentally:

\begin{align}
P\!\left(T_{ijks}=1 \mid \boldsymbol{G},\boldsymbol{\eta}\right)
=& p^T_{ijks} + \Bigl(1-p^T_{ijks}\Bigr)
      P\!\left(T_{ijks}^{incidental}=1 \mid \boldsymbol{G},\boldsymbol{\eta}\right)
\nonumber\\[0.3em] %%%
=& p^T_{ijks}
\nonumber\\
&\, \, + \Bigl(1-p^T_{ijks}\Bigr) \nonumber \\
& \quad \times \underset{\left(\iota \kappa\right)\in\left\{ (i,j),(i,k),(j,k)\right\} }{\Pi}\left(1-\left(1-p^L_{\iota \kappa s}\right)\underset{\mu\in\varOmega_{s}\setminus\{i,j,k\}}{\Pi}\left(1-p^T_{\iota \kappa \mu s}\right)\right)
\label{eq:Tdirobs}
\end{align}

\subsection{Probability of forming links and triangles, conditional on instruments and first-stage errors}
Next we derive expressions for the probability of seeing links and triangles conditional on the vector of instruments and first-stage errors.We start by inspecting the conditional probabilities $p^L_{\iota \kappa s}$ and $p^T_{\iota \kappa \mu s}$. Note that these are random variables due to their dependence on $\boldsymbol{G},\boldsymbol{\eta}$. 

Using the definition of $\xi_{ijs}$ and $\xi_{ijks}$ as well as \eqref{eq:ldirect} and \eqref{eq:tdirect} we can write:
\begin{align*}
    p^L_{ijs}&=\beta_{0}+\beta_{1}G_{ijs}+h_{L}(u_{ijs})+\xi_{ijs} \\
p^T_{ijks}&=\theta_{0}+\theta_{1}\sum_{\left(\iota\kappa\right)\in\left\{ (i,j),(i,k),(j,k)\right\} }G_{\iota\kappa s}+h_{T}(u_{ijs},u_{iks,}u_{jks})+\xi_{ijks}
\end{align*}
From the first-stage, $G_{ijs}$ is a deterministic function of $\boldsymbol{Z}$ and $\boldsymbol{u}$. Combining this with the \eqref{eq:sugmstrucCF}, it follows that $\left\{ p^L_{ijs}\right\} _{(i,j)\in \Omega_s^L} \cup \left\{ p^T_{ijks}\right\} _{(i,j,k)\in \Omega_s^T}$ is mutually independent conditional on $\boldsymbol{Z},\boldsymbol{u}$. Defining $\tilde{p}^L_{ijs}=E[p^L_{ijs}|\boldsymbol{Z},\boldsymbol{u}]$ and $\tilde{p}^T_{ijks}=E[p^T_{ijks}|\boldsymbol{Z},\boldsymbol{u}]$ this implies the following for the conditional expectations of observed link and triangle indicators:
\begin{align}
E\left[L_{ijs}|\boldsymbol{Z},\boldsymbol{u}\right] = & E\left[P\left(L_{ijs}=1|\boldsymbol{G},\boldsymbol{\eta}\right)|\boldsymbol{Z},\boldsymbol{u}\right] \nonumber \\
= &\tilde{p}^L_{ijs}+\left(1-\tilde{p}^L_{ijs}\right)\left(1-\underset{k\in\varOmega_{s}\setminus\left\{ i,j\right\} }{\Pi}\left(1-\tilde{p}^T_{ijks}\right)\right)& \label{eq:LcondZu}
\end{align}

\begin{align}
   E\left[T_{ijks} \mid \boldsymbol{Z},\boldsymbol{u}\right] =& E\left[ P\!\left(T_{ijks}=1 \mid \boldsymbol{G},\boldsymbol{\eta}\right) \mid \boldsymbol{Z},\boldsymbol{u} \right]
\nonumber \\
=& \tilde{p}^T_{ijks}
\nonumber\\
&\, \, + \Bigl(1-\tilde{p}^T_{ijks}\Bigr) \nonumber \\
& \quad \times \underset{\left(\iota \kappa\right)\in\left\{ (i,j),(i,k),(j,k)\right\} }{\Pi}\left(1-\left(1-\tilde{p}^L_{\iota \kappa s}\right)\underset{\mu\in\varOmega_{s}\setminus\{i,j,k\}}{\Pi}\left(1-\tilde{p}^T_{\iota \kappa \mu s}\right)\right) \label{eq:TcondZu}
\end{align}

\subsection{Parametric restriction on the conditional expectation of the unobservables}
To set up our control function estimation, we next restrict the conditional expectation of the unobservables to be parametric functions, with parameter vectors $\boldsymbol{\rho}^L$ and $\boldsymbol{\rho}^T$ :
$$h_{L}(u_{ijs})=\tilde{h}_{L}(u_{ijs};\boldsymbol{\rho}^L)-E\left[\tilde{h}_{L}(u_{ijs};\boldsymbol{\rho}^L)\right]$$
$$h_{T}(u_{ijs},u_{iks},u_{jks})=\tilde{h}_{T}(u_{ijs},u_{iks},u_{jks};\boldsymbol{\rho}^T)-E\left[\tilde{h}_{T}(u_{ijs},u_{iks},u_{jks};\boldsymbol{\rho}^T)\right]$$
Note that subtracting off expectated value in these definitions is simply a convenient way to impose a mean of zero.

With these parametric restrictions we can write:
\begin{align}
\widetilde{p}^L_{ijs}&=\beta_{0}+\beta_{1}G_{ijs}+h_{L}(u_{ijs}) \nonumber \\
&=\psi_L +\beta_1 G_{ijs} +\tilde{h}_{L}(u_{ijs};\boldsymbol{\rho}^L) \nonumber\\
&=f_L(G_{ijs},u_{ijs};\psi_L,\beta_1,\boldsymbol{\rho}^L) \label{eq:ptildL}
\end{align}

the second to last line defines the parameter $\psi_L=\beta_{0}-E\left[\tilde{h}_{L}(u_{ijs};\boldsymbol{\rho}^L)\right]$ while the last line defines the function $f_L$. Completely analogously, we can write:

\begin{align}
\widetilde{p}^T_{ijks} &=\theta_{0}+\theta_{1}\sum_{\left(\iota\kappa\right)\in\left\{ (i,j),(i,k),(j,k)\right\} }G_{\iota\kappa s}+h_{T}(u_{ijs},u_{iks},u_{jks})\nonumber\\
&=\psi_T+\theta_{1}\sum_{\left(\iota\kappa\right)\in\left\{ (i,j),(i,k),(j,k)\right\} }G_{\iota\kappa s}+\tilde{h}_{T}(u_{ijs},u_{iks},u_{jks};\boldsymbol{\rho}^T) \nonumber\\
&=f_T(G_{ijs},G_{iks},G_{jks},u_{ijs},u_{iks},u_{jks};\psi_T,\theta_1,\boldsymbol{\rho}^T) \label{eq:ptildT}
\end{align}

\subsection{Control function estimation}
Combining  \eqref{eq:LcondZu}, \eqref{eq:TcondZu}, \eqref{eq:ptildL} and \eqref{eq:ptildT} we can write the conditional expectations $E\left[L_{ijs} \mid  \boldsymbol{Z}, \boldsymbol{u} \right]$ and $E\left[T_{ijks} \mid \boldsymbol{Z}, \boldsymbol{u}\right]$ as functions of first-stage error terms, group membership and parameters:
\begin{align*}
    E\left[L_{ijs} \mid  \boldsymbol{Z}, \boldsymbol{u} \right] & =g_L \left(i,j,s,\boldsymbol{G},\boldsymbol{u};\psi_L,\psi_T,\beta_1,\theta_1,\boldsymbol{\rho}^L,\boldsymbol{\rho}^T)\right) \\
    E\left[T_{ijks} \mid \boldsymbol{Z}, \boldsymbol{u}\right] & = g_T \left(i,j,k,s,\boldsymbol{G},\boldsymbol{u};\psi_L,\psi_T,\beta_1,\theta_1,\boldsymbol{\rho}^L,\boldsymbol{\rho}^T \right)
\end{align*}
where
\begin{align}
g_L & \left(i,j,s,\boldsymbol{G},\boldsymbol{u};\psi_L,\psi_T,\beta_1,\theta_1,\boldsymbol{\rho}^L,\boldsymbol{\rho}^T)\right) \nonumber \\  &= f_L(G_{ijs},u_{ijs};\psi_L,\beta_1,\boldsymbol{\rho}^L)\nonumber \\
 & \qquad\qquad+\left(1-f_L(G_{ijs},u_{ijs};\psi_L,\beta_1,\boldsymbol{\rho}^L)\right) \nonumber \\ & \qquad \qquad \qquad \times \left(1-\underset{k\in\varOmega_{s}\setminus\left\{ i,j\right\} }{\Pi} \left( 1- f_T(G_{ijs},G_{iks},G_{jks},u_{ijs},u_{iks},u_{jks};\psi_T,\theta_1,\boldsymbol{\rho}^T)\right) \right) \label{eq:Moment1} 
\end{align}
and

\begin{align}
g_T & \left(i,j,k,s,\boldsymbol{G},\boldsymbol{u};\psi_L,\psi_T,\beta_1,\theta_1,\boldsymbol{\rho}^L,\boldsymbol{\rho}^T \right) \nonumber
\\
&= f_T\bigl(
    G_{ijs}, G_{iks}, G_{jks},
    u_{ijs}, u_{iks}, u_{jks};
    \psi_T, \theta_1, \boldsymbol{\rho}^T
\bigr) \nonumber\\
&\quad + \Bigl(1 - f_T\bigl(
    G_{ijs}, G_{iks}, G_{jks},
    u_{ijs}, u_{iks}, u_{jks};
    \psi_T, \theta_1, \boldsymbol{\rho}^T
\bigr)\Bigr) \nonumber\\
&\qquad \times
\Pi_{(\iota,\kappa)\in\{(i,j),(i,k),(j,k)\}}
\Biggl[
1 - \Bigl(1 - f_L\bigl(
    G_{\iota\kappa s}, u_{\iota\kappa s};
    \psi_L, \beta_1, \boldsymbol{\rho}^L
\bigr)\Bigr) \nonumber\\
&\qquad\qquad \times
\Pi_{\mu\in\varOmega_s\setminus\{i,j,k\}}
\Bigl(
1 - f_T\bigl(
    G_{\iota\kappa s}, G_{\iota\mu s}, G_{\kappa\mu s},
    u_{\iota\kappa s}, u_{\iota\mu s}, u_{\kappa\mu s};
    \psi_T, \theta_1, \boldsymbol{\rho}^T
\bigr)
\Bigr)
\Biggr]
\label{eq:Moment2}
\end{align}

This leads to the two-step control function estimator that we use which entails the following two steps:

\begin{enumerate}
    \item Estimate the first stage regression, \eqref{eq:1sttriad}, using data on all pairs, to obtain the residuals $\widehat{\boldsymbol{u}}$.
    \item Use the residuals as estimates of the true first stage errors and estimate the parameters of interest $\beta_1,\theta_1$ as well as $\psi_L,\psi_T,\boldsymbol{\rho}^L,\boldsymbol{\rho}^T$ using (weighted) nonlinear least squares on the sample of all pairs and all triads:
    $$(\hat{\beta}_1,\hat{\theta}_1,\hat{\psi}_L,\hat{\psi}_T,\widehat{\boldsymbol{\rho}}^L,\widehat{\boldsymbol{\rho}}^T)=\textrm{argmin}_{\beta_1,\theta_1,\psi_L,\psi_T,\boldsymbol{\rho}^L,\boldsymbol{\rho}^T}\quad Q(\beta_1,\theta_1,\psi_L,\psi_T,\boldsymbol{\rho}^L,\boldsymbol{\rho}^T)$$ 
    where 
    \begin{align}
Q(&\beta_1,\theta_1,\psi_L,\psi_T,\boldsymbol{\rho}^L,\boldsymbol{\rho}^T) \\
        =&\sum_s\sum_{ij\in\varOmega_s^L} \left(L_{ijs}-g_L \left(i,j,s,\boldsymbol{G},\widehat{\boldsymbol{u}};\psi_L,\psi_T,\beta_1,\theta_1,\boldsymbol{\rho}^L,\boldsymbol{\rho}^T\right) \right)^2 \nonumber \\
        & \qquad +\omega_T \sum_s \sum_{ijk\in\varOmega_s^T} \left(T_{ijks}- g_T  \left(i,j,k,s,\boldsymbol{G},\widehat{\boldsymbol{u}};\psi_L,\psi_T,\beta_1,\theta_1,\boldsymbol{\rho}^L,\boldsymbol{\rho}^T \right) \right)^2 \nonumber
    \end{align}
\end{enumerate}

In the above, $\omega_T$ is a deterministic constant that governs how much weight is given to triads in estimation. For our baseline estimates we simply set $\omega_T=1$ (another obvious choice would be to set it to the ratio of pairs to triads in the data so as to give equal weight, $\omega_T=\frac{\sum_s |\varOmega_s^L|}{\sum_s |\varOmega_s^T|}$). To obtain standard errors and conduct inference, we apply a bootstrap procedure that resamples study programs and repeats both steps above on the the bootstrap samples.

In implementing the estimator, we of course need to settle on a particular functional form for $\tilde{h}_L$ and $\tilde{h}_T$. A natural baseline choice would be second order polynomials that are symmetric in the arguments. This corresponds to setting $\boldsymbol{\rho}^L=(\rho_1^L,\rho_2^L)$,$\boldsymbol{\rho}^T=(\rho_1^T,\rho_2^T,\rho_3^T)$ and setting
\begin{align*}
\tilde h_L(u;\boldsymbol{\rho}^L)&=\rho_1^L u+\rho_2^L u^2
\\
    \tilde h_T(u_1,u_2,u_3;\boldsymbol{\rho}^T)
&=
\rho_1^T(u_1+u_2+u_3)
+\rho_2^T(u_1^2+u_2^2+u_3^2)
+\rho_3^T(u_1u_2+u_1u_3+u_2u_3)
\end{align*}

Other choices could be simply (symmetric) linear functions
\begin{align*}
\tilde h_L(u;\boldsymbol{\rho}^L)&=\rho_1^L u
\\
\tilde h_T(u_1,u_2,u_3;\boldsymbol{\rho}^T)&=\rho_1^T(u_1+u_2+u_3),
\end{align*}
with $\boldsymbol{\rho}^L=(\rho_1^L)$ and $\boldsymbol{\rho}^T=(\rho_1^T)$, or a higher-order approach based on monomials such as
\begin{align*}
\tilde h_L(u;\rho_L)=&\rho_1^L u+\rho_2^L u^2+\rho_3^L u^3
\\
    \tilde h_T(u_1,u_2,u_3;\rho^T)
=
&\rho_1^T(u_1+u_2+u_3)
+\rho_2^T(u_1^2+u_2^2+u_3^2)
+\rho_3^T(u_1u_2+u_1u_3+u_2u_3)
\\
&\quad +\rho_4^T(u_1^3+u_2^3+u_3^3)
\\
&\qquad+\rho_5^T(u_1^2u_2+u_1^2u_3+u_2^2u_1+u_2^2u_3+u_3^2u_1+u_3^2u_2)
\\
& \qquad \quad +\rho_6^T u_1u_2u_3
\end{align*}
with $\boldsymbol{\rho}^L=(\rho_1^L,\rho_2^L,\rho_{3}^L)$ and $\boldsymbol{\rho}^T=(\rho_1^T,\rho_2^T,\rho_3^T,\rho_4^T,\rho_5^T,\rho_6^T)$.

Note that after implementing the estimator above, we can also recover estimates of the constant terms in the original link and triangle formation equations by simply inverting the definitions of $\psi_L$ and $\psi_T$, and plugging in estimates and sample averages for expectations and true parameters:
    \begin{align*}
\hat\beta_0
& =
\hat\psi_L
+
\frac{1}{\sum_s |\Omega_s^L|}
\sum_s \sum_{(i,j)\in \Omega_s^L}
\tilde h_L(\hat u_{ijs};\hat{\boldsymbol{\rho}}^L),
\\
\hat\theta_0
& =
\hat\psi_T
+
\frac{1}{\sum_s |\Omega_s^T|}
\sum_s \sum_{(i,j,k)\in \Omega_s^T}
\tilde h_T(\hat u_{ijs},\hat u_{iks},\hat u_{jks};\hat{\boldsymbol{\rho}}^T)
\end{align*}

\putbib[bibliography]
\end{bibunit}
%\bibliographystyle{unsrtnat}
%\bibliography{bibliography}  %%% Uncomment this line and comment out the ``thebibliography'' section below to use the external .bib file (using bibtex) .

%%% Uncomment this section and comment out the \bibliography{references} line above to use inline references.
% \begin{thebibliography}{1}

% 	\bibitem{kour2014real}
% 	George Kour and Raid Saabne.
% 	\newblock Real-time segmentation of on-line handwritten arabic script.
% 	\newblock In {\em Frontiers in Handwriting Recognition (ICFHR), 2014 14th
% 			International Conference on}, pages 417--422. IEEE, 2014.

% 	\bibitem{kour2014fast}
% 	George Kour and Raid Saabne.
% 	\newblock Fast classification of handwritten on-line arabic characters.
% 	\newblock In {\em Soft Computing and Pattern Recognition (SoCPaR), 2014 6th
% 			International Conference of}, pages 312--318. IEEE, 2014.

% 	\bibitem{hadash2018estimate}
% 	Guy Hadash, Einat Kermany, Boaz Carmeli, Ofer Lavi, George Kour, and Alon
% 	Jacovi.
% 	\newblock Estimate and replace: A novel approach to integrating deep neural
% 	networks with existing applications.
% 	\newblock {\em arXiv preprint arXiv:1804.09028}, 2018.

% \end{thebibliography}

\clearpage
\part*{Supplementary Material}
\addcontentsline{toc}{part}{Supplementary Material}   % optional divider in ToC

\appendix
\setcounter{section}{0}
\renewcommand{\thesection}{\Alph{section}}
\renewcommand{\thesubsection}{\thesection.\arabic{subsection}}

%%%%%%%%%%%%%%%%%%%%%%%%%%%%%%%%%%%%%%%%%%%%%%%%%%%%%%%%%%%%%%%%%%%%%%%%%
%%%% Main text entry area:                                             %%
%%%%%%%%%%%%%%%%%%%%%%%%%%%%%%%%%%%%%%%%%%%%%%%%%%%%%%%%%%%%%%%%%%%%%%%%%
\renewcommand{\thesection}{\Alph{section}}

\renewcommand{\thetable}{\Alph{section}\arabic{table}}
\setcounter{table}{0}
\renewcommand{\thefigure}
{\Alph{section}\arabic{figure}}
\setcounter{figure}{0}

\begin{bibunit}[ecta-fullname]
\section{Balance Tests}
\setcounter{table}{0}
\setcounter{figure}{0}
\label{app:balance}
In addition to the data described in the main paper, we were able to access data from registries at Statistics Denmark which was used to conduct a balance test. Further, these variables are used in an additional analysis on homophily patterns reported  in the Supplemental Material (Section \ref{app:homophily}). In the following we first describe the additional variables obtained and then report the results of the balance test.

\subsection{Register Data Variables}

\textbf{High School Graduation GPA.} Based on data of the Danish registry on upper secondary education (UDGK), the high school graduation GPA is computed as the mean of oral and written exam grades. The Danish system uses a 7-point grading scale with values 12, 7, 4, 02, 00 and -3. 12 is the highest possible grade, while 00 and -3 indicate that the student has failed the exam.

\textbf{Parental Education.} Based on the Danish registry on highest completed education (UDDA), we compute the duration of education (including primary, secondary and higher education) for both parents. The exact mapping from the highest completed education to duration of education in years is shown in Table \ref{tab:education}. For our analysis we use the duration of the higher educated parent.

\begin{table}[t]
\caption{Mapping of education types to years of education}
\label{tab:education}
\begin{tabular}{lcc}  
\hline  
Education (Danish) & Education (English translation) & Years\\
Grundskole & Primary school & 9 \\
Erhvervsfaglige praktik- og hovedforløb & Vocational training and main course & 11\\
Erhvervsfaglige grundforløb & Vocational foundation course & 10\\
Forberedende uddannelser & Preparatory education & 11\\
Efteruddannelse af specialarb/faglærte & Continuing education for specialists/skilled workers & 11\\
Almengymnasial uddannelser & General upper secondary education & 12\\
Erhvervsgymnasial uddannelser & Vocational upper secondary education & 12 \\
Korte videregående uddannelser & Short-cycle higher education & 14\\
Mellemlange videregående uddannelser & Medium-cycle higher education & 15\\
Bachelor & Bachelor's degree & 15\\
Lange videregående uddannelser & Long-cycle higher education & 17\\
Forskeruddannelser & PhD/Doctoral education & 20\\
\hline
\end{tabular}

\end{table}

\textbf{Parental Income \& Wealth.} Income and wealth data for both parents are obtained from the Danish income registry (INDK). Data is available on a yearly basis for the years 2004 to 2012. The income measure is based on the gross household income (QBRUKOR2), comprising personal taxable income (such as salary, pensions from pension funds, transfer income from public funds, fees etc.), income as self-employed (i.e income from self-employed business including consulting fees), capital income (i.e. interest income etc.) and income from abroad \citep{DSTFAMBRUTTOINDK}. For our measure we first calculate the mean household income across parents (note: parental household incomes may often be identical if both parents live in the same household) for each year in our data and  then average over years. The wealth measure for an individual parent is defined as the sum of the market value of bond assets (OBLAKT) \citep{DSTOBLAKT}, market value of Danish shares and investment fund certificates (KURSAKT) \citep{DSTKURSAKT} and deposits in banks (BANKAKT) \citep{DSTBANKAKT}. For our wealth measure we calculate the joint wealth of both parents for each year and then average over years.

\textbf{Immigration Background.} Data on immigration background is obtained from the Danish population registry (BEF). The variable differentiates between three categories: (1) immigrants (individuals born abroad where neither parent is both a Danish citizen and born in Denmark), (2) descendants of immigrants (individuals born in Denmark where none of the parents are both Danish citizens and born in Denmark) and (3) persons of Danish origin (individuals who have at least one parent who is both a Danish citizen and born in Denmark). For our analysis we combine immigrants and descendants of immigrants into a single category termed \emph{immigration background}.

\textbf{Sex.} Indicates whether the person is male or female based on their personal registry number. This is collected as a binary variable in the Danish population registry (BEF).

\subsection{Balance Test Results}
Pairs in the same  assigned classroom (social group) and pairs not in the same assigned classroom (social group) are similar across most pre-determined features such as age, high school graduation GPA, parental wealth, parental income, immigration background and duration of parental education. Thus, overall the balance tests provide confirmatory evidence that the randomization was successful.
\input{tables_appendix/A/01_balance}

\section{Measuring links}
\label{app:pca}
\setcounter{table}{0}
\setcounter{figure}{0}
To create a binary indicator of friendship, we employ principal component analysis \citep{wold_principal_1987} on the dyadic dataset. The input variables are the rescaled versions of (1) the average weekly call duration, (2) the average weekly number of exchanged text messages and (3) the average weekly number of meetings outside of the university campus for the second semester. We conduct the PCA on the sample of all pairs within the same study major where we have data on all three measures which has 4,636 observations. Loadings on the first principal component (FPC) were 0.470 for call duration, 0.628 for meetings and 0.619 for SMS, reflecting that call duration has slightly less influence on our PCA based friendship measure than the other two variables. The FPC captures 63.3\% of the total variance of the three individual measures and ranges from -0.281 to 75.392 with 77.9\% of pairs having negative scores. This illustrates that the distribution is highly left-skewed, with a few very extreme observations. 

Based on this we define all pairs as friends whose score of the first principal component is higher than 90\% of the scores among pairs observed within their study major. This cutoff reflects the intuition that close friends usually interact significantly more with each other than average students within the same major and acquaintances. This cutoff results in a median number of one friend within the sample. However, our sample only consists of between 18.6\% and 55.7\% of all students within a study major ($mean = 35.1\%$). If we were to assume that our sub sample is essentially a random draw from the overall student body, this implies that any individual student has the same probability of being friends with any other study major member, no matter if they are part of our sample or not. If we adjust for this and multiply the average number of friends observed per study major with the inverse probability of a student within this major being part of our sample, the estimated average number of friends within study major ranges from 4.1 to 6.9. Table \ref{tab:mean_friends} reports detailed statistics for each study major. A survey conducted at the University of British Columbia, \citeauthor{whillans_misperception_2017} (\citeyear{whillans_misperception_2017}) found that the average first year student reported to have 3.63 close friends (defined as “someone who [they] would likely to confide in/talk to about [themselves] and [their] problems.”). Our measure could thus be regarded as similar, but potentially slightly broader in the definition of friendship.  

\begin{figure}[h!]
    \centering
  \includegraphics[width=0.7\textwidth]{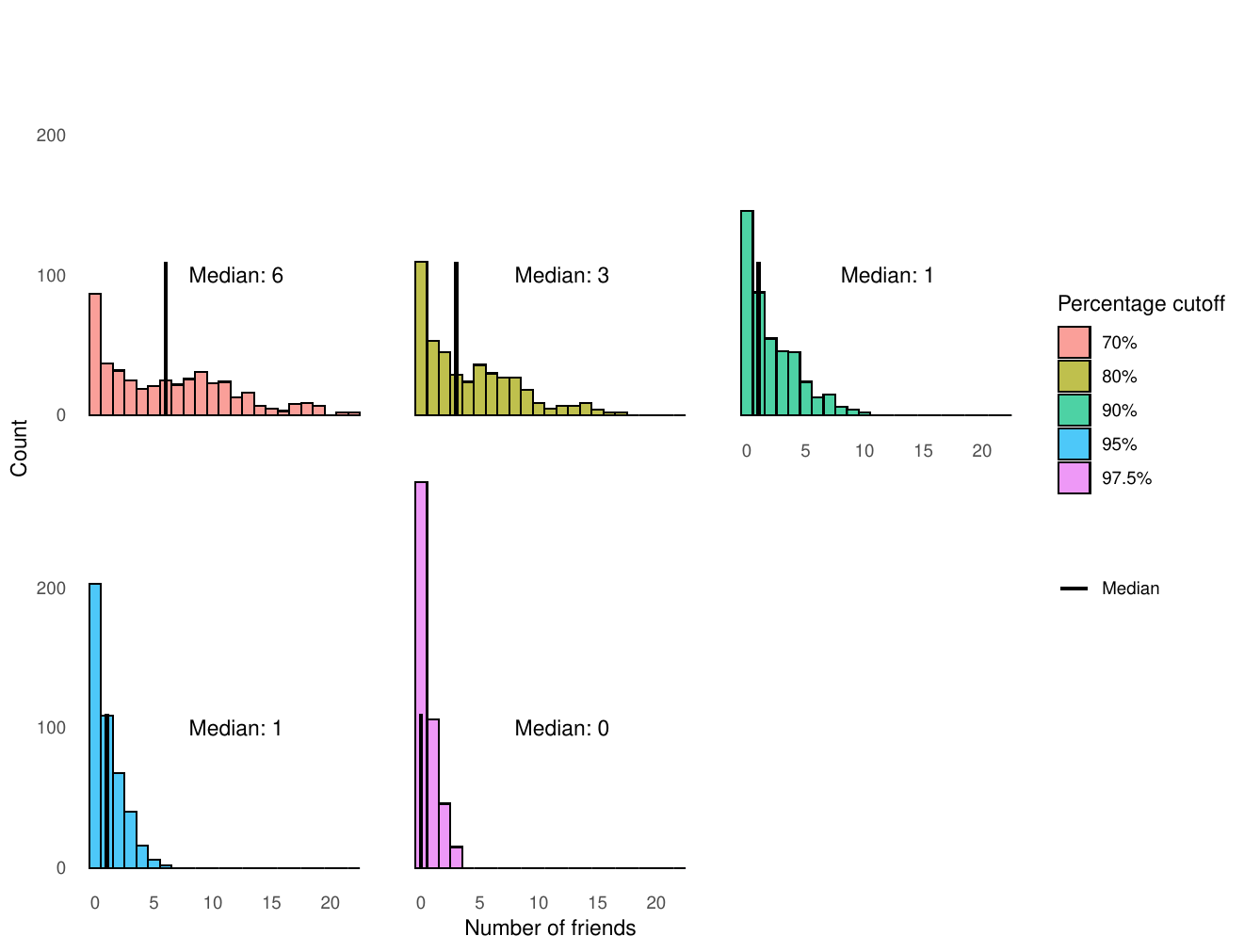}
    \caption{Histogram of the number of friends per individual based on different percentage cutoffs.}
    \label{fig:app_cutoff_dist}
\end{figure}

Figure \ref{fig:app_cutoff_dist} shows the distribution and median number of friends resulting from different choices of the cutoff percentage and Table \ref{tab:num_friends_cutoff} lists the percentage of individuals that have a given number of friends.

\begin{figure}[h!]
    \centering
  \includegraphics[width=\textwidth]{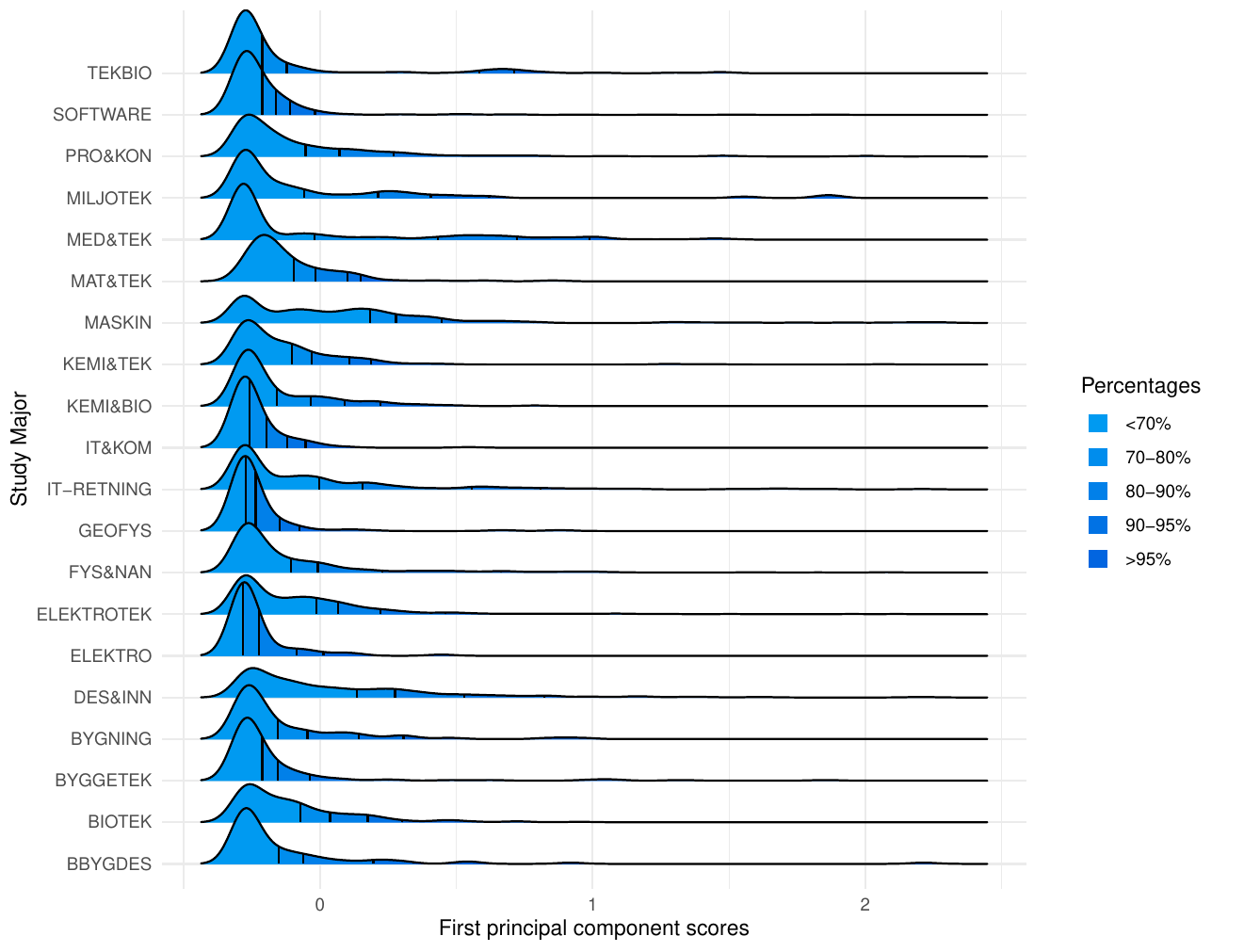}
    \caption{Distribution of the observed scores of the first principal component for each study major. \\ \footnotesize \emph{Note: Due to some extreme observations in the right tail of the distribution, we only plot FPC scores smaller than 2.5. The highest value of the 95th percentile observed in any study program is 2.25 - thus the figure covers at least 95\% of the distribution for all study programs.}}

    \label{fig:app_pca_dist}

\end{figure}

Figure \ref{fig:app_pca_dist} shows the distribution of the first principal component scores within each study major. 

\input{tables_appendix/F/01_study_major_friends}

\input{tables_appendix/F/02_number_of_friends_per_cutoff}

We further report the results of the IV regression on the binary friendship indicator for different cutoff values in Table \ref{tab:iv_binary_pca}. The effect of being in the same social group is significant for all tested thresholds, with the relative effect size in relation to the baseline increasing for higher cutoffs. For classrooms we only find a significant effect for the 90\% and the 80\% cutoffs, indicating that classrooms don't exert an influence on extremly close friendship ties.
\input{tables_appendix/F/03_iv_binary_results}

\input{tables_appendix/F/04_network_stats_per_program}

\FloatBarrier

\FloatBarrier
\section{Instrument Recentering}
\label{app:recentering}
\setcounter{table}{0}
\setcounter{figure}{0}

\subsection{Details on instrument recentering procedure}

We are interested in estimating 
\begin{equation}
\label{eq:main}
y_{ijs} = \beta_0 + \beta_1 SameGroup_{ijs} + v_{ijs}
\end{equation}

As $SameGroup_{ijs}$s is endogeneous we identified a candidate instrument, i.e. \\$SameAssigendGroup_{ijs}$ which, following \cite{borusyak_nonrandom_2023}, we can define as follows:

$$SameAssignedGroup_{ijs} = f_{ijs}(g;w)$$

For simplicity of notation, we denote a dyad by $d := ijs$ and the set of all possible dyads within study programs as $D$. Then $f_1(\cdot), \dots ,f_D(\cdot)$ are a known list of non-stochastic functions, depending on the individual random assignments of both members of the pair, i.e. $g = ((RandomAssignment_i, RandomAssignment_j))_{d=1}^D$ and the relevant characteristics of the individuals in the pair, i.e. $w =((Sex_i, Sex_j, Diet_i, Diet_j, StudyProgram_{ij}))_{d=1}^D$.

We can further rewrite the equation of interest, \ref{eq:main} in terms of the observable moments:

\begin{align*}
\mathbf{E}\left[ \frac{1}{D} \sum_d SameAssignedGroup_d y_d \right]
&= \beta \mathbf{E}\left[ \frac{1}{D} \sum_d SameAssignedGroup_d SameGroup_d \right] \\
&\quad + \mathbf{E}\left[ \frac{1}{D} \sum_d SameAssignedGroup_d v_d \right].
\end{align*}

Thus $\beta$ can be estimated as 
$$\hat{\beta} = \frac{\frac{1}{D}\sum_d SameAssignedGroup_d \times y_d}{\frac{1}{D} \sum_d SameAssignedGroup_d \times SameGroup_d}$$
if $SameAssignedGroup_d$ is a relevant instrument, i.e. $\mathbf{E}[\frac{1}{N} \sum_d SameAssignedGroup_d \times SameGroup_d] \neq 0$ and the instrument is independent of the error term i.e. \\ $\mathbf{E}[\frac{1}{N} \sum_d SameAssignedGroup_d \times v_d] = 0$.

However, as $SameAssignedGroup_d$ is not fully random but dependent on $g$ and $w$, $\mathbf{E}[\frac{1}{N} \sum_d SameAssignedGroup_d \times v_d] = 0$ might not hold, rather

\begin{equation*}
\mathbf{E} \left[ \frac{1}{N} \sum_d SameAssignedGroup_d \times v_d \right] = \mathbf{E} \left[ \frac{1}{N} \sum_d \mu_d \times v_d \right]
\end{equation*}

with $\mu_d = \textbf{E}[f_d(g;w)]$. This follows from the law of iterated expectations, based on the assumption of shock exogeneity, i.e. $\mathbf{E}[SameAssignedGroup_d \times v_d] = \mathbf{E}[\mathbf{E}[f_d(g;w) \times v_d | w]] = \mathbf{E}[\mu_d \mathbf{E}[v_d|w]] = \mathbf{E}[\mu_dv_d]$ for all $d$.

Thus the recentered instrument can be defined as $\widetilde{SameAssignedGroup_d} = SameAssignedGroup_d - \mu_d$.

\begin{align*}
\mathbf{E} \left[ \frac{1}{N} \sum_d \widetilde{SameAssignedGroup_d} \times v_d \right]
&= \mathbf{E} \left[ \frac{1}{N} \sum_d SameAssignedGroup_d \times v_d \right] \\
&\quad - \mathbf{E} \left[ \frac{1}{N} \sum_d \mu_d v_d \right] = 0
\end{align*}

Subsequently, if $\widetilde{SameAssignedGroup_d}$ remains relevant for $SameGroup_d$, $\beta$ can be recovered using the recentered instrument.

\label{app:randomization}
\subsection{Implementation of the Recentering Procedure}
To calculate the random assignments for the calculation of the re-centered instrument, we first define strata, which encode all possible combinations of the variables that were conditioned on in the actual randomization process, i.e. study major and sex for the classroom assignment and study major, sex and diet in the social group assignment. Then we follow the algorithm presented below to randomly draw 1,000 counterfactual assignments.
\\

\newpage
\SetKwFunction{KwFunction}{Function}
\begin{algorithm}[H]
\SetAlgoLined
Create a list of all groups\;
\While{list of all groups is not empty}{
    Select a group\;
    strata\_counts $\leftarrow$ GetCountPerStratum(group)\;
    
    \ForEach{stratum \textbf{in} strata\_counts}{
        count $\leftarrow$ strata\_counts[stratum]\;
        selected\_individuals $\leftarrow$ RandomlyDraw(count, overall\_dataset[stratum])\;
        Assign(selected\_individuals, group)\;
        Remove(selected\_individuals, overall\_dataset)\;
    }
    Remove(group, list of all groups)\;
}
\caption{Stratified Random Assignment to Groups}
\end{algorithm}
\bigskip
\bigskip
\subsection{Distribution of Baseline Same-Group Assignment Probabilities}
\begin{figure}[h]
    \centering
  \includegraphics[width=0.47\textwidth]{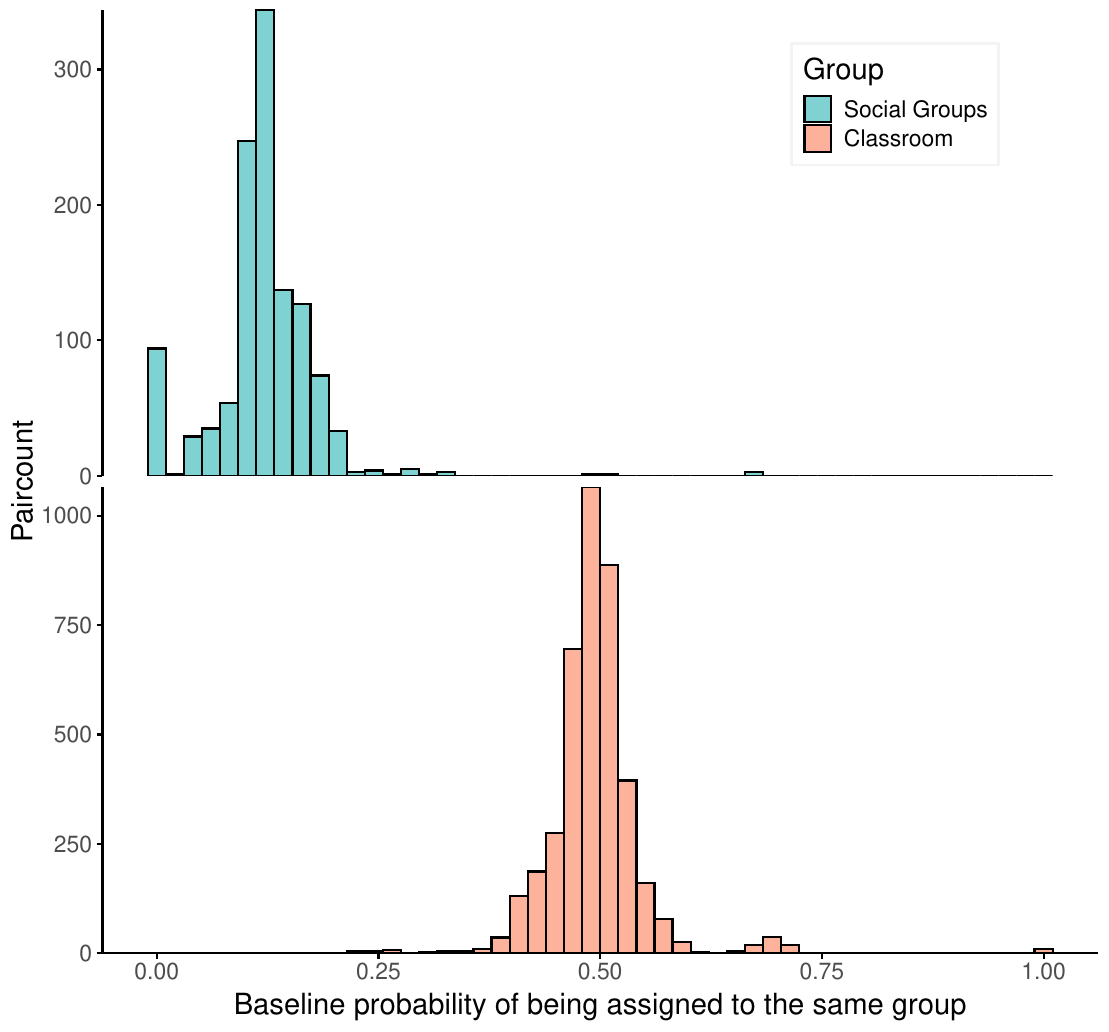}
    \caption{Distribution of pairwise probabilities for being assigned to the same group}
    \label{fig:prob_distribution}
\end{figure}

Figure \ref{fig:prob_distribution} shows the baseline distribution of probabilities for pairs to be in the same group. The distribution clearly shows that, even after correcting for dietary restrictions, sex and study major, pairs do not necessarily have the same likelihood of ending up in the same classroom or social group. For classrooms the median probability of being assigned to the same group is 49.3\%, which is driven by the fact that most study majors consist of two classrooms. There are two study majors (Physics \& Nanotechnologie, Software engineering) where from one of the classrooms we have only males in our sample,  thus leading all females to have a probability of 1 of being assigned to the other classroom. Similarly, in the study major machine engineering, we only have two females in our sample.
For social groups, the median probability of being in the same group is 11.8\%. Here, some individuals (7.9\% of possible pairs) have a 0\% chance of being assigned to the same group due to non-compatible dietary preferences.

\section{First stage and reduced form estimates}
\label{app:first_stage_reduced}
\setcounter{table}{0}
\setcounter{figure}{0}

\subsection{First Stage}
Table \ref{tab:first_stages} includes first stages for all samples used in the main analysis. Sample sizes vary across outcomes due to different students participating.
\input{tables_appendix/C/01_all_first_stage}

\FloatBarrier
\subsection{Reduced form results}

Table \ref{tab:reduced_form} provides reduced form results for the main specification.

\input{tables_appendix/C/02_reduced_form}

\FloatBarrier
\section{Robustness Checks}
\label{app:robustness_checks}
\setcounter{table}{0}
\setcounter{figure}{0}

\subsection{Alternative Definition of Classroom Membership}
The administrative data on TA classroom membership include not only the actual TA classroom assignments but also other classes that students attend. Most users therefore have multiple entries—some corresponding to TA classrooms and others to different course memberships. In our main specification, we adopt a conservative approach and classify a dyad as being in the same group whenever at least one administratively assigned group membership overlaps.

Below, we report results from an alternative specification in which two individuals are considered to be in the same group only if their actual TA classroom assignments coincide. This procedure yields a much smaller sample, as many students lack group membership information that can be matched to classroom names. The estimated effects remain similarly insignificant. For physical meetings and Facebook friendship, we even observe slightly negative effects.

\input{tables_appendix/E/iv_results_alternative_ta}

\subsection{Boostrapped standard errors}
\label{app:bootstrapped}
\begin{figure}[h!]
    \centering

      \centering
  % First subfigure
  \begin{subfigure}[b]{0.48\textwidth}
    \centering
    \includegraphics[width=\textwidth]{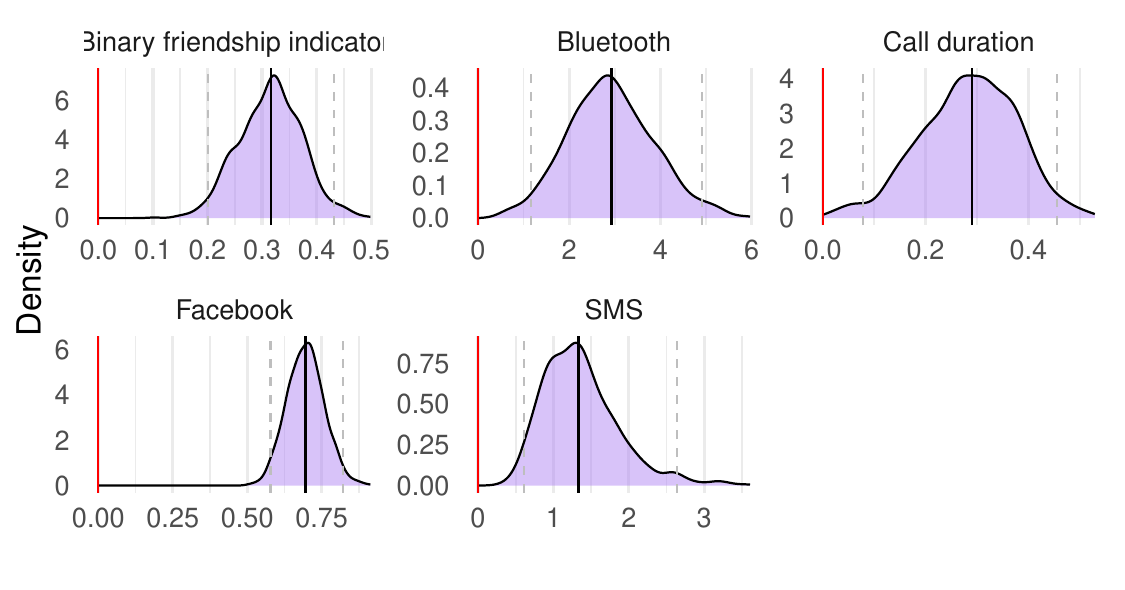} % Replace with your image file
    \caption{Social groups}
    \label{fig:bootstrap_vec}
  \end{subfigure}
  \hfill
    \begin{subfigure}[b]{0.48\textwidth}
    \centering
    \includegraphics[width=\textwidth]{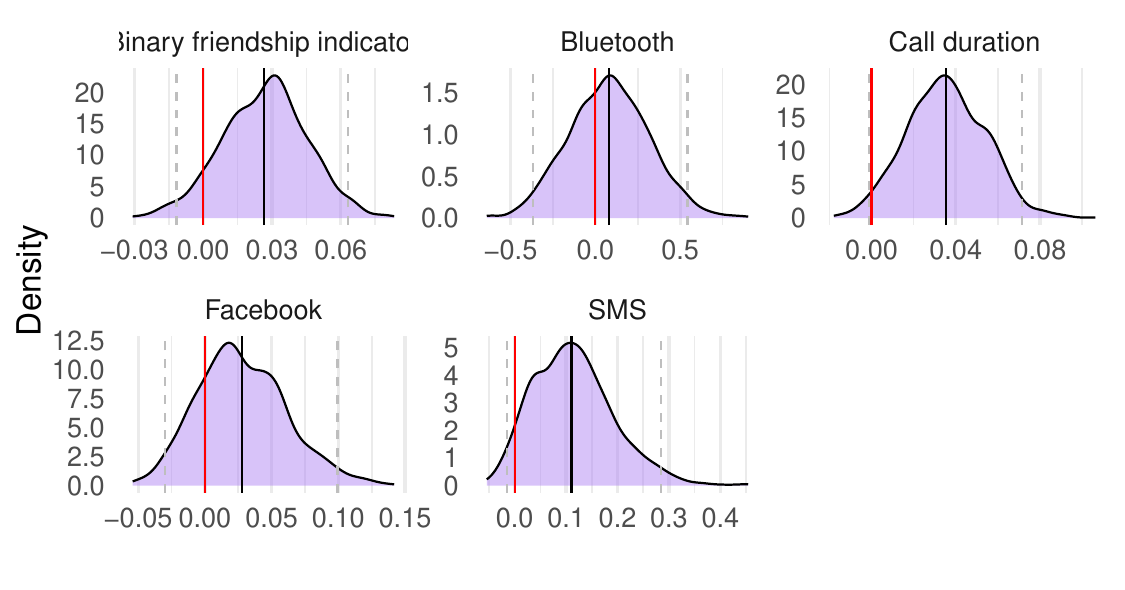} % Replace with your image file
    \caption{TA classrooms}
    \label{fig:bootstrap_ta}
  \end{subfigure}
    \caption{Distribution of estimated effect sizes of group membership across 1,000 random draws based on 2SLS estimation with re-centered group assignment instrument}
    \label{fig:app_bootstrap}
\end{figure}

In the main specification standard errors are computed using a dyad robust sandwich-type estimator \citep{samii_cluster-robust_2015}. To check the robustness of the estimates, we apply a bootstrap approach to estimate standard errors. To compute bootstraped standard errors, we draw 1,000 random samples with replacement from the set of of all study majors $S$ samples of size $|S|$ and estimate the regression coefficients based on the equations reported in the main manuscript. The estimated bootstrap standard error is then the standard deviation of the coefficients across all draws. The resulting distributions of the group membership coefficients are shown in Figure \ref{fig:app_bootstrap} and bootstrapped standard errors are reported in Table \ref{tab:iv_results_bootstrap}.

\input{tables_appendix/E/01_iv_results_bootstrap_se}

\subsection{Study program fixed effects}
\label{app:study_ff}
As yet another robustness check we also ran our analysis with additional fixed effects accounting for the study program.

\input{tables_appendix/E/02_iv_results_with_study_ff}

\subsection{Alternative bluetooth measures}
\label{app:alternative_bluetooth}
For our main specification we only count physical meetings happening outside of the university campus in order to avoid including any meetings that are initiated by the university e.g. tutoring sessions or organized social activities. As reported in Table \ref{tab:bluetooth_no_class} we only exclude interactions within classes (inferred based on the university's class schedule), we find even larger effects, significant at the 1\% level for the first two semesters. As expected, the same is true when analyzing the effect on across all types of physical meetings (see Table \ref{tab:bluetooth_all}).

\input{tables_appendix/E/03_bluetooth_alternative_no_class}
\input{tables_appendix/E/04_bluetooth_alternative_all}

\section{Results over time}
\label{app:dynamic}
\setcounter{table}{0}
\setcounter{figure}{0}
\input{tables_appendix/D/01_iv_results_dynamic}
\FloatBarrier

\section{Additional results on homophily patterns}
\label{app:homophily}
\setcounter{table}{0}
\setcounter{figure}{0}
In these additional analyses we examine interactions of group assignment with homophily effects, i.e. the tendency for links to form particularly among people who are alike in some relevant dimension.

For each pair in our data, we are able to construct measures of similarity in a range of different dimensions using our linked administrative data. For our main analysis, we focus on similarity in four dimensions: academic ability, parental wealth, parental education and sex. We measure academic ability using students GPA from the nationally standardized high school exam. Parental wealth is measured as the average joint wealth stated on the tax returns of both parents over the years 2004-2012, while parental education is measured as the years of schooling of the most highly educated parent. Our measure of sex stems from the official Danish population (birth) records and so is a binary measure of biological sex at birth. %More details on those variables can be found in Appendix \ref{app:admin_data}.

To simplify the exposition and maximize power, our main analysis converts all four homophily dimension to single binary measures of similarity. Specifically, for each of the continuous dimensions (GPA, parental education and parental wealth) we say that two students are \emph{similar} in this dimension if they are both above or both below the median within their study program. For sex, we simply say that students are similar if they are of the same sex.

For each of the four homophily dimensions and for both social groups and classrooms we then re-estimate a version of our baseline 2SLS specification which adds a dummy, $Similar_{ijs}$, for whether the student pair is similar along the given dimensions, as well as the corresponding interaction term with $SameGroup_{ijs}$ and the natural corresponding first stage. In all cases, we use our summary measure of whether the student pair has formed a link, $Link_{ijs}$, as the outcome variable:

\begin{align} 
\label{eq:homomain}
Link_{ijs} &= \beta_0 + \beta_1 \, SameGroup_{ijs} + \beta_2 \, Similar_{ijs} \\ \nonumber
&\quad + \beta_3 \, (SameGroup \cdot Similar)_{ijs} + v_{ijs}  \\[5pt]
\label{eq:homo1st1}
SameGroup_{ijs} &= \gamma_0 + \gamma_1 \, \widetilde{SameAssignedGroup}_{ijs} + \gamma_2 \, Similar_{ijs} \\ \nonumber
&\quad + \gamma_3 \, (\widetilde{SameAssignedGroup} \cdot Similar)_{ijs} + u_{ijs} \\[5pt] 
\label{eq:homo1st2}
(SameGroup \cdot Similar)_{ijs} &= \pi_0 + \pi_1 \, \widetilde{SameAssignedGroup}_{ijs} + \pi_2 \, Similar_{ijs} \\ \nonumber
&\quad + \pi_3 \, (\widetilde{SameAssignedGroup} \cdot Similar)_{ijs} + e_{ijs}
\end{align}

The coefficients of interests here are $\beta_1,\beta_2$ and $\beta_3$. $\beta_2$ measures the extent to which similar students are more likely to have formed a link and thus provides a descriptive measure of baseline homophily in our setting.\footnote{Students who are similar in one of the four dimensions we consider, are likely to be similar along many unobserved dimensions as well. As a result, $\beta_2$ should not be interpreted as the causal effect of similarity on link formation.} $\beta_1$ measures the causal effect of being in the same social group/classroom on the likelihood of forming a link for people who are not similar. Finally, $\beta_3$ measures the additional increase or decrease in the likelihood of link formation for individuals in the same social group/classroom, if individuals are similar in terms of the homophily dimension in question. Mechanically $\beta_3 \ne 0$ implies that our social foci interventions interact with homophily effects in shaping students final observed networks.

\begin{figure}[h]
    \centering
    \begin{subfigure}{0.45\textwidth}
        \centering
        \includegraphics[width=\linewidth]{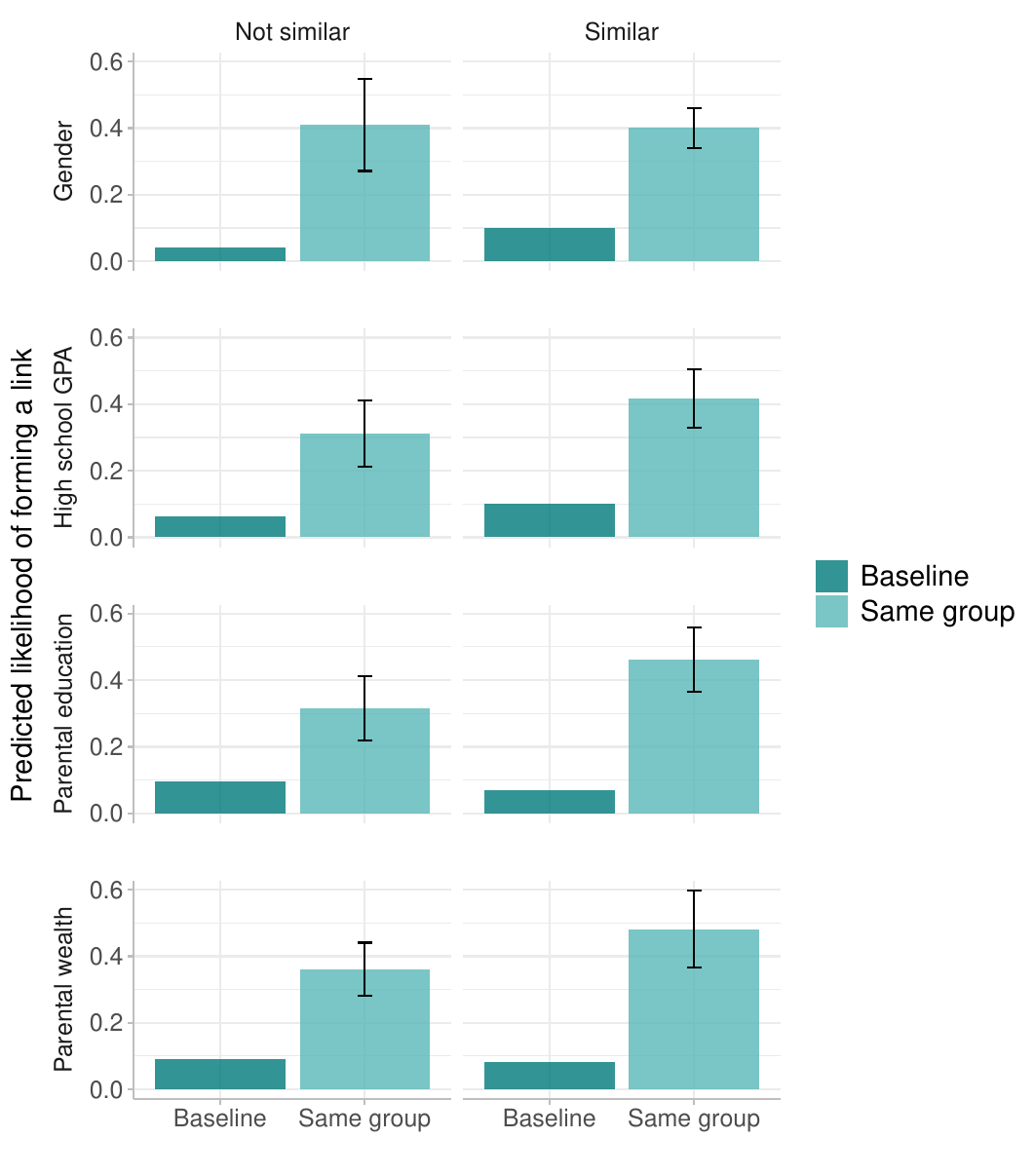}
        \caption{Social groups}
        \label{fig:hom_vec}
    \end{subfigure}
    \hfill
    \begin{subfigure}{0.45\textwidth}
        \centering
        \includegraphics[width=\linewidth]{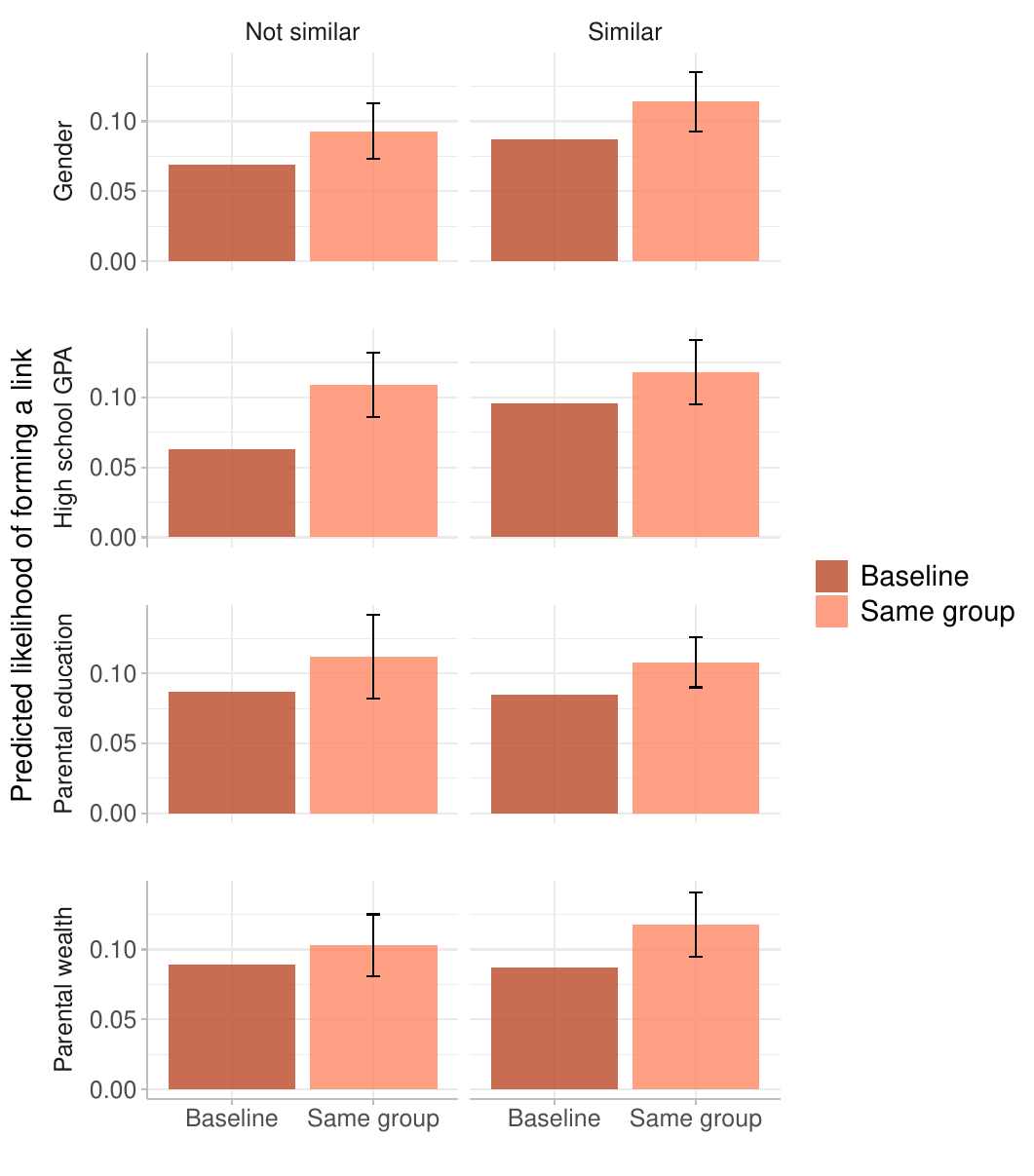}
        \caption{Classrooms}
        \label{fig:hom_ta}
    \end{subfigure}
    \caption{Likelihood of forming a link for non-similar pairs not sharing a group vs. non similar pairs sharing a group (left panel) and for similar pairs not sharing 
 a  group vs. similar pairs sharing a group (right panel), based on 2SLS with recentered instrument and interaction terms.}
    \label{fig:hom_results}
\end{figure}

Figure \ref{fig:hom_results} summarizes our 2SLS estimates regarding homophily interactions. Figure \ref{fig:hom_vec} shows results for social groups while Figure \ref{fig:hom_ta} shows results for classrooms.  The figures each contain four rows corresponding to the four homophily dimensions we consider. Within each row, the two bars on the left concerns individuals who are not similar on the relevant dimension, while the two bars on the right concern similar individuals. The bars then compare the predicted baseline probability of forming a link for pairs not in the same group ($\beta_0$ or $\beta_0+\beta_3$ respectively) vs the probability for pairs who are in the same group ($\beta0+\beta_1$ or $\beta_0+\beta_1+\beta_2+\beta_3$). Error bars reflect the standard error on the difference between same group and non-same group respectively. Table \ref{tab:iv_hom} shows the corresponding numeric regression results.

The results show little indication that membership of social groups or classrooms interact with homophily in our setting. Mirroring the overall results, being in the same group or classroom leads to a substantial increase in the likelihood of forming a link, both among students who are similar in some dimension and those who are not. Accordingly, looking at the numeric estimates in Table \ref{tab:iv_hom}, the coefficient on the interaction term between group membership and similarity is never statistically significant at conventional levels. At the same time, however, it is worth noting that many of the corresponding confidence intervals do allow for substantial interaction effects to exist.

One possible reason that we see little evidence of a homophily interaction in our setting may be that we also see relatively limited baseline homophily. Looking at the likelihood of forming a link among similar pairs compare to non-similar ones, significant baseline homophily is only observed for biological sex and academic ability. This may reflect that Danish society---and more specifically, our sample of first-year students at DTU---is relatively homogeneous in terms of socio-economic status (especially compared to most non-Scandinavian countries). This may make differences in parental wealth and education less noticeable to students in our setting.

\input{tables_appendix/H/01_iv_results_homophily}

\FloatBarrier

\section{SUGM - Standard Errors and specification with non-negative constraings}
\label{app:boostrap_triangle}
\subsection{Boostrapped standard errors}
\begin{figure}[h!]
    \centering
    % First subfigure
    \begin{subfigure}[b]{0.48\textwidth}  % Adjusted width
        \centering
        \includegraphics[width=\textwidth]{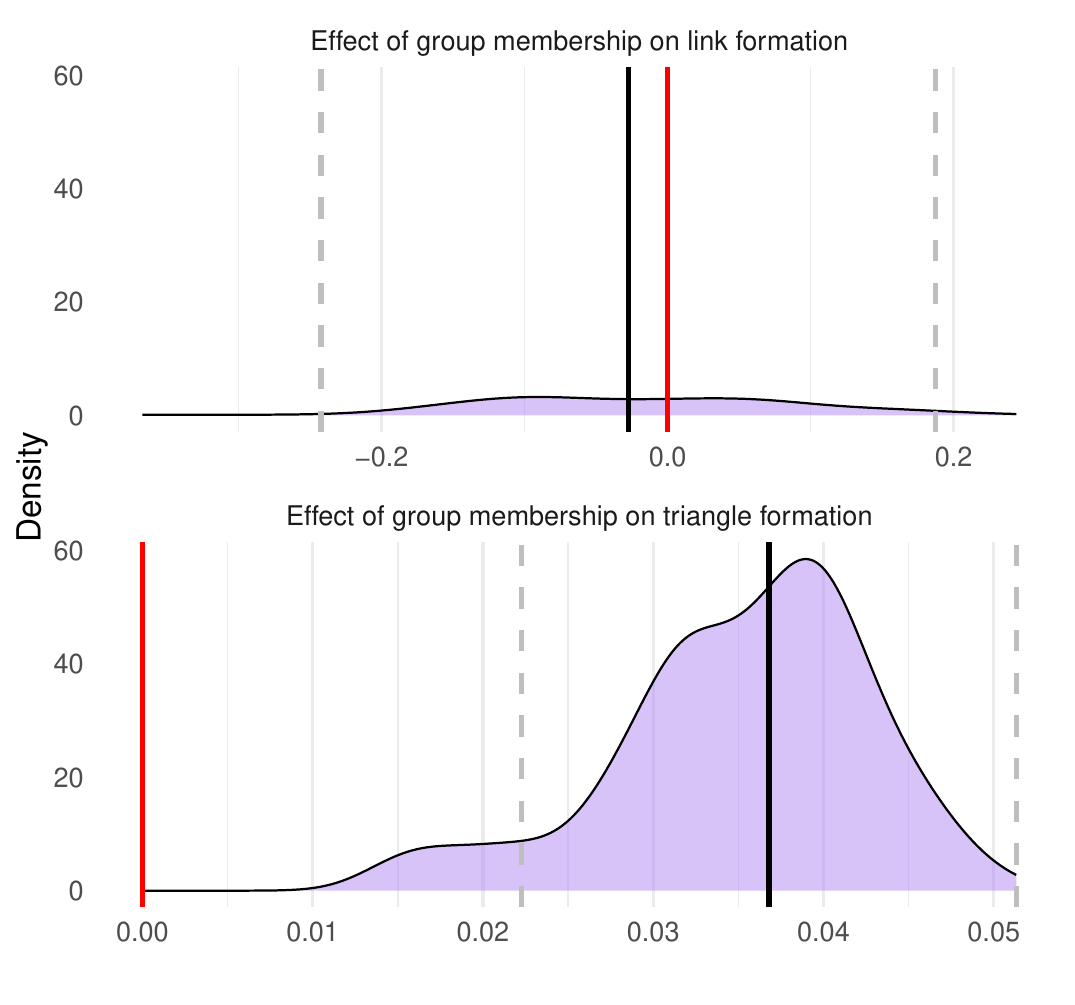}
        \caption{Social groups}
        \label{fig:bootstrap_triangle_vec}
    \end{subfigure}
    \hfill
    % Second subfigure
    \begin{subfigure}[b]{0.48\textwidth}  % Adjusted width
        \centering
        \includegraphics[width=\textwidth]{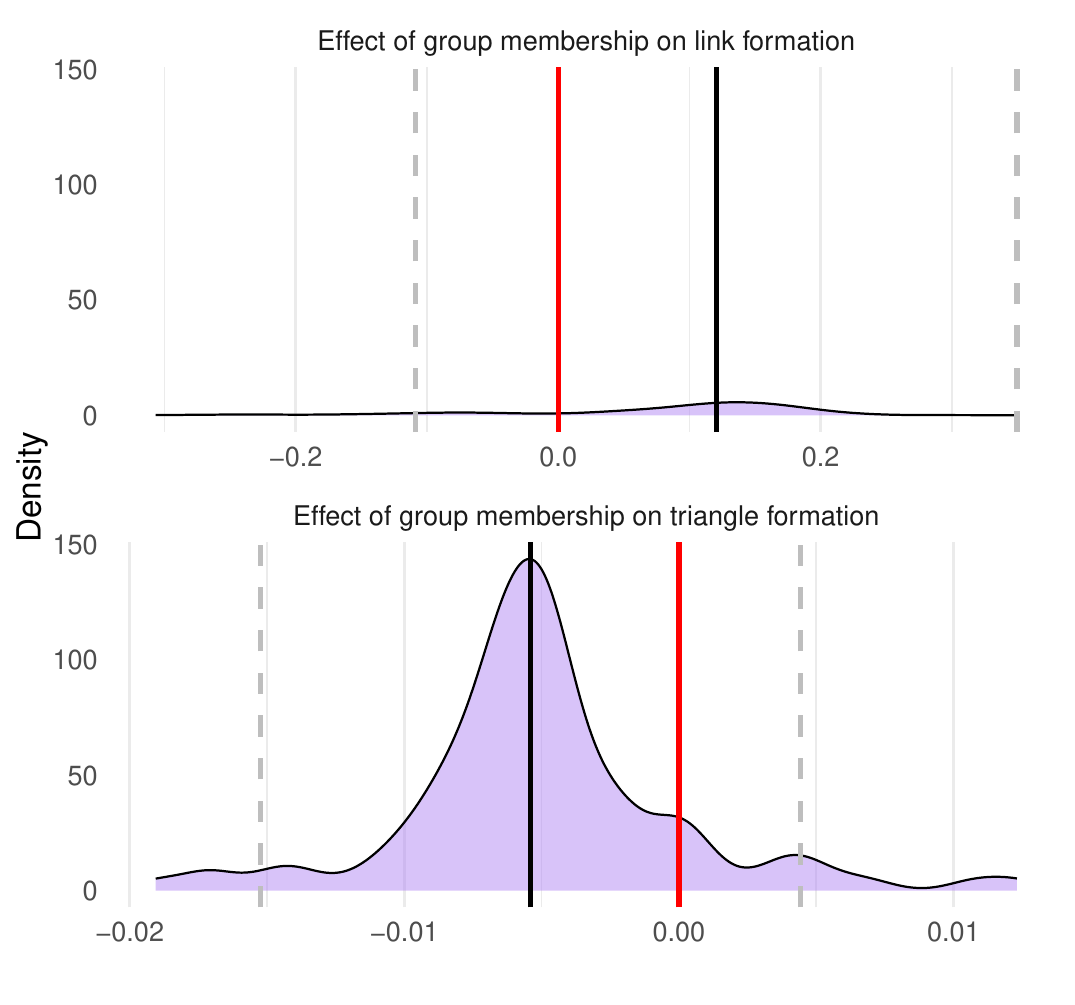}
        \caption{Classrooms}
        \label{fig:bootstrap_triangle_ta}
    \end{subfigure}
    
    \caption{Distribution of estimated effect sizes of group membership on link and triangle formation across 100 random draws based on the generalized non-linear least squares estimator for the SUGM.}
    \label{fig:app_bootstrap_triangle}
\end{figure}

Figures \ref{fig:bootstrap_triangle_vec} and \ref{fig:bootstrap_triangle_ta} show the distribution of the estimated effect of social groups or classroom membership on link and triad formation as estimated using the SUGM and a generalized non-linear least squares estimation procedure.

\subsection{SUGM Estimates with Triad Weighting}
As a robustness check, we re-estimate the model setting the triad weight to
$\omega_T = \frac{\sum_s |\Omega_s^L|}{\sum_s |\Omega_s^T|}$, the ratio of the
total number of dyads to the total number of triads in the estimation sample.
Because triads vastly outnumber dyads, the unweighted objective
($\omega_T = 1$) is dominated by the triad residuals. This choice of $\omega_T$
instead equalizes the aggregate contribution of the dyad and triad moments,
giving the link- and triangle-formation equations equal influence in estimation. In our sample $\omega_T = 0.23$.

Table \ref{tab:triad_results_omega} shows the results for the SUGM estimation with triad weighting.

\input{tables_appendix/I/02_triads_omega}

\subsection{SUGM Estimates with Linear and Quadratic Control Functions}

Our baseline estimates employ a second-order (quadratic) polynomial control
function. To assess sensitivity to this choice, we re-estimate the model using
the two alternative specifications discussed in the appendix: a linear control
function and a higher-order cubic control function.
Table~\ref{tab:triad_results_linear} reports the linear specification and
Table~\ref{tab:triad_results_cubic} the cubic specification. Across all three specifications, shared group membership has a positive and
statistically significant effect on triangle formation among social groups,
with $\theta_1$ ranging from roughly $0.03$ to $0.04$.  The estimated effect on direct link formation
($\beta_1$) is not statistically distinguishable from zero in any
specification; its point estimate is small and varies in sign across
specifications, and its large standard errors preclude any firm conclusion
about a direct dyadic channel.
The classroom estimates are likewise broadly consistent across specifications,
showing no robust effect on either link or triangle formation. The only
exception is a marginally significant effect on link formation in the linear
specification ($\beta_1 = 0.064$, significant at the 10\% level), which does not
persist in the quadratic or cubic specifications and which we therefore do not
interpret as evidence of a robust direct dyadic channel. 

\input{tables_appendix/I/03_triads_linear}

\input{tables_appendix/I/04_triads_cubic}

\subsection{SUGM Estimates with Non-Negativity Constraints}
\label{app:sugm_non_neg}

To ensure positive link and triad forming probabilities we re-estimate the SUGM model while constraining (1) the intercept in both equations, (2) the sum of the intercept and the same group coefficient in the dyad equation and (3) the sum of the intercept and three times the same group coefficient in the triangle equation to be between 0 and 1.

The resulting estimates are shown in Table \ref{tab:triad_results_non_neg}.

\input{tables_appendix/I/01_triads_non_neg}

\putbib[bibliography]
\end{bibunit}

%%%%%%%%%%%%%%%%%%%%%%%%%%%%%%%%%%%%%%%%%%%%%%
%% Bibliography:                            %%
%%%%%%%%%%%%%%%%%%%%%%%%%%%%%%%%%%%%%%%%%%%%%%
%% IMPORTANT: References in the bibliography should be complete, 
%% including the first and last names, and date of publication.

%% If your bibliography is in bibtex format, uncomment commands:

%% Or include bibliography directly:

% \begin{thebibliography}{}
% \bibitem{b1}
% \end{thebibliography}

\end{document}

%% file: tables/01_overview.tex
\begin{table}[ht]
\centering
\caption{Overview of the sample totals for the social group sample and the classroom sample} % Table title
\label{tab:overview}
\begin{tabularx}{\linewidth}{lYYY} % l = left, c = center, r = right alignment for columns
\hline % Horizontal line at the top
 & Social Group Sample& Classroom Sample  \\ % Column headers
\hline % Horizontal line below headers
Initial group assignment information available  & 616 & 1,372 \\
Corrected group assignment information available  & 606& 1,143 \\
Data on sex available  & 578 & 1,103\\
Participation in Copenhagen Network Study  & 223 &409\\
Part of a study major with min. 10 observations  & 171 &392\\

\hline % Horizontal line at the bottom

\end{tabularx}

\end{table}

%% file: tables/02_network_stats.tex
\begin{table}
\centering  
\small
\caption{Summary of network topology measures across all 20 study programs based on the composite social links measure and a cutoff at the 90\textsuperscript{th} percentile within in each study program. Each study program constitutes its own network as we only look at pairs within the same study program.}
\label{tab:network_stats}
\begin{threeparttable}[t]
\centering
%\rotatebox{90}{
%\adjustbox{width=\textheight}{
\begin{tabularx}{\linewidth}{lYYYYY}
\hline
& Min. & 25\textsuperscript{th} percentile & Median & 75\textsuperscript{th} percentile & Max.\\
\hline
Number of pairs & 66 &136 & 200 & 253 & 595 \\
Number of links & 7 & 14 & 20 & 26 & 60\\
Average degree & 2.33 & 3.29 & 3.90& 4.52 & 6.86 \\
Clustering coefficient & 0.00 & 0.31 & 0.41 & 0.50& 0.80\\
Diameter & 2 & 3 & 4&5&6\\
Average path length &1.30&1.71&2.26&2.36&2.74\\
\hline
$n=20$\\
\hline
\end{tabularx}
\end{threeparttable}

\end{table}

%% file: tables/03_iv_results.tex
\begin{table}
\centering  
\small
\begin{threeparttable}[t]
%\rotatebox{90}{
%\adjustbox{width=\textheight}{

 \caption{Group membership effects on social interactions: Results from 2SLS estimation based on re-centered group assignment instrument}
\label{tab:iv_results}
\begin{tabularx}{\linewidth}{lYYYYYY}
\hline                                        
                                   
Outcome variable &  Facebook & Weekly call duration & Weekly SMS count & Weekly physical meetings & Binary friendship indicator\\
& (1) & (2) & (3) &(4) &(5)\\
\hline                                               
\multicolumn{3}{l}{\emph{Panel A. Social Groups}} &&\\
Same group & 0.696***&0.290***&1.334**&2.928*** & 0.317***\\
&(0.063)&(0.096)&(0.529)&(1.052) & (0.067)\\[.5cm]
Intercept & 0.207***&0.022**&0.083**&1.775*** & 0.085***\\
&(0.029)&(0.010)&(0.035)&(0.304) & (0.015)\\ [.5cm]
Dyads &1,002 & 1,057 & 1,057 & 1,082 & 1,057\\
Students &156 & 160 & 160 & 162 & 160\\[.5cm]
1\textsuperscript{st} stage $F$& 1,311.75 & 1,539.29 & 1,539.29&1,511.12& 1,539.29\\                                    
\hline    
\multicolumn{3}{l}{\emph{Panel B. Classrooms}} \\
Same group & 0.028 & 0.035** & 0.110 &0.079 & 0.026*\\
&(0.030)& (0.017)&(0.069)&(0.209) & (0.015)\\ [.5cm]
Intercept & 0.274***&0.035***&0.101***&1.637*** & 0.083***\\
&(0.027)&(0.011)&(0.038)&(0.222) & (0.014)\\[.5cm]
Dyads & 3,468 & 3,459 & 3,459 & 3,459 & 3,491\\
Students & 362 & 260 & 260 & 262 & 364\\ [.5cm]
1\textsuperscript{st} stage $F$ & 2,667.30 &2,848.18&2,848.18&2,864.81,2,841.31\\

\hline
\end{tabularx}

  \begin{tablenotes}[para]

            \item \textit{Notes:
            This table shows instrumental variable estimates of the effect of being in the same group on various interaction outcomes during the second semester, i.e. spring term 2014.  The baseline mean
refers to the average level of interactions among pairs who don't share a group. The coefficients represent the effect of being in the same group. Group membership is instrumented by group assignment. Dyad robust standard errors are in parentheses. }\\
\item \textit{  * p $<$ .1, ** p $<$ .05, *** p $<$ .01.}
   
        \end{tablenotes}
%}
%}

\end{threeparttable}

  \end{table}

%% file: tables/04_triad_results.tex
\begin{table}
\centering  
\small
\begin{threeparttable}[t]
%\rotatebox{90}{
%\adjustbox{width=\textheight}{
 \caption{Control-function SUGM estimates of link and triad formation}
\label{tab:triad_results}
\begin{tabularx}{\linewidth}{lYYYY}
\hline
 & \multicolumn{2}{c}{Social groups} & \multicolumn{2}{c}{Classrooms} \\
 & Pairs & Triads & Pairs & Triads\\
 & (1) &(2)&(3)&(4)\\
 \hline
Same group ($\beta_1$ or $\theta_1$)& -0.027 & 0.037***& 0.120 & -0.005 \\
& (0.110)&(0.007) & (0.117)&(0.005)\\
Intercept ($\beta_0$ or $\theta_0$) & 0.031 & -0.004* & -0.007 & 0.012\\
&(0.036)&(0.003)&(0.078)&(0.010)\\

\hline
\end{tabularx}
%}
%}

  \begin{tablenotes}[para]

            \item \textit{Notes:
            This table shows the estimates of the effect of being in the same group on link and triad formation, based on nonlinear least-squares estimation of the parameters in the SUGM. The baseline mean
refers to the likelihood of a given pair (triad) forming a link (triangle) when none of the individuals share a group. The coefficient in the link equation represents the effect of a pair sharing a group on their likelihood of forming a link. The coefficient in the triangle equation represents the effect of a triad forming a triangle if one additional individual shares a group with any of the other members of the triad. Standard errors are based on 100 bootstrap samples.}\\
\item \textit{  * p $<$ .1, ** p $<$ .05, *** p $<$ .01.}
        \end{tablenotes}
\end{threeparttable}

  \end{table}

%% file: tables_appendix/A/01_balance.tex
\begin{table}[!htbp]  
\small
\centering

\begin{threeparttable}
\caption{Balance tests on pair characteristics for social groups and classroom samples}                  
\label{tab:balance_vec} 

\begin{tabular}{lcccc}

\hline                      
Variable \tnote{3} & Different social group & Same social group & s.e. & Pr( \textgreater  \textbar t \textbar )  \\   
\hline
\hline
Absolute number of pairs & 1,054 & 143 \\

\% both  immigration background &$0.001$ & $-0.002$ & $0.004$ & $0.532$ \\ 
\% both Danish & 0.865 & 0.875 & 0.028 & 0.731 \\

Mean $\Delta$ age & $1.204$ & $0.960$ & $0.137$ & $0.075$ \\               
Mean $\Delta$ High School GPA & $1.356$ & $1.256$ & $0.120$ & $0.404$ \\                                                     
Mean $\Delta$ parental income\tnote{1}& $506,753$ & $485,147$ & $30,436$ & $0.478$ \\                                                      
Mean $\Delta$ parental wealth\tnote{1}& $1,017,109$ & $1,045,095$ & $224,491$ & $0.984$ \\                                                 
Mean $\Delta$ parental education\tnote{2}  & $2.687$ & $2.646$ & $0.206$ & $0.842$ \\                                                         
                                                               N = 1,197\\
\hline    

Variable \tnote{3} & Different classroom & Same classroom & s.e. & Pr( \textgreater  \textbar t \textbar )  \\   
\hline
\hline
Absolute number of pairs & 2,056 & 2,008\\

\% both immigration background &0.005 & 0.006&0.0002&0.629\\
\% both Danish & 0.872&0.865&0.011&0.607\\

Mean $\Delta$ age & $2.153$&$2.236$& $0.0825$&$0.317$\\
Mean $\Delta$ High School GPA & $1.569$&$1.594$&$0.045$&$0.585$\\                      
Mean $\Delta$ parental income\tnote{1}& $453,818$&$441,050$&$13,560$&$0.346$\\                     
Mean $\Delta$ parental wealth\tnote{1}&$754,747$&$773,128$&$57,356$&$0.749$\\                             
Mean $\Delta$ parental education\tnote{2}  & $2.774$&$2.732$&$0.077$&$0.583$     \\                                 
                                                                N = 4,064\\
\hline

\end{tabular}

\begin{tablenotes}
      
        \item [1] Average of the yearly joint income (wealth) of mother and father between 2004 and 2012, in DKK.
        \item [2] Number of years spent in education by the parent with the longer time spent in education.
         \item [3] Since the randomisation of the social groups was conducted conditional on sex and dietary requirements, these were variables were also controlled for in the balance tests.
     \end{tablenotes}
\end{threeparttable}                   
\end{table}

%% file: tables_appendix/F/01_study_major_friends.tex
\begin{table}[t]
\caption{Average number of friends per individual within each study major when choosing a 90\% cutoff }
\label{tab:mean_friends}
\begin{tabularx}{\linewidth}{lYYYYY}  
\hline     
Study program & Study program  & Sample & Coverage & Mean \# of  & Overall mean \# \\ 
&size&size& in \%&friends in sample & of friends\\
\hline
BBYGDES & 57 & 13 &22.8 & 1.23 &5.40\\
BIOTEK & 61 & 21 & 34.4 & 2.00 & 5.81\\
BYGGETEK & 56 & 18 & 32.1 & 1.67 & 5.19 \\
BYGNING & 70 & 13 & 18.6 &1.23 & 6.63 \\
DES\&INN& 64 & 22 & 32.1 & 2.09 & 6.08\\
ELEKTRO & 55 &18 & 32.7 & 1.78 & 5.43 \\
ELEKTROTEK &61 & 34 & 55.7 & 3.24 & 5.80 \\
FYS\&NAN & 64 & 30 & 46.9 & 2.93 & 6.26 \\
GEOFYS & 46 & 22 & 47.8 & 2.09 & 4.37 \\

IT\&KOM & 46 & 17 & 37.0 & 1.65 & 4.46 \\
IT-RETNING& 71 & 23 &32.4 & 2.26 & 6.98 \\
KEMI\&BIO& 62 & 17 & 27.4 & 1.65 & 6.01 \\
KEMI\&TEK & 65 & 24 & 36.9 & 2.33 & 6.32 \\
MASKIN & 70 & 21 & 30.0 & 2.00& 6.67 \\
MAT\&TEK& 59 & 15 & 25.4 & 1.47 & 5.77 \\
MED\&TEK& 58 & 20 & 34.5 & 1.90 & 5.51 \\
MILJOTEK & 42 & 12 & 28.6 & 1.17 & 4.08 \\
PRO\&KON& 70 & 35 & 50.0 & 3.43 & 6.86 \\
SOFTWARE&60 & 26 & 43.3 & 2.54 & 5.86 \\
TEKBIO& 59 & 18 & 30.5 & 1.78 & 5.83 \\
\hline
\end{tabularx}

\end{table}

%% file: tables_appendix/F/02_number_of_friends_per_cutoff.tex
\begin{table}[t]
\small
\caption{Percentage of individuals in our data that have a given number of friends depending on the chosen cutoff value for the score of the composite interaction measure}
\label{tab:num_friends_cutoff}
\begin{tabularx}{\linewidth}{lYYYYYYYYYYYC{1cm}}  
\hline     
& \multicolumn{11}{c}{Number of friends}\\
Cutoff &  0&1&2&3&4&5&6&7&8&9&10& $>10$\\
\hline     
97.5\% & 62.02& 23.82 & 10.37 & 0.22 & 0.22 \\
95\%& 45.62 & 24.49 & 15.28 & 3.60 & 1.35 & 0.45 & 0.22 \\
90\%& 32.81 & 19.76 & 12.36 & 10.34 & 10.11 & 5.39 & 2.91 & 3.37 & 1.35 & 0.90 & 0.45 & 0.22\\
80\% & 24.72 & 11.91 & 10.11 & 6.52 & 5.39 & 8.09 & 6.74 & 6.07 & 6.07 & 4.04 & 2.02 & 8.31 \\
70\% & 19.55 & 8.31 & 7.19 & 5.62 & 4.27 & 4.72 & 5.62 & 4.94 & 5.84 & 6.97 & 5.17 & 21.80 \\
\hline
\end{tabularx}

\end{table}

%% file: tables_appendix/F/03_iv_binary_results.tex
\begin{table}
\small
\centering  
 \caption{Group membership effects on binary friendship measures based on different thresholds: Results from 2SLS estimation based on re-centered group assignment instrument}
\label{tab:iv_binary_pca}
\begin{threeparttable}[t]
%\rotatebox{90}{
%\adjustbox{width=\textheight}{
\begin{tabularx}{\linewidth}{lYYYYY}
\hline                                        

Threshold & 97.5\% & 95\% & 90\% & 80\% & 70\% \\
                                       
\hline       
\emph{Panel A. Social Groups} &&\\
Group membership & 0.174***&0.272***&0.317***&0.305***&0.232***\\
&(0.047)&(0.056)&(0.067)&(0.070)&(0.066)\\[.5cm]
Intercept & 0.014**&0.032***&0.085***&0.210***&0.334***\\
&(0.006)&(0.008)&(0.015)&(0.027)&(0.037)\\[.5cm]
Dyads & 1,057 & 1,057 & 1,057 &1,057&1,057\\
Students & 160 & 160 &160&160&160 \\
\hline
\rule{0pt}{4ex}    

\emph{Panel B. Classrooms}&&&&\\
Group membership &-0.006 & 0.004 & 0.026*&0.034*&0.034\\
&(0.008)&(0.011)&(0.015)&(0.019)&(0.023)\\[.5cm]
Intercept & 0.033***&0.049***&0.083***&0.173***&0.279**\\
&(0.008)&(0.009)&(0.014)&(0.021)&(0.028)\\[.5cm]
Dyads & 3,417  & 3,417 & 3,417 & 3,417& 3,417\\
Student & 360 & 360 & 360 &360&360\\

\hline

\end{tabularx}

  \begin{tablenotes}[para]

            \item \textit{Notes:
            This table shows instrumental variable estimates of the effect of being in the same group on our composite measure of friendship when using different thresholds for the transformation into a binary indicator. The baseline refers to the likelihood of a pair that does not share a group to form a link. The coefficients represent the effect of being in the same group on the likelihood of forming a link. Group membership is instrumented by group assignment. Dyad robust standard errors are in parentheses. }\\
\item \textit{  * p $<$ .1, ** p $<$ .05, *** p $<$ .01.}
   
        \end{tablenotes}
%}
%}

\end{threeparttable}

  \end{table}

%% file: tables_appendix/F/04_network_stats_per_program.tex
\begin{table}[t]
\caption{Network measures for each study program network using the composite interaction measure with a cutoff at the 90\textsuperscript{th} percentile per study program. }
\label{tab:detailed_network_stats}
\begin{tabularx}{\linewidth}{lYYYYYYY}  
\hline     
Study program & Pairs & Links &  Avg.degree & Clustering coefficient & Diameter & Avg. path length\\

\hline
BBYGDES & 78 & 8 & 2.462 & 0.000 & 5 & 2.357 \\
BIOTEK & 210 & 21 & 4.000 & 0.312 & 4 & 2.370 \\
BYGGETEK & 153 & 15  & 3.333 & 0.387 & 4 & 1.868 \\
BYGNING & 78 & 8 & 2.462 & 0.250  & 5 & 2.393 \\
DES\&INN& 231 & 23 & 4.182 & 0.222 &6 & 2.743 \\
ELEKTRO & 153 & 16 & 3.556 & 0.750 & 3 & 1.400\\
ELEKTROTEK & 561 & 55 & 6.471 & 0.443 & 5 & 2.362 \\
FYS\&NAN & 435 & 44 1 & 5.867 & 0.372 & 5 & 2.265 \\
GEOFYS & 231 & 23  & 4.182 & 0.527 & 3 & 1.709 \\

IT\&KOM & 136 & 14  & 3.294 &0.429&5&2.382\\
IT-RETNING& 253 & 26 & 4.522 & 0.250 & 4 & 2.275\\
KEMI\&BIO& 136 & 14  & 3.294 & 0.477 & 3 & 1.722\\
KEMI\&TEK & 276 & 28 & 4.667 & 0.303 &4 & 2.231 \\
MASKIN & 210 & 21 & 4.000& 0.488&4& 2.250 \\
MAT\&TEK& 105 & 11  & 2.933 & 0.375 & 5 & 2.378 \\
MED\&TEK& 190 & 19 &3.800&0.592&2&1.486\\
MILJOTEK & 66 & 6 & 2.333 & 0.800 & 2 & 1.300\\
PRO\&KON& 595 & 60& 6.857 & 0.500&4&2.090\\
SOFTWARE& 325 & 33 & 5.077 &0.360 & 5 & 2.331\\
TEKBIO& 153 & 16 & 3.556 & 0.783 & 3 & 1.655\\
\hline
\end{tabularx}

\end{table}

%% file: tables_appendix/C/01_all_first_stage.tex
\begin{table}
\centering  
\small
 \caption{First stage effects based on the recentered instrument for the social groups and the classroom samples, subset according to data availability of the respective outcome measures.}
\label{tab:first_stages}
\begin{threeparttable}[t]
%\rotatebox{90}{
%\adjustbox{width=\textheight}{
\begin{tabular}{p{4cm}C{2cm}C{2cm}C{2cm}}
\hline

Sample &  Facebook & Calls \& SMS  & Physical meetings \\
& (1) & (2) & (3) \\
\hline
\multicolumn{3}{l}{\emph{Panel A. Social Groups}}\\
Recentered  & 0.768***&0.781***&0.775***\\
random assignment&(0.055)&(0.053)&(0.053)\\[.5cm]
Intercept & 0.123*** & 0.120***&0.118***\\
&(0.006)&(0.006)&(0.006)\\[.5cm]
Dyads & 1,002 & 1,057 & 1,082\\
Student & 156 & 160 & 162\\[.5cm]
F-statistic & 1,311.75 & 1,539.29 & 1,511.12\\
                                        
\hline    
\multicolumn{3}{l}{\emph{Panel B. Classrooms}} \\
Recentered  & 0.619***&0.637***&0.635***\\
random assignment&(0.040)&(0.038)&(0.038)\\[.5cm]
Intercept & 0.677***&0.668***&0.670***\\
&(0.020)&(0.019)&(0.019)\\[.5cm]
Dyads & 3,468&3,459&3,491\\
Students & 362&362&364\\
F-statistic & 2,667.30&2,848.18&2,864.81\\

\hline

\end{tabular}
%}
%}

  \begin{tablenotes}[para]

            \item \textit{Notes:
            This table shows the first stage resulting from regressing the recentered random group assignment on the actual group membership. Note that the instrument can take on negative values and the coefficients should thus not be interpreted as increases in likelihood.
            Dyad robust standard errors are in parentheses. }\\
\item \textit{  * p $<$ .1, ** p $<$ .05, *** p $<$ .01.}
   
        \end{tablenotes}

\end{threeparttable}

  \end{table}

%% file: tables_appendix/C/02_reduced_form.tex
  \begin{table}
\small
\centering  
\begin{threeparttable}

\caption{Group membership effects on social interactions: Results from reduced-form regressions}
\label{tab:reduced_form}
\begin{tabularx}{\linewidth}{lYYYYY}%{p{1.7cm}C{1.7cm}C{1.7cm}C{1.7cm}C{1.7cm}C{1.7cm}}\hline                                           
\hline
Independent variable & Facebook & Weekly call duration & Weekly SMS count & Weekly physical meetings & Binary indicator\\
& (1) & (2) & (3) &(4) &(5) \\
                                
\hline       
\multicolumn{6}{l}{\emph{Panel A. Social Groups}}\\
Recentered  & 0.534*** &0.226***&
1.042***&
2.269***& 0.247***\\
random assignment & (0.048)&(0.072)&(0.421)&(0.841)&(0.052)\\[.5cm]
Intercept & 0.292*** &0.057***&0.243***&2.122***&0.122***\\
&(0.026)&(0.012)&(0.057)&(0.331)&(0.018)\\[.5cm]

Dyads & 1,002 & 1,057 & 1,057 & 1,082 & 1,057\\
Students & 156 & 160 & 160 &162 & 160\\

\hline
\multicolumn{6}{l}{\emph{Panel B. Classrooms}}\\

Recentered  & 0.017 & 0.023***&0.070&0.050 &0.017*\\
random assignment& (0.019)&(0.011)&(0.044)&(0.133)&(0.010)\\[.5cm]
Intercept & 0.293***&0.059***&0.175***&1.690***&0.100***\\
&(0.210)&(0.010)&(0.027)&(0.178)&(0.010)\\[.5cm]
Dyads & 3,468&3,459&3,459&3,491&3,417\\
Students &  362 &362&362&364&360\\
\hline

\end{tabularx}

   \begin{tablenotes}[para]

\item \textit{Notes:
This table shows the results from the reduced form regression based on the recentered \\ instrument. Dyad robust standard errors are in parentheses. }\\
\item \textit{  * p $<$ .1, ** p $<$ .05, *** p $<$ .01.}
   
        \end{tablenotes}

\end{threeparttable}

  \end{table}

%% file: tables_appendix/E/iv_results_alternative_ta.tex
\begin{table}
\centering  
\small
\begin{threeparttable}
 \caption{Group membership effects on social interactions: Results from 2SLS estimation based on re-centered group assignment instrument with dyad clustered standard errors and an alternative definition of being in the same TA classroom.}
\label{tab:iv_results_alternative}

\begin{tabularx}{\linewidth}{lYYYYY}

%{{p{1.7cm}C{1.7cm}C{1.7cm}C{1.7cm}C{1.7cm}C{1.7cm}}
\hline

Outcome variable &  Facebook & Weekly call duration & Weekly SMS count & Weekly physical meetings & Binary indicator\\
& (1) & (2) & (3) &(4) & (5) \\
                    
\hline    
\multicolumn{6}{l}{\emph{Panel B. Classrooms}} \\
Same group &-0.006&0.014&0.092&-0.167&0.000\\
&(0.027)&(0.025)&(0.111)&(0.261)&(0.016)\\
\rule{0pt}{4ex} 
Intercept & 0.245***&0.049***&0.126**&1.737***&0.098***\\
&(0.030)&(0.016)&(0.049)&(0.327)&(0.018)\\
\rule{0pt}{4ex} 
Dyads & 1,315 & 1,444&1,444&1,462&1,424\\
Students & 155&162&162&163&161\\
\hline

\end{tabularx}
%}
%}

  \begin{tablenotes}[para]

            \item \textit{Notes:
            This table shows instrumental variable estimates of the effect of being in the same group on various interaction outcomes during the second semester, i.e. spring term 2014.  The baseline mean
refers to the average level of interactions among pairs who don't share a group. The coefficients represent the effect of being in the same group. Group membership is instrumented by group assignment. Bootstrapped standard errors are in parentheses. }\\
\item \textit{  * p $<$ .1, ** p $<$ .05, *** p $<$ .01.}
   
        \end{tablenotes}
%}

\end{threeparttable}

  \end{table}

%% file: tables_appendix/E/01_iv_results_bootstrap_se.tex
\begin{table}
\centering  
\small
\begin{threeparttable}
 \caption{Group membership effects on social interactions: Results from 2SLS estimation based on re-centered group assignment instrument with bootstrapped standard errors}
\label{tab:iv_results_bootstrap}

\begin{tabularx}{\linewidth}{lYYYYY}

%{{p{1.7cm}C{1.7cm}C{1.7cm}C{1.7cm}C{1.7cm}C{1.7cm}}
\hline

Outcome variable &  Facebook & Weekly call duration & Weekly SMS count & Weekly physical meetings & Binary friendship indicator\\
& (1) & (2) & (3) &(4) &(5)\\
\hline                                        
      
\multicolumn{3}{l}{\emph{Panel A. Social Groups}} &&\\
Same group & 0.696*** &0.290*** & 1.334*** &2.928***&  0.317***\\
& (0.062) & (0.094)&(0.508)&(0.943)&(0.058)\\[.5cm]
Intercept & 0.207*** & 0.022*&0.083***&0.1775**&0.085***\\
&(0.035)&(0.012)&(0.031)&(0.355)&(0.010)\\[.5cm]
                           
\hline    
\multicolumn{3}{l}{\emph{Panel B. Classrooms}} \\
Same group &0.027 & 0.045* & 0.132& 0.212 & 0.033\\
&(0.033)&(0.018)&(0.078)&(0.238)&(0.019)\\[.5cm]
Intercept & 0.312***&0.045**&0.140**&1.751***&0.088***\\
&(0.037)&(0.014)&(0.049)&(0.242)&(0.014)\\
\hline

\end{tabularx}
%}
%}

  \begin{tablenotes}[para]

            \item \textit{Notes:
            This table shows instrumental variable estimates of the effect of being in the same group on various interaction outcomes during the second semester, i.e. spring term 2014.  The baseline mean
refers to the average level of interactions among pairs who don't share a group. The coefficients represent the effect of being in the same group. Group membership is instrumented by group assignment. Bootstrapped standard errors are in parentheses. }\\
\item \textit{  * p $<$ .1, ** p $<$ .05, *** p $<$ .01.}
   
        \end{tablenotes}
%}

\end{threeparttable}

  \end{table}

%% file: tables_appendix/E/02_iv_results_with_study_ff.tex
\begin{table}
\centering  
\small
\begin{threeparttable}[t]
 \caption{Group membership effects on social interactions: Results from 2SLS estimation based on re-centered group assignment instrument incl. study major fixed effects (fixed effect coefficients not shown).}

\label{tab:iv_results_ff}
%\rotatebox{90}{
%\adjustbox{width=\textheight}{
% \begin{tabular}{p{2cm}C{2cm}C{2cm}C{2cm}C{2cm}C{2cm}C{2cm}}
\begin{tabularx}{\linewidth}{lYYYYYY}
\hline

Outcome variable &  Facebook & Weekly call duration & Weekly SMS count & Weekly physical meetings & Binary friendship indicator\\
& (1) & (2) & (3) &(4) &(5)\\
\hline                                           
\multicolumn{3}{l}{\emph{Panel A. Social Groups}} &&\\

Same group & 0.694***&0.289***&1.328**&2.917***&0.316***\\
&(0.075)&(0.095)&(0.522)&(1.051)&(0.066)\\[.5cm]
Intercept & 0.359***&0.026&0.148&2.345**&0.094*\\
&(0.085)&(0.027)&(0.160)&(0.923)&(0.049)\\[.5cm]
Dyads &1,002&1,057&1,057&1,082&1,057\\
Students & 156 & 160 & 160 & 162&160\\

\hline    
\multicolumn{3}{l}{\emph{Panel B. Classrooms}} \\
Same group & 0.025 &0.035**&0.109&0.058&0.026*\\
&(0.027)&(0.017)&(0.069)&(0.209)&(0.015)\\[.5cm]
Intercept & 0.388***&0.043&0.098&1.094**&0.143***\\
&(0.117)&(0.041)&(0.093)&(0.463)&(0.049)\\[.5cm]
Dyads & 3,468&3,459&3,459&3,491&3,417\\
Students &362&362&362&364&360\\

\hline
\end{tabularx}

  \begin{tablenotes}[para]

            \item \textit{Notes:
            This table shows instrumental variable estimates of the effect of being in the same group on various interaction outcomes during the second semester, i.e. spring term 2014.  The baseline mean
refers to the average level of interactions among pairs who don't share a group. The coefficients represent the effect of being in the same group. Group membership is instrumented by group assignment. All regressions control for study program through a dummy variable. Dyad robust standard errors are in parentheses.}\\
\item \textit{  * p $<$ .1, ** p $<$ .05, *** p $<$ .01.}
   
        \end{tablenotes}
%}

\end{threeparttable}

  \end{table}

%% file: tables_appendix/E/03_bluetooth_alternative_no_class.tex
\begin{table}
\small
\centering  
\begin{threeparttable}
\centering
\captionsetup{justification=centering,singlelinecheck=false}

 \caption{Group membership effects on physical meetings outside of class over time: Results from two stage least squares estimation based on re-centered group assignment instrument.} 
\label{tab:bluetooth_no_class}

\begin{tabularx}{\linewidth}{lYYYY}
\hline                                        

Independent variable & \multicolumn{4}{c}{Average weekly interactions } \\
&Semester 1 & Semester 2 & Semester 3 & Semester 4 \\
& (1) & (2) & (3) &(4)  \\
\hline                                        
   
\multicolumn{5}{l}{\emph{Panel A. Social Groups}}\\
Same group &5.206***&5.030***&0.472 & -0.238\\
&(1.670)&(1.516)&(0.620)&(0.477)\\[.5cm]
Intercept & 7.818***&4.138***&2.259***&1.324***\\
&(0.757)&(0.576)&(0.435)&(0.335)\\[.5cm]
Dyads & 1,082 & 1,082 & 1,082 & 1,082\\
Students & 162 & 162 & 162& 162\\
\hline    
\multicolumn{5}{l}{\emph{Panel B. Classrooms}}\\
  
Same group & 0.564&0.046&0.078&0.549**\\
&(0.378)&(0.227)&(0.289)&(0.259)\\[.5cm]
Intercept & 7.228***&3.602***&1.961***&0.659***\\
&(0.584)&(0.406)&(0.301)&(0.155)\\[.5cm]
Dyads & 3,491 &3,491&3,491&3,491 \\
Students & 364&364& 364&364\\

\hline
\end{tabularx}

\begin{tablenotes}[flushleft]
\item \textit{Notes:
This table shows instrumental variable estimates of the effect of being in the same group on \textbf{average weekly physical interactions outside of class} over the four semesters of the study. The baseline mean
refers to the average level of interactions among pairs who don't share a group. The coefficients represent the effect of being in the same group. Group membership is instrumented by group assignment. Dyad robust standard errors are in parentheses.}\\
\item \textit{  * p $<$ .1, ** p $<$ .05, *** p $<$ .01.}
\end{tablenotes}
\end{threeparttable}
\end{table}

%% file: tables_appendix/E/04_bluetooth_alternative_all.tex
\begin{table}
\small
\centering  
\begin{threeparttable}
\centering
\captionsetup{justification=centering,singlelinecheck=false}

 \caption{Group membership effects on aggregate physical meetings: Results from two stage least squares estimation based on re-centered group assignment instrument}
 \label{tab:bluetooth_all}

\begin{tabularx}{\linewidth}{lYYYY}
\hline

Independent variable & \multicolumn{4}{c}{Average weekly interactions outside of class} \\
& Semester 1 & Semester 2 & Semester 3 & Semester 4 \\
& (1) & (2) & (3) &(4)  \\
                                     
\hline       
\emph{Panel A. Social Groups} &&\\
Same group &6.380***&7.791***&0.885&-0.474\\
&(2.302)&(2.340)&(0.989)&(0.775)\\[.5cm]
Intercept & 13.947***&9.168***&4.944***&2.913***\\
&(1.144)&(1.149)&(0.849)&(0.625)\\[.5cm]

Dyads & 1,082 & 10,82 & 1,082 & 1,082\\
Students & 162 & 162 & 162& 162\\
\hline   
\rule{0pt}{4ex}    

\emph{Panel B. Classrooms}&&&&\\
Same group &0.907&0.001&-0.063&0.648\\
&(0.561)&(0.283)&(0.349)&(0.404)\\[.5cm]
Intercept & 13.078***&7.631***&4.288***&1.688***\\
&(0.943)&0.722)&(0.605)&0.329)\\[.5cm]
Dyads & 3,491 &3,491&3,491&3,491 \\
Students & 364&364& 364&364\\

\hline
\end{tabularx}

\begin{tablenotes}[flushleft]
\item \textit{Notes:
This table shows instrumental variable estimates of the effect of being in the same group on \textbf{average weekly aggregate physical interactions} over the four semesters of the study. The baseline mean
refers to the average level of interactions among pairs who don't share a group. The coefficients represent the effect of being in the same group. Group membership is instrumented by group assignment. Dyad robust standard errors are in parentheses.}\\
\item \textit{  * p $<$ .1, ** p $<$ .05, *** p $<$ .01.}
\end{tablenotes}
\end{threeparttable}
\end{table}

%% file: tables_appendix/D/01_iv_results_dynamic.tex
  \begin{sidewaystable}
\small
\centering  
\resizebox*{\textheight}{!}{%
\begin{threeparttable}[t]
%\rotatebox{90}{
%\adjustbox{width=\textheight}{
 \caption{Group membership effects on social interactions over time: Results from two stage least squares estimation based on re-centered group assignment instrument}
\label{tab:iv_results_dynamic}
\begin{tabular}{lccccccccccccc}  
\hline                                        

Independent variable & Facebook & \multicolumn{4}{c}{Weekly call duration}  & \multicolumn{4}{c}{Weekly SMS count} & \multicolumn{4}{c}{Weekly physical meetings}\\
&&Semester 1 & Semester 2 & Semester 3 & Semester 4 &Semester 1 & Semester 2 & Semester 3 & Semester 4 & Semester 1 & Semester 2 & Semester 3 & Semester 4  \\
& (1) & (2) & (3) &(4) &(5) &(6) & (7) & (8)&(9)&(10) &(11)&(12)&(13) \\
\hline                                        
\hline       
\emph{Panel A. Social Groups} &&\\
Same group & 0.696***&0.220***&0.290***&0.142***&0.127***&1.453***&1.334**&0.466***&0.111&1.823**&2.928***&0.228&-0.201\\
&(0.063)&(0.082)&(0.096)&(0.047)&(0.044)&(0.515)&(0.529)&(0.139)&(0.086)&(0.814)&(1.052)&(0.423)&(0.306)\\[.5cm]
Intercept & 0.207***&0.035&0.022**&0.017**&0.012**&0.097***&0.083**&0.066***&0.059**&2.845***&1.775***&1.204***&0.682***\\
&(0.029)&(0.022)&(0.010)&(0.007)&(0.006)&(0.033)&(0.035)&(0.023)&(0.024)&(0.400)&(0.304)&(0.287)&(0.226)\\[.5cm]
Dyads & 1,002 & 1,057 & 1,057 &1,057&1,057&1,057&1,057&1,057&1,057&1,082&1,082&1,082&1,082\\
Student & 156 & 160 & 160 & 160 & 160 &160&160&160&160&162&162&162&162\\

\hline
\rule{0pt}{4ex}    

\emph{Panel B. Classrooms}&&&&& \\
Same group & 0.028 &0.033&0.035**&0.026&0.033***&-0.026&0.110&0.011&0.054&0.250&0.079&0.080&0.396**\\
&(0.030)&(0.029)&(0.017)&(0.017)&(0.013)&(0.092)&(0.069)&(0.039)&(0.031)&(0.247)&(0.209)&(0.247)&(0.199)\\[.5cm]
Intercept & 0.274*** & 0.037 & 0.035*** & 0.027** & 0.008 & 0.235*** & 0.101*** & 0.096*** & 0.030* & 2.715*** & 1.637*** & 0.993*** & 0.248*** \\

&(0.027)&(0.020)&(0.011)&(0.10)&(0.006)&(0.079)&(0.038)&(0.027)&(0.017)&(0.273)&(0.222)&(0.188)&(0.081)\\[.5cm]
Dyads &3,459&3,459&3,459&3,459&3,459&3,459&3,459&3,459&3,459&3,491&3,491&3,491&3,491\\
Students & 362&362&362&362&362&362&362&362&362&364&364&364&364\\

\hline

\end{tabular}
%}
%}

  \begin{tablenotes}[para]

            \item \textit{Notes:
            This table shows instrumental variable estimates of the effect of being in the same group on various interaction outcomes over the four semesters of the study. The baseline mean
refers to the average level of interactions among pairs who don't share a group. The coefficients represent the effect of being in the same group. Group membership is instrumented by group assignment. Dyad robust standard errors are in parentheses. }\\
\item \textit{  * p $<$ .1, ** p $<$ .05, *** p $<$ .01.}
   
        \end{tablenotes}
%}

\end{threeparttable}
}

\end{sidewaystable}

%% file: tables_appendix/H/01_iv_results_homophily.tex
\begin{sidewaystable}
\small
\centering  
\resizebox*{0.8\textheight}{!}{%
\begin{threeparttable}
\caption{Group membership and homophily effects on likelihood of link formation: Results from 2SLS estimation based on recentered group assignment instrument.}
\label{tab:iv_hom}

%\rotatebox{90}{
%\adjustbox{width=\textheight}{

\begin{tabular}{lcccc|cccc}  

\hline

Independent variable & \multicolumn{8}{c}{Binary friendship indicator - 90\% threshold} \\
\hline
& \multicolumn{4}{c|}{Social Groups } & \multicolumn{4}{c}{Classrooms}\\
\hline                                            

Same group &0.249**&0.271***&0.046**&0.014&0.025&0.024\\
&(0.100)&(0.080)&(0.097)&(0.138)&(0.023)&(0.022)&(0.030)&(0.024)\\[.3cm]
% high school GPA
Similar high School GPA & 0.037* &&&& 0.033\\
& (0.022)&&&&(0.023)\\
Similar high School GPA X Group membership & 0.068 &&&& -0.024\\
& (0.120) &&&& (0.032)\\[.3cm]
% wealth
Similar parental wealth  && -0.009 &&&&-0.002\\
&&(0.025) &&&&(0.028)\\
Similar parental wealth  X Group membership && 0.129 &&&& 0.017\\
&& (0.138) &&&& (0.032)\\[.3cm]
% education
Similar parental education level &&& -0.027 &&&&-0.002 \\
&&&(0.026) &&&& (0.028))\\
Similar parental education  level X Group Membership &&& 0.173 &&&&-0.002\\
& &&(0.138) &&&& (0.036)\\[.3cm]
%sex
Same sex &&&&0.059*** &&&& 0.018\\
&&&&(0.023) &&&&(0.020)\\
Same sex X Group membership &&&&-0.068 &&&&0.003\\
&&&&(0.126) &&&&(0.031)\\[.3cm]
%intercept
Intercept & 0.063***&0.090***&0.097***&0.040**&0.063***&0.089***&0.087***&0.069***\\
&(0.015)&(0.017)&(0.018)&(0.018)&(0.018)&(0.017)&(0.023)&(0.014)\\[.3cm]
Students & 153 & 154 & 153 & 160 & 328&341&342&360\\
Dyads & 960 & 978 & 956 & 1,057 & 2,842&3,090&3,100&3,417\\
\hline

\hline

\end{tabular}
%}
%}

  \begin{tablenotes}[para]

            \item \textit{Notes:
            This table shows instrumental variable estimates of the effect of being in the same group on the likelihood of forming a link as defined through our aggregate measure, using a cutoff at the 90\textsuperscript{th} percentile within each study program.  The baseline mean
refers to the probability of a pair that doesn't share a group and is not similar, to form a link. The coefficients with regards to the ``same group'' variable, represent the effect of being in the same group. The coefficients with regards to the ``similar'' terms, represent the effect of a pair either both having values above the median or both having values below the median of the respective dimension, or in the case of sex, of having the same sex. The interaction coefficients represent the added effect of being both similar and sharing a group. Group membership is instrumented by group assignment. Dyad robust standard errors are in parentheses. }\\
\item \textit{  * p $<$ .1, ** p $<$ .05, *** p $<$ .01.}
   
        \end{tablenotes}

\end{threeparttable}
}
\end{sidewaystable}

%% file: tables_appendix/I/02_triads_omega.tex
\begin{table}
\centering  
\small
\begin{threeparttable}[t]
%\rotatebox{90}{
%\adjustbox{width=\textheight}{
 \caption{Weighted control-function SUGM estimates of link and triad formation; $\omega_T=0.23$}
\label{tab:triad_results_omega}
\begin{tabularx}{\linewidth}{lYYYY}
\hline
 & \multicolumn{2}{c}{Social groups} & \multicolumn{2}{c}{Classrooms} \\
 & Pairs & Triads & Pairs & Triads\\
 & (1) &(2)&(3)&(4)\\
 \hline
Same group ($\beta_1$ or $\theta_1$)& -0.066 & 0.038***& 0.116 & -0.007 \\
& (0.122)&(0.008) & (0.116)&(0.006)\\
Intercept ($\beta_0$ or $\theta_0$) & 0.043 & -0.006** & -0.008 & 0.014\\
&(0.040)&(0.003)&(0.084)&(0.013)\\

\hline
\end{tabularx}
%}
%}

  \begin{tablenotes}[para]

            \item \textit{Notes:
            This table shows the estimates of the effect of being in the same group on link and triad formation, based on nonlinear least-squares estimation of the parameters in the SUGM. The baseline mean
refers to the likelihood of a given pair (triad) forming a link (triangle) when none of the individuals share a group. The coefficient in the link equation represents the effect of a pair sharing a group on their likelihood of forming a link. The coefficient in the triangle equation represents the effect of a triad forming a triangle if one additional individual shares a group with any of the other members of the triad. Standard errors are based on 100 bootstrap samples.}\\
\item \textit{  * p $<$ .1, ** p $<$ .05, *** p $<$ .01.}
        \end{tablenotes}
\end{threeparttable}

  \end{table}

%% file: tables_appendix/I/03_triads_linear.tex
\begin{table}
\centering  
\small
\begin{threeparttable}[t]
%\rotatebox{90}{
%\adjustbox{width=\textheight}{
 \caption{Control-function SUGM estimates of link and triad formation with linear control function}
\label{tab:triad_results_linear}
\begin{tabularx}{\linewidth}{lYYYY}
\hline
 & \multicolumn{2}{c}{Social groups} & \multicolumn{2}{c}{Classrooms} \\
 & Pairs & Triads & Pairs & Triads\\
 & (1) &(2)&(3)&(4)\\
 \hline
Same group ($\beta_1$ or $\theta_1$)& -0.047 & 0.0393***& 0.064* & -0.002 \\
& (0.108)&(0.009) & (0.036)&(0.002)\\
Intercept ($\beta_0$ or $\theta_0$) & 0.035 & -0.006** & 0.027 & 0.006\\
&(0.035)&(0.002)&(0.030)&(0.005)\\

\hline
\end{tabularx}
%}
%}

  \begin{tablenotes}[para]

            \item \textit{Notes:
            This table shows the estimates of the effect of being in the same group on link and triad formation, based on nonlinear least-squares estimation of the parameters in the SUGM. The baseline mean
refers to the likelihood of a given pair (triad) forming a link (triangle) when none of the individuals share a group. The coefficient in the link equation represents the effect of a pair sharing a group on their likelihood of forming a link. The coefficient in the triangle equation represents the effect of a triad forming a triangle if one additional individual shares a group with any of the other members of the triad. Standard errors are based on 100 bootstrap samples.}\\
\item \textit{  * p $<$ .1, ** p $<$ .05, *** p $<$ .01.}
        \end{tablenotes}
\end{threeparttable}

  \end{table}

%% file: tables_appendix/I/04_triads_cubic.tex
\begin{table}
\centering  
\small
\begin{threeparttable}[t]
%\rotatebox{90}{
%\adjustbox{width=\textheight}{
 \caption{Control-function SUGM estimates of link and triad formation with cubic control function}
\label{tab:triad_results_cubic}
\begin{tabularx}{\linewidth}{lYYYY}
\hline
 & \multicolumn{2}{c}{Social groups} & \multicolumn{2}{c}{Classrooms} \\
 & Pairs & Triads & Pairs & Triads\\
 & (1) &(2)&(3)&(4)\\
 \hline
Same group ($\beta_1$ or $\theta_1$)& 0.024 & 0.033**& 0.017 & -0.012 \\
& (0.172)&(0.013) & (0.153)&(0.010)\\
Intercept ($\beta_0$ or $\theta_0$) & 0.018 & -0.003 & 0.060 & 0.027\\
&(0.033)&(0.003)&(0.111)&(0.021)\\

\hline
\end{tabularx}
%}
%}

  \begin{tablenotes}[para]

            \item \textit{Notes:
            This table shows the estimates of the effect of being in the same group on link and triad formation, based on nonlinear least-squares estimation of the parameters in the SUGM. The baseline mean
refers to the likelihood of a given pair (triad) forming a link (triangle) when none of the individuals share a group. The coefficient in the link equation represents the effect of a pair sharing a group on their likelihood of forming a link. The coefficient in the triangle equation represents the effect of a triad forming a triangle if one additional individual shares a group with any of the other members of the triad. Standard errors are based on 100 bootstrap samples.}\\
\item \textit{  * p $<$ .1, ** p $<$ .05, *** p $<$ .01.}
        \end{tablenotes}
\end{threeparttable}

  \end{table}

%% file: tables_appendix/I/01_triads_non_neg.tex
\begin{table}
\centering  
\small
\begin{threeparttable}[t]
 \caption{Control-function SUGM estimates of link and triad formation with non-negativity constraints}
\label{tab:triad_results_non_neg}

%\rotatebox{90}{
%\adjustbox{width=\textheight}{
\begin{tabularx}{\linewidth}{lYYYY}
\hline
 & \multicolumn{2}{c}{Social groups} & \multicolumn{2}{c}{Classrooms} \\
 & Pairs & Triads & Pairs & Triads\\
 & (1) &(2) &(3)&(4)\\
 \hline
Same group ($\beta_1$ or $\theta_1$) & 0.057 &0.025*** &0.059*** & -0.002**\\ % & 0.028 & 0.000 \\
&(0.073)&(0.007) & (0.021)&(0.001)\\
Intercept ($\beta_0$ or $\theta_0$) & 0.016 & 1.33e-13 &0.0298 & 0.005**\\ %& 0.039&0.002\\
&(0.020)&(0.001) &(0.023)&(0.002)\\

\hline
\end{tabularx}
%}
%}

  \begin{tablenotes}[para]

            \item \textit{Notes:
            This table shows the estimates of the effect of being in the same group on link and triad formation, based on nonlinear least-squares estimation of the parameters in the SUGM with constraints. The baseline mean
refers to the likelihood of a given pair (triad) forming a link (triangle) when none of the individuals share a group. The coefficient in the link equation represents the effect of a pair sharing a group on their likelihood of forming a link. The coefficient in the triangle equation represents the effect of a triad forming a triangle if one additional individual shares a group with any of the other members of the triad. Standard errors are based on 100 bootstrap samples.}\\
\item \textit{  * p $<$ .1, ** p $<$ .05, *** p $<$ .01.}
   
        \end{tablenotes}
\end{threeparttable}

  \end{table}

%% file: template.bbl
\begin{thebibliography}{40}
\newcommand{\enquote}[1]{``#1''}
\expandafter\ifx\csname natexlab\endcsname\relax\def\natexlab#1{#1}\fi

\bibitem[\protect\citeauthoryear{Algan, Dalvit, Do, Le~Chapelain, and Zenou}{Algan et~al.}{2023}]{Algan2023}
\textsc{Algan, Yann, Nicolo Dalvit, Quoc-Anh Do, Alexis Le~Chapelain, and Yves Zenou} (2023): \enquote{Friendship Networks and Political Opinions: A Natural Experiment among Future French Politicians,} \emph{CESifo Working Ppars}.

\bibitem[\protect\citeauthoryear{Altmejd, Barrios-Fern{\'a}ndez, Drlje, Goodman, Hurwitz, Kovac, Mulhern, Neilson, and Smith}{Altmejd et~al.}{2021}]{altmejd2021brother}
\textsc{Altmejd, Adam, Andr{\'e}s Barrios-Fern{\'a}ndez, Marin Drlje, Joshua Goodman, Michael Hurwitz, Dejan Kovac, Christine Mulhern, Christopher Neilson, and Jonathan Smith} (2021): \enquote{O brother, where start thou? Sibling spillovers on college and major choice in four countries,} \emph{The Quarterly Journal of Economics}, 136 (3), 1831--1886.

\bibitem[\protect\citeauthoryear{Barrios-Fern{\'a}ndez}{Barrios-Fern{\'a}ndez}{2022}]{barrios2022neighbors}
\textsc{Barrios-Fern{\'a}ndez, Andr{\'e}s} (2022): \enquote{Neighbors' effects on university enrollment,} \emph{American Economic Journal: Applied Economics}, 14 (3), 30--60.

\bibitem[\protect\citeauthoryear{Bjerre-Nielsen}{Bjerre-Nielsen}{2020}]{bjerre2020assortative}
\textsc{Bjerre-Nielsen, Andreas} (2020): \enquote{Assortative matching with network spillovers,} \emph{Journal of Economic Theory}, 187, 105028.

\bibitem[\protect\citeauthoryear{Bjerre-Nielsen, Andersen, Minor, and Lassen}{Bjerre-Nielsen et~al.}{2020}]{bjerre2020negative}
\textsc{Bjerre-Nielsen, Andreas, Asger Andersen, Kelton Minor, and David~Dreyer Lassen} (2020): \enquote{The negative effect of smartphone use on academic performance may be overestimated: Evidence from a 2-year panel study,} \emph{Psychological Science}, 31 (11), 1351--1362.

\bibitem[\protect\citeauthoryear{Booij, Leuven, and Oosterbeek}{Booij et~al.}{2017}]{booij_ability_2017}
\textsc{Booij, Adam~S, Edwin Leuven, and Hessel Oosterbeek} (2017): \enquote{Ability {Peer} {Effects} in {University}: {Evidence} from a {Randomized} {Experiment},} \emph{The Review of Economic Studies}, 84 (2), 547--578.

\bibitem[\protect\citeauthoryear{Borusyak and Hull}{Borusyak and Hull}{2023}]{borusyak_nonrandom_2023}
\textsc{Borusyak, Kirill and Peter Hull} (2023): \enquote{Nonrandom {Exposure} to {Exogenous} {Shocks},} \emph{Econometrica}, 91 (6), 2155--2185, \_eprint: https://onlinelibrary.wiley.com/doi/pdf/10.3982/ECTA19367.

\bibitem[\protect\citeauthoryear{Bramoull{\'e}, Currarini, Jackson, Pin, and Rogers}{Bramoull{\'e} et~al.}{2012}]{bramoulle2012homophily}
\textsc{Bramoull{\'e}, Yann, Sergio Currarini, Matthew~O Jackson, Paolo Pin, and Brian~W Rogers} (2012): \enquote{Homophily and long-run integration in social networks,} \emph{Journal of Economic Theory}, 147 (5), 1754--1786.

\bibitem[\protect\citeauthoryear{Breza and Chandrasekhar}{Breza and Chandrasekhar}{2019}]{breza2019social}
\textsc{Breza, Emily and Arun~G Chandrasekhar} (2019): \enquote{Social networks, reputation, and commitment: evidence from a savings monitors experiment,} \emph{Econometrica}, 87 (1), 175--216.

\bibitem[\protect\citeauthoryear{Bursztyn and Jensen}{Bursztyn and Jensen}{2017}]{bursztyn_social_2017}
\textsc{Bursztyn, Leonardo and Robert Jensen} (2017): \enquote{Social {Image} and {Economic} {Behavior} in the {Field}: {Identifying}, {Understanding}, and {Shaping} {Social} {Pressure},} \emph{Annual Review of Economics}, 9 (Volume 9, 2017), 131--153, publisher: Annual Reviews.

\bibitem[\protect\citeauthoryear{Chandrasekhar and Jackson}{Chandrasekhar and Jackson}{2025}]{chandrasekhar_network_res}
\textsc{Chandrasekhar, Arun~G and Matthew~O Jackson} (2025): \enquote{A network formation model based on subgraphs,} \emph{Review of Economic Studies}, 92 (6), 3741--3787.

\bibitem[\protect\citeauthoryear{Duncan, Boisjoly, Kremer, Levy, and Eccles}{Duncan et~al.}{2005}]{duncan_peer_2005}
\textsc{Duncan, Greg~J., Johanne Boisjoly, Michael Kremer, Dan~M. Levy, and Jacque Eccles} (2005): \enquote{Peer {Effects} in {Drug} {Use} and {Sex} {Among} {College} {Students},} \emph{Journal of Abnormal Child Psychology}, 33 (3), 375--385.

\bibitem[\protect\citeauthoryear{Eagle, Pentland, and Lazer}{Eagle et~al.}{2009}]{eagle2009inferring}
\textsc{Eagle, Nathan, Alex Pentland, and David Lazer} (2009): \enquote{Inferring friendship network structure by using mobile phone data,} \emph{Proceedings of the national academy of sciences}, 106 (36), 15274--15278.

\bibitem[\protect\citeauthoryear{Easley and Kleinberg}{Easley and Kleinberg}{2010}]{easley_networks_2010}
\textsc{Easley, David and Jon Kleinberg} (2010): \emph{Networks, {Crowds}, and {Markets}: {Reasoning} about a {Highly} {Connected} {World}}, Cambridge: Cambridge University Press.

\bibitem[\protect\citeauthoryear{Epple and Romano}{Epple and Romano}{2011}]{epple_chapter_2011}
\textsc{Epple, Dennis and Richard~E. Romano} (2011): \enquote{Chapter 20 - {Peer} {Effects} in {Education}: {A} {Survey} of the {Theory} and {Evidence},} in \emph{Handbook of {Social} {Economics}}, ed. by Jess Benhabib, Alberto Bisin, and Matthew~O. Jackson, North-Holland, vol.~1, 1053--1163.

\bibitem[\protect\citeauthoryear{Feld}{Feld}{1981}]{feld_focused_1981}
\textsc{Feld, Scott~L.} (1981): \enquote{The {Focused} {Organization} of {Social} {Ties},} \emph{American Journal of Sociology}, 86 (5), 1015--1035, publisher: University of Chicago Press.

\bibitem[\protect\citeauthoryear{Goette, Huffman, and Meier}{Goette et~al.}{2012}]{goette2012impact}
\textsc{Goette, Lorenz, David Huffman, and Stephan Meier} (2012): \enquote{The impact of social ties on group interactions: Evidence from minimal groups and randomly assigned real groups,} \emph{American Economic Journal: Microeconomics}, 4 (1), 101--115.

\bibitem[\protect\citeauthoryear{Golub and Jackson}{Golub and Jackson}{2012}]{golub2012homophily}
\textsc{Golub, Benjamin and Matthew~O Jackson} (2012): \enquote{How homophily affects the speed of learning and best-response dynamics,} \emph{The Quarterly Journal of Economics}, 127 (3), 1287--1338.

\bibitem[\protect\citeauthoryear{Graham, Ridder, Thiemann, and Zamarro}{Graham et~al.}{2020}]{graham_teacher--classroom_2020}
\textsc{Graham, Bryan~S., Geert Ridder, Petra Thiemann, and Gema Zamarro} (2020): \enquote{Teacher-to-{Classroom} {Assignment} and {Student} {Achievement},} .

\bibitem[\protect\citeauthoryear{Griffith}{Griffith}{2024}]{griffith2024random}
\textsc{Griffith, Alan} (2024): \enquote{Random assignment with nonrandom peers: A structural approach to counterfactual treatment assessment,} \emph{Review of Economics and Statistics}, 106 (3), 859--871.

\bibitem[\protect\citeauthoryear{Hallinan and S{\o}rensen}{Hallinan and S{\o}rensen}{1985}]{hallinan1985ability}
\textsc{Hallinan, Maureen~T and Aage~B S{\o}rensen} (1985): \enquote{Ability grouping and student friendships,} \emph{American Educational Research Journal}, 22 (4), 485--499.

\bibitem[\protect\citeauthoryear{Harmon, Fisman, and Kamenica}{Harmon et~al.}{2019}]{harmon_peer_2019}
\textsc{Harmon, Nikolaj, Raymond Fisman, and Emir Kamenica} (2019): \enquote{Peer {Effects} in {Legislative} {Voting},} \emph{American Economic Journal: Applied Economics}, 11 (4), 156--180.

\bibitem[\protect\citeauthoryear{Imbens and Angrist}{Imbens and Angrist}{1994}]{ImbensAngrist1994}
\textsc{Imbens, Guido~W. and Joshua~D. Angrist} (1994): \enquote{Identification and Estimation of Local Average Treatment Effects,} \emph{Econometrica}, 62 (2), 467--475.

\bibitem[\protect\citeauthoryear{Jackson}{Jackson}{2008}]{jackson_social_2008}
\textsc{Jackson, Matthew~O.} (2008): \emph{Social and {Economic} {Networks}}, Princeton University Press.

\bibitem[\protect\citeauthoryear{Jackson, Rodriguez-Barraquer, and Tan}{Jackson et~al.}{2012}]{jackson2012social}
\textsc{Jackson, Matthew~O, Tomas Rodriguez-Barraquer, and Xu~Tan} (2012): \enquote{Social capital and social quilts: Network patterns of favor exchange,} \emph{American Economic Review}, 102 (5), 1857--1897.

\bibitem[\protect\citeauthoryear{Kossinets}{Kossinets}{2006}]{KOSSINETS2006247}
\textsc{Kossinets, Gueorgi} (2006): \enquote{Effects of missing data in social networks,} \emph{Social Networks}, 28 (3), 247--268.

\bibitem[\protect\citeauthoryear{Kossinets and Watts}{Kossinets and Watts}{2006}]{kossinets_empirical_2006}
\textsc{Kossinets, Gueorgi and Duncan~J. Watts} (2006): \enquote{Empirical {Analysis} of an {Evolving} {Social} {Network},} \emph{Science}, 311 (5757), 88--90, publisher: American Association for the Advancement of Science.

\bibitem[\protect\citeauthoryear{Kremer and Levy}{Kremer and Levy}{2003}]{kremer_peer_2003}
\textsc{Kremer, Michael and Dan~M. Levy} (2003): \enquote{Peer {Effects} and {Alcohol} {Use} {Among} {College} {Students},} .

\bibitem[\protect\citeauthoryear{Marmaros and Sacerdote}{Marmaros and Sacerdote}{2002}]{marmaros_peer_2002}
\textsc{Marmaros, David and Bruce Sacerdote} (2002): \enquote{Peer and social networks in job search,} \emph{European Economic Review}, 46 (4), 870--879.

\bibitem[\protect\citeauthoryear{Marmaros and Sacerdote}{Marmaros and Sacerdote}{2006}]{marmaros2006friendships}
---\hspace{-.1pt}---\hspace{-.1pt}--- (2006): \enquote{How do friendships form?} \emph{The Quarterly Journal of Economics}, 121 (1), 79--119.

\bibitem[\protect\citeauthoryear{McPherson, Smith-Lovin, and Cook}{McPherson et~al.}{2001}]{mcpherson_birds_2001}
\textsc{McPherson, Miller, Lynn Smith-Lovin, and James~M Cook} (2001): \enquote{Birds of a {Feather}: {Homophily} in {Social} {Networks},} \emph{Annual Review of Sociology}, 27 (1), 415--444, \_eprint: https://doi.org/10.1146/annurev.soc.27.1.415.

\bibitem[\protect\citeauthoryear{Mehta, Stinebrickner, and Stinebrickner}{Mehta et~al.}{2019}]{mehta_time-use_2019}
\textsc{Mehta, Nirav, Ralph Stinebrickner, and Todd Stinebrickner} (2019): \enquote{Time-{Use} and {Academic} {Peer} {Effects} in {College},} \emph{Economic Inquiry}, 57 (1), 162--171, \_eprint: https://onlinelibrary.wiley.com/doi/pdf/10.1111/ecin.12730.

\bibitem[\protect\citeauthoryear{Onnela, Saram{\"a}ki, Hyv{\"o}nen, Szab{\'o}, Lazer, Kaski, Kert{\'e}sz, and Barab{\'a}si}{Onnela et~al.}{2007}]{onnela2007structure}
\textsc{Onnela, J-P, Jari Saram{\"a}ki, Jorkki Hyv{\"o}nen, Gy{\"o}rgy Szab{\'o}, David Lazer, Kimmo Kaski, J{\'a}nos Kert{\'e}sz, and A-L Barab{\'a}si} (2007): \enquote{Structure and tie strengths in mobile communication networks,} \emph{Proceedings of the national academy of sciences}, 104 (18), 7332--7336.

\bibitem[\protect\citeauthoryear{Rivera, Soderstrom, and Uzzi}{Rivera et~al.}{2010}]{rivera2010dynamics}
\textsc{Rivera, Mark~T, Sara~B Soderstrom, and Brian Uzzi} (2010): \enquote{Dynamics of dyads in social networks: Assortative, relational, and proximity mechanisms,} \emph{annual Review of Sociology}, 36 (1), 91--115.

\bibitem[\protect\citeauthoryear{Sacerdote}{Sacerdote}{2001}]{sacerdote_peer_2001}
\textsc{Sacerdote, Bruce} (2001): \enquote{Peer {Effects} with {Random} {Assignment}: {Results} for {Dartmouth} {Roommates}*,} \emph{The Quarterly Journal of Economics}, 116 (2), 681--704.

\bibitem[\protect\citeauthoryear{Samii}{Samii}{2015}]{samii_cluster-robust_2015}
\textsc{Samii, Cyrus} (2015): \enquote{Cluster-{Robust} {Variance} {Estimation} for {Dyadic} {Data},} Artwork Size: 3670, 4784, 1699, 17157, 1005, 2546, 3749751, 5192296, 806, 4772, 2540, 1841, 1573, 1396 Pages: 3670, 4784, 1699, 17157, 1005, 2546, 3749751, 5192296, 806, 4772, 2540, 1841, 1573, 1396.

\bibitem[\protect\citeauthoryear{Sapiezynski, Stopczynski, Lassen, and Lehmann}{Sapiezynski et~al.}{2019}]{sapiezynski_interaction_2019}
\textsc{Sapiezynski, Piotr, Arkadiusz Stopczynski, David~Dreyer Lassen, and Sune Lehmann} (2019): \enquote{Interaction data from the {Copenhagen} {Networks} {Study},} \emph{Scientific Data}, 6 (1), 315, number: 1 Publisher: Nature Publishing Group.

\bibitem[\protect\citeauthoryear{Stadtfeld and Pentland}{Stadtfeld and Pentland}{2015}]{stadtfeld2015partnership}
\textsc{Stadtfeld, Christoph and Alex Pentland} (2015): \enquote{Partnership ties shape friendship networks: A dynamic social network study,} \emph{Social Forces}, 94 (1), 453--477.

\bibitem[\protect\citeauthoryear{Stopczynski, Sekara, Sapiezynski, Cuttone, Madsen, Larsen, and Lehmann}{Stopczynski et~al.}{2014}]{stopczynski2014measuring}
\textsc{Stopczynski, Arkadiusz, Vedran Sekara, Piotr Sapiezynski, Andrea Cuttone, Mette~My Madsen, Jakob~Eg Larsen, and Sune Lehmann} (2014): \enquote{Measuring large-scale social networks with high resolution,} \emph{PloS one}, 9 (4), e95978.

\bibitem[\protect\citeauthoryear{Watts and Strogatz}{Watts and Strogatz}{1998}]{Watts1998}
\textsc{Watts, Duncan~J. and Steven~H. Strogatz} (1998): \enquote{Collective dynamics of `small-world' networks,} \emph{Nature}, 393 (6684), 440--442.

\end{thebibliography}


\begin{thebibliography}{8}
\newcommand{\enquote}[1]{``#1''}
\expandafter\ifx\csname natexlab\endcsname\relax\def\natexlab#1{#1}\fi

\bibitem[\protect\citeauthoryear{Borusyak and Hull}{Borusyak and Hull}{2023}]{borusyak_nonrandom_2023}
\textsc{Borusyak, Kirill and Peter Hull} (2023): \enquote{Nonrandom {Exposure} to {Exogenous} {Shocks},} \emph{Econometrica}, 91 (6), 2155--2185, \_eprint: https://onlinelibrary.wiley.com/doi/pdf/10.3982/ECTA19367.

\bibitem[\protect\citeauthoryear{{Denmark Statistics}}{{Denmark Statistics}}{2025{\natexlab{a}}}]{DSTBANKAKT}
\textsc{{Denmark Statistics}} (2025{\natexlab{a}}): \enquote{{BANKAKT},} Accessed: 2025-03-09.

\bibitem[\protect\citeauthoryear{{Denmark Statistics}}{{Denmark Statistics}}{2025{\natexlab{b}}}]{DSTFAMBRUTTOINDK}
---\hspace{-.1pt}---\hspace{-.1pt}--- (2025{\natexlab{b}}): \enquote{FAMBRUTTOINDK,} Accessed: 2025-03-09.

\bibitem[\protect\citeauthoryear{{Denmark Statistics}}{{Denmark Statistics}}{2025{\natexlab{c}}}]{DSTOBLAKT}
---\hspace{-.1pt}---\hspace{-.1pt}--- (2025{\natexlab{c}}): \enquote{OBLAKT,} Accessed: 2025-03-09.

\bibitem[\protect\citeauthoryear{Samii}{Samii}{2015}]{samii_cluster-robust_2015}
\textsc{Samii, Cyrus} (2015): \enquote{Cluster-{Robust} {Variance} {Estimation} for {Dyadic} {Data},} Artwork Size: 3670, 4784, 1699, 17157, 1005, 2546, 3749751, 5192296, 806, 4772, 2540, 1841, 1573, 1396 Pages: 3670, 4784, 1699, 17157, 1005, 2546, 3749751, 5192296, 806, 4772, 2540, 1841, 1573, 1396.

\bibitem[\protect\citeauthoryear{Statistics}{Statistics}{2025}]{DSTKURSAKT}
\textsc{Statistics, Denmark} (2025): \enquote{KURSAKT,} Accessed: 2025-03-09.

\bibitem[\protect\citeauthoryear{Whillans, Christie, Cheung, Jordan, and Chen}{Whillans et~al.}{2017}]{whillans_misperception_2017}
\textsc{Whillans, Ashley~V., Chelsea~D. Christie, Sarah Cheung, Alexander~H. Jordan, and Frances~S. Chen} (2017): \enquote{From {Misperception} to {Social} {Connection}: {Correlates} and {Consequences} of {Overestimating} {Others}’ {Social} {Connectedness},} \emph{Personality and Social Psychology Bulletin}, 43 (12), 1696--1711, publisher: SAGE Publications Inc.

\bibitem[\protect\citeauthoryear{Wold, Esbensen, and Geladi}{Wold et~al.}{1987}]{wold_principal_1987}
\textsc{Wold, Svante, Kim Esbensen, and Paul Geladi} (1987): \enquote{Principal component analysis,} \emph{Chemometrics and Intelligent Laboratory Systems}, 2 (1), 37--52.

\end{thebibliography}
